\documentclass[sigconf, noacm]{acmart}

\usepackage{hyperref}
\usepackage{amsmath}
\usepackage{graphicx}
\usepackage{textcomp}
\usepackage{xcolor}
\usepackage{listings}
\usepackage[ruled, linesnumbered]{algorithm2e}
\usepackage{lipsum}
\usepackage{multirow}
\usepackage{booktabs}
\usepackage{threeparttable}
\usepackage{url}
\usepackage{fancyhdr}
\usepackage{indentfirst}
\usepackage{resizegather}
\usepackage{colortbl}
\usepackage{dblfloatfix}
\usepackage{makecell}
\usepackage{array}
\usepackage{enumitem}
\usepackage{hhline}
\usepackage{float}
\usepackage[binary-units]{siunitx}
\usepackage[export]{adjustbox}
\usepackage{subcaption}
\usepackage[hyphens]{xurl}
\usepackage{array}
\usepackage{soul}
\usepackage{tcolorbox}
\usepackage{xspace}
\usepackage{tikz}
\usepackage{tablefootnote}
\usepackage{tabularx}
\usepackage{booktabs}

\AtBeginDocument{%
  }

\definecolor{LightBlue}{rgb}{0.83, 0.91, 1}
\definecolor{Red}{rgb}{1, 0, 0}

\renewcommand{\small}{\fontsize{9pt}{9pt}\selectfont}

\setcopyright{none}
\renewcommand\footnotetextcopyrightpermission[1]{} 
\newcommand{\SPECO}{\textit{SPEC CPU17}\xspace}
\newcommand{\SPEC}{\textit{SPEC CPU26}\xspace}

\title{SPEC CPU2026: Characterization, Representativeness, and Cross-Suite Comparison}

\author{Ruihao Li \enskip Andrew Jacob \enskip Neeraja J. Yadwadkar \enskip Lizy K. John }
\affiliation{The University of Texas at Austin \country{}}

\begin{document}
\pagestyle{plain}

\begin{abstract}
Specialized accelerators dominate AI workloads, but CPUs remain critical for latency-sensitive workloads, agentic AI, and many other everyday services. 
Their performance therefore shapes end-to-end system efficiency, raising the question of whether the latest \textit{SPEC CPU} benchmarks change the architectural conclusions drawn from prior generations.
We comprehensively characterize \textit{SPEC CPU2026} across nine recent Intel, AMD, Ampere, and Nvidia platforms to understand how the new suite differs from its predecessors and what new capabilities it introduces. 
We observe that compared with \textit{SPEC CPU2017}, \textit{SPEC CPU2026} increases instruction volume, memory footprint, and instruction-cache stress. \par

We also compare \textit{SPEC CPU2026} with \textit{SPEC CPU2017}, recent datacenter and machine learning suites DCPerf and MLPerf, and agentic AI probes, using microarchitectural metrics to understand how specialized suites resemble and differ from SPEC CPU suites.
We note that \textit{SPEC CPU2026} remains a complementary general-purpose suite: closer to datacenter-like frontend pressure than prior CPU benchmark generations, yet less vector-intensive than MLPerf and less frontend-extreme than DCPerf.
Importantly, the expanded frontend envelope closely matches emerging CPU-centric agentic AI workloads: an agentic pipeline's features fall inside \textit{SPEC CPU2026}'s behavioral spread, making the suite a ready-made evaluation proxy for this fast-growing workload class.
Furthermore, case studies on page sizes and memory allocators, prefetching, compilers, ISA sensitivity, many-core scaling, and rolling round-robin (RRR) runs (new in \textit{SPEC CPU2026}) demonstrate the suite's utility beyond aggregate scores.
Finally, using cluster-based representativeness analysis, we identify subsets of 4-5 workloads per group that preserve 96.4-99.9\% of full-suite behavior.
Overall, \textit{SPEC CPU2026} updates the standardized general-purpose CPU baseline for the next decade of architecture evaluation.
\end{abstract}

\maketitle

\section{Introduction}
\label{section_introduction}
General-purpose CPUs remain indispensable: they run desktop and datacenter services~\cite{kanev2015profiling, sriraman2019softsku, zhao2023contiguitas, gonzalez2023profiling, nasr2025concorde}, coordinate heterogeneous accelerators~\cite{zhao2022understanding, um2023fastflow, jiang2025neo, na2025flexinfer, yu2024twinpilots}, execute latency-sensitive control and systems code~\cite{yahya2022agilewatts, antoniou2024agile, park2025ecocore}, and support workloads that map poorly to specialized hardware~\cite{jain2023optimizing, wang2024atrec, zahran2019heterogeneous, na2024understanding}.
Although the research community has focused on machine learning (ML) models~\cite{mattson2020mlperf, reddi2020mlperf, janapa2022mlperf, tschand2025mlperf}, and GPUs~\cite{rasley2020deepspeed, aminabadi2022deepspeed, kwon2023efficient, patel2024splitwise, shan2024guser, zhao2025insights, chu2025scaling}, FPGAs~\cite{zhang2015optimizing, alwani2016fused, shen2017maximizing, fowers2018configurable, chen2024understanding}, and custom domain-specific accelerators~\cite{jouppi2017datacenter, jouppi2023tpu, firoozshahian2023mtia, coburn2025meta, imani2019floatpim, chen2016eyeriss, xu2025wsc, ishida2020supernpu}, CPU benchmarking is not a ``solved'' problem or less relevant in an accelerator-dominated era.
Industry reports point to renewed CPU demand from agentic AI deployment~\cite{cpunews, armagi, raj2025cpu}.
With many emerging trends, rigorous CPU evaluation is increasingly important for the computer architecture community. \par

\textit{SPEC CPU} has long been the de facto standard for CPU performance evaluation in both industry and academia~\cite{phansalkar2007analysis, henning2006spec, spradling2007spec, ye2006performance, henning2002spec, phansalkar2005measuring,panda2018wait, limaye2018workload, bucek2018spec, singh2019memory, sr2019battle, navarro2019memory, schmitt2020performance, hebbar2019spec}.
There are more than 40,000 score reports on \SPECO and about 300 on \SPEC in the three months since release~\cite{spec}.
But ML, cloud/datacenter, and agentic AI workloads~\cite{hendrycks2021apps, zhuo2024bigcodebench, vu2024freshllms, khot2020qasc} have become mainstream, and suites such as MLPerf~\cite{reddi2020mlperf} and DCPerf~\cite{su2025dcperf} have emerged.
This raises a timely question: what does the newest \textit{SPEC CPU} release -- \textit{SPEC CPU2026} (\SPEC) -- contribute beyond \SPECO, and how should it be interpreted relative to modern domain-focused suites? \par

\textit{In this paper, we comprehensively characterize \SPEC, compare it to other suites, and draw implications for modern CPU architecture evaluation.}
We answer four questions that matter to the computer architecture community:
\begin{itemize}
\item How does \SPEC differ from \SPECO in microarchitectural behavior and coverage?
\item How does it compare with datacenter suite DCPerf and machine learning benchmark MLPerf? Does its coverage extend to agentic AI? Can adding a dense matrix-multiply kernel based benchmark reduce its distance to MLPerf?
\item What other evaluations can \SPEC support beyond score reporting?
\item Can compact subsets preserve \SPEC's behavioral diversity at lower evaluation cost?
\end{itemize}

\begin{table*}[t]
    \caption{\small{Hardware configurations of the nine machines used in the experiments (machines are either dual/single-socket systems, we report per-socket configurations in this table).}}
    \label{table_machine_configurations}
    \centering
    \footnotesize
    \setlength{\tabcolsep}{1pt}
    \renewcommand{\arraystretch}{1.1}
    \begin{tabular}{|| m{1.8cm} | m{1.5cm} | m{1.5cm} | m{2.2cm} | m{2.2cm} | m{1.4cm} | m{1.4cm} | m{1.4cm} | m{1.4cm} | m{2cm}
        ||}
        \hline
        Machines & CPU-A & CPU-B & CPU-C & CPU-D & CPU-E & CPU-F & CPU-G & CPU-H & CPU-I \\
         \hline \hline
        Vendor \& Model & Intel Skylake & Intel Icelake & Intel Sapphire Rapids & Intel Sapphire Rapids & AMD Milan & AMD Genoa & AMD Turin & Ampere Altra & Nvidia Grace \\
        \hline
        Release Year & 2017 & 2021 & 2023 & 2023 & 2021 & 2022 & 2024 & 2020 & 2023 \\
        \hline
        Core & Platinum 8160 & Platinum 8380 & Platinum 8468 & MAX 9480 & EPYC 7763 & EPYC 9454 & EPYC 9555 & Neoverse-N1 & Neoverse-V2 \\
        \hline
        \# of Cores & 24 & 40 & 48 & 56 & 64 & 48 & 64 & 80 & 72 \\
        \hline
        L1-D/I cache & 32/32KB & 48/32KB & 48/32KB & 48/32KB & 32/32KB & 32/32KB & 48/32KB & 64/64KB & 64/64KB \\
        \hline
        L2-cache & 1MB & 1.25MB & 2MB & 2MB & 512KB & 1MB & 1MB & 1MB & 1MB \\
        \hline
        L3-cache/core & 1.38MB & 1.5MB & 2.19 MB & 2MB & 4MB & 4MB & 4MB & 0.4MB & 1.63MB \\
        \hline
        \multirow{3}{*}{DIMM}
        & $6\times16$ GB & $8\times16$ GB & $8\times96$ GB & $4\times16$ GB & $8\times16$ GB & $12\times16$ GB & $12\times32$ GB & $8\times16$ GB & 224GB 32-Channel \\
        & DDR4 & DDR4 & DDR5 & HBM2e & DDR4 & DDR5 & DDR5 & DDR4 & LPDDR5X \\
        & 2666 MT/s & 3200 MT/s & 4800 MT/s & 3200 MT/s & 3200 MT/s & 4800 MT/s & 6000 MT/s & 3200 MT/s & 8533 MT/s \\
        \hline
    \end{tabular}
\end{table*}

Using microarchitectural characterization and cross-generation similarity analysis, we show that \SPEC updates the general-purpose CPU evaluation baseline: it provides meaningful new stress coverage over \SPECO while remaining complementary to DCPerf and MLPerf; and its scope already covers agentic AI orchestration. 
MLPerf is more dissimilar to \SPEC than DCPerf, and a dense matrix-multiply kernel narrows this distance for MLPerf's dense kernels.
\SPEC shifts stress toward larger instruction footprints, larger working sets, and higher many-core scalability, while reducing dependence on several extreme \SPECO outliers. Broader compiler and ISA sensitivity is also observed.
These changes matter because modern CPU research increasingly studies high-core-count scaling~\cite{stojkovic2023mumanycore, stojkovic2026dorado}, chiplet and non-chiplet organization~\cite{stojkovic2026accelflow, kim2026phaseweave, agarwal2026tina}, heterogeneous memory~\cite{sun2023demystifying, song2025hybridtier}, service co-location~\cite{chen2019parties, margaritov2019stretch},  compiler/runtime configuration~\cite{zhang2022ocolos, bhuiyan2026wax}, and agentic AI~\cite{cpunews, armagi, raj2025cpu}.

In summary, this paper makes the following contributions:
\begin{itemize}
\item We provide the first detailed characterization of \SPEC across nine modern platforms, showing how it differs from \SPECO in all four suite groups (\S~\ref{section_overview}).
\item We compare \SPEC with \SPECO, DCPerf, and MLPerf (\S~\ref{section_compare}): \SPEC narrows the gap to datacenter workloads through greater frontend pressure while remaining distinct from extreme service and ML inference behaviors; it already captures agentic AI orchestration, and a dense matrix-multiply kernel mitigates the dissimilarity to MLPerf's dense kernels.
\item We demonstrate the use cases of \SPEC for circa 2026 CPU research (\S~\ref{section_case}): compared to \SPECO, \SPEC sustains higher many-core scaling, is more sensitive to compiler optimizations, shows broader ISA-dependent variation, and enables proxy workload generation via its Rolling Round-Robin (RRR) mode.
\item We perform redundancy analysis and representative subset selection (\S~\ref{section_subset}), showing that 4-5 workloads per group preserve 96.4-99.9\% of suite behavior.
\end{itemize}

Overall, \SPEC is a timely suite that updates standardized, comparable CPU evaluation for the next decade.

\section{Overview of \SPEC}
\label{section_overview}
This section summarizes \SPEC's organization, instruction volume, and high-level performance behavior.

\subsection{Suite Overview of \SPEC}
\label{section_benchmark_overview}

Following \SPECO, \SPEC is organized into four categories -- INT Rate, INT Speed, FP Rate, and FP Speed -- with 14, 13, 12, and 13 benchmarks, respectively, totaling 52 benchmarks, 9 more than \SPECO. 
It also retains C, C++, and Fortran. 
However, all \SPEC workloads were developed from scratch rather than inherited from previous suites. 
Counting Rate and Speed variants as a single workload, only 3 of \SPEC's 19 integer workloads preserve \SPECO names (\textit{omnetpp}, \textit{gcc}, and \textit{xz}), and 6 of its 19 floating-point workloads do (\textit{cactus}, \textit{fotonik3d}, \textit{lbm}, \textit{nab}, \textit{namd}, and \textit{roms}).
Despite sharing these names, the corresponding \SPEC workloads use entirely new codebases and inputs. 
Recent work~\cite{madhav2026spec} describes the development process of \SPEC, whereas this paper provides a deeper characterization of the suite across multiple modern processor architectures. \par

\begin{table}[t]
    \caption{\small{Performance metrics used in our analysis.}}
    \label{table_perf_metrics}
    \footnotesize
    \setlength{\tabcolsep}{1pt}
    \renewcommand{\arraystretch}{1.1}
    \begin{tabular}{|| m{1.8cm} | m{6cm} 
        ||}
        \hline
        Category & Perf Metrics \\
         \hline \hline
         Inst/Cycle & IPC \\
         \hline
         Cache & L1I\$ MPKI, L1D\$ MPKI, L2\$ MPKI, L3\$ MPKI \\
         \hline
         TLB & L1 iTLB MPMI, L1 dTLB MPMI, L2 TLB MPMI \\
         \hline
         Branch Predictor & Branch MPKI \\
         \hline
         Pipeline & Frontend stall \%, Backend stall \%  \\
         \hline
         Inst Mix & Kernel\%, User\%, Load\%, Store\%, Branch\%, FP\%, Vector\% \\
         \hline
         DRAM & Mem access (bytes per cycle) \\
         \hline
\end{tabular}
\end{table}

\begin{figure}[t]
    \centerline{\includegraphics[width=\columnwidth, trim = 1mm 1mm 1mm 1mm, page=1, clip=true]{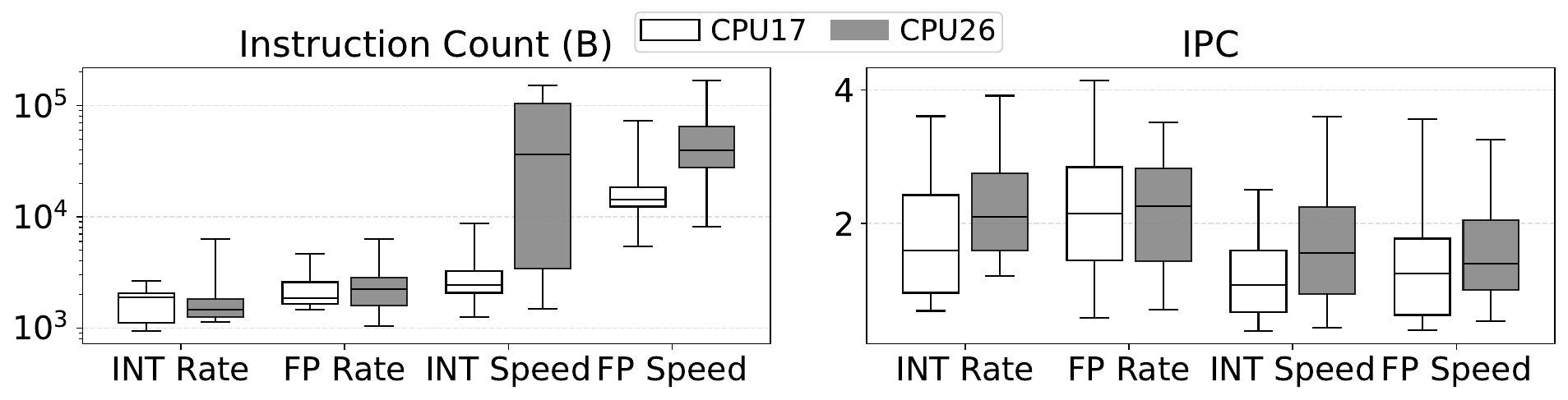}}
    \caption{\small{Per-workload dynamic instruction count (log scale) and IPC distributions for \SPECO and \SPEC across INT and FP, Speed and Rate workloads, pooled over all nine platforms. \SPEC increases total instruction count by $23.2\times$ for INT Speed and $3.4\times$ for FP Speed, while Rate footprints grow modestly.}}
	\label{fig_counter_overview}
\end{figure}

\subsection{Performance Characterization}
\label{section_performance_characterization}
\noindent{\textbf{Platforms and metrics.}}
We evaluate \SPEC on nine systems (Table~\ref{table_machine_configurations}) spanning \texttt{x86\_64} and \texttt{AArch64} ISAs, recent Intel, AMD, Ampere, and Nvidia processors, and diverse cache, core-count, and memory configurations. 
All benchmarks are compiled with gcc -O3, and hardware counters are collected with Linux perf~\cite{perf}.
All systems use their default deployment configurations, including dynamic frequency scaling, since our analysis relies primarily on cycle- and instruction-normalized metrics.
Table~\ref{table_perf_metrics} lists our used 19 metrics, covering IPC, cache/TLB behavior, branch prediction, frontend/backend stalls, instruction mix, and DRAM access intensity.
This feature space follows established architectural characterization practice~\cite{phansalkar2007analysis, panda2018wait, su2025dcperf, kanev2015profiling, sriraman2019softsku} and captures the major frontend, execution, memory-system, and control-flow bottleneck dimensions of general-purpose CPU workloads. \par

\begin{figure*}[t]  \centerline{\includegraphics[width=2.1\columnwidth, trim = 2mm 2mm 2mm 2mm, page=1, clip=true]{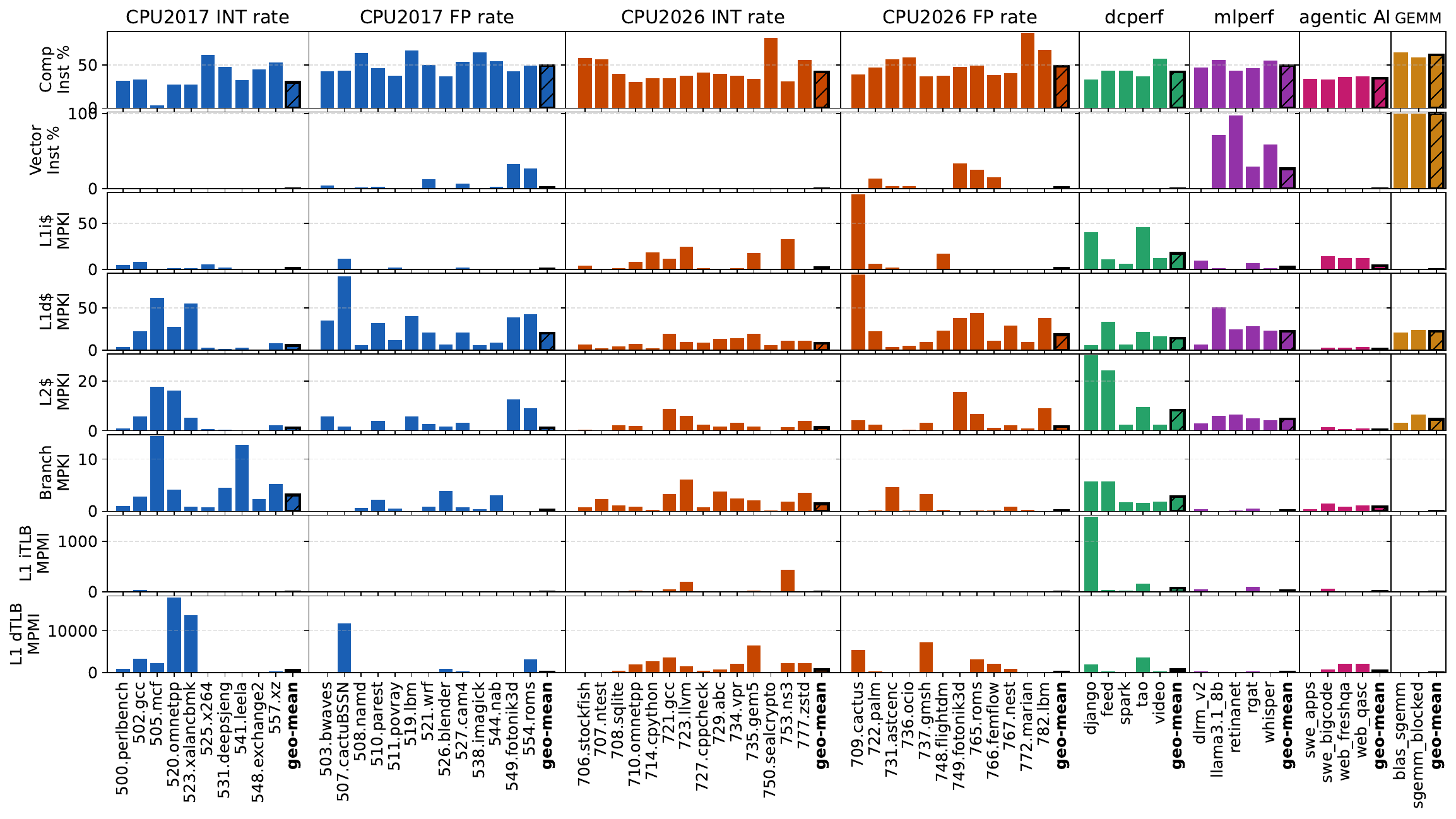}}
    \caption{\small{Comparison of key metrics between \SPEC, \SPECO, MLPerf, DCPerf, an agentic AI suite, and a GEMM kernel; each per-workload value is the median across the nine machines in Table~\ref{table_machine_configurations}. 
    \SPEC offers broader coverage than \SPECO and the agentic AI suite, while DCPerf places far more stress on the frontend (L1I\$ and iTLB) and MLPerf on vector instructions -- with the GEMM kernel further out still, at the vector-throughput extreme. }}
    \label{fig_overview_uarch}
\end{figure*}

Figure~\ref{fig_counter_overview} shows the dynamic instruction count and IPC distributions (per-workload detail in Appendix Table~\ref{table_spec_overview}) for \SPEC and \SPECO across the four suite groups, aggregated across the nine platforms.
Relative to \SPECO, \SPEC increases total dynamic instruction count by $1.61\times$, $1.03\times$, $23.21\times$, and $3.38\times$ for INT Rate, FP Rate, INT Speed, and FP Speed, respectively.
This growth reflects modern processor scaling, especially higher core counts in Speed runs~\cite{papazian2020new, nassif2022sapphire, soltis2023next, evers2022amd, bhargava2024amd, singh2025zen, ampere, grace}, but also increases detailed simulation cost~\cite{panda2018wait, binkert2011gem5, perelman2003using, sherwood2002automatically, gottschall2023balancing, karandikar2018firesim}.
Median IPC also rises in every group (e.g., $1.60\rightarrow2.10$ for INT Rate). \par

\noindent{\underline{\textbf{Takeaway:}}}
\SPEC substantially increases dynamic instruction volume, especially for Speed workloads, increasing simulation and evaluation cost.

\section{\SPEC vs. \SPECO, DCPerf, MLPerf, Agentic AI, GEMM}
\label{section_compare}
This section compares \SPEC with its predecessor \SPECO and two production-driven suites, DCPerf and MLPerf (Table~\ref{table_benchmarks}).
DCPerf targets hyperscale datacenter services with large instruction footprints and high frontend pressure~\cite{su2025dcperf}, whereas MLPerf Inference emphasizes ML-centric tensor/vector and floating-point behavior under standardized model, dataset, and accuracy rules~\cite{reddi2020mlperf}. 
But how similar are these suites to \SPEC?
Since \SPECO primarily consists of standalone general-purpose CPU programs, we explore whether \SPEC tilts this general-purpose nature toward modern datacenter-adjacent behavior, while remaining distinct from domain-specific suites such as DCPerf and MLPerf.
To probe the edges of that scope, we add two widely used agentic AI pipeline benchmarks~\cite{raj2025cpu} (an SWE agent on APPS/BigCodeBench tasks~\cite{hendrycks2021apps, zhuo2024bigcodebench} and a web-augmented agent on FreshQA/QASC queries~\cite{vu2024freshllms, khot2020qasc}), motivated by the CPU demand~\cite{cpunews, armagi, raj2025cpu}. 
We also a dense fp32 GEMM ($8192^3$), run as both a hand-written cache-blocked kernel~\cite{goto2008anatomy} and OpenBLAS~\cite{openblas}, in some ways similar to matrix300 from \textit{SPEC89}.
For the agentic pipeline we profile the \textit{orchestration} rather than the LLM serving backend (Llama-3.1-8B~\cite{grattafiori2024llama}), since MLPerf's \textit{llama3.1\_8b} already represents LLM inference in this comparison.
We analyze detailed microarchitectural metrics (\S~\ref{section_03_01}) and workload-level similarity (\S~\ref{section_03_02}). \par
 
\begin{table}[t]
    \caption{\small{Benchmarks used in our analysis.}}
    \label{table_benchmarks}
    \centering
    \footnotesize
    \setlength{\tabcolsep}{1pt}
    \renewcommand{\arraystretch}{1.1}
    \begin{tabular}{|| m{1.8cm} | m{6cm} 
        ||}
        \hline
        Category & Workloads \\
         \hline \hline
         \SPECO & Workloads indicated by numbers 5xx in Figure~\ref{fig_overview_uarch}\\
         \hline
         \SPEC & Workloads indicated by numbers 7xx in Figure~\ref{fig_overview_uarch}\\
         \hline
         DCPerf & django, feed, spark, tao, video \\
         \hline
         MLPerf & dlrm\_v2, llama3.1\_8b, retinanet, rgat, whisper \\
         \hline
         Agentic AI & swe\_apps, swe\_bigcode, web\_freshqa, web\_qasc \\
         \hline
         GEMMs & blas\_sgemm, sgemm\_blocked \\
         \hline
\end{tabular}
\end{table}

\subsection{Detailed Microarchitectural Analysis}
\label{section_03_01}
We first compare \SPEC, \SPECO, DCPerf, MLPerf, the agentic AI suite, and GEMM on raw performance metrics/features.
Figure~\ref{fig_overview_uarch} summarizes compute intensity, cache and branch behavior, and address-translation pressure for every workload of the six groups; per-workload medians across all nine machines are robust to machine-specific outliers. \par

\begin{figure*}[t]
    \centerline{\includegraphics[width=2.1\columnwidth, trim = 0mm 0mm 0mm 0mm, page=1, clip=true]{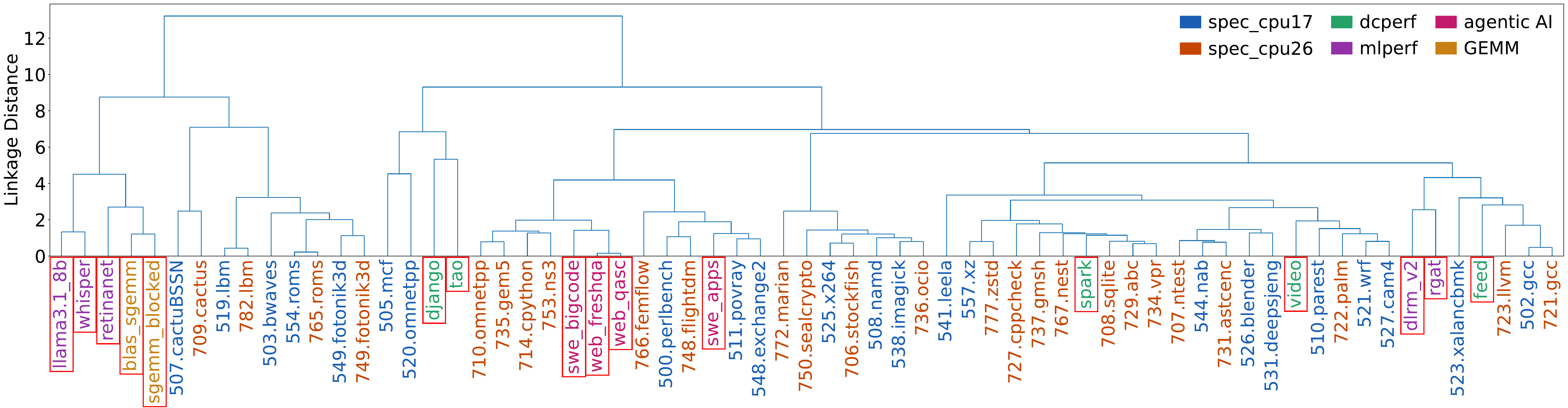}}
	\caption{\small{Dendrogram over \SPECO/\SPEC Rate, DCPerf, MLPerf, an agentic AI suite, and a dense GEMM kernel; red boxes mark the non-\textit{SPEC} workloads.
    DCPerf and MLPerf remain behaviorally distinct from \SPECO/\SPEC; 
    the agentic AI workloads land within \SPEC's clusters, while GEMM merges into MLPerf's dense-inference cluster (leftmost in the figure).}}
	\label{fig_dendrogram_all}
\end{figure*}

\noindent{\textit{Compute and vector intensity:}}
For INT Rate, \SPEC has higher fraction of non-load/store/branch instructions over \SPECO (41.83\% vs.\ 29.77\%, or $1.40\times$), matching DCPerf (42.04\%); FP Rate shows a similar compute-intensity range across all three suites ($\sim$49\%).
MLPerf, however, remains clearly distinct in vector instruction stress: its vector share reaches 26.48\%, versus 1.72\% and 1.81\% in \SPECO and \SPEC FP Rate (and effectively zero in both INT Rate suites).
The GEMM kernels form the extreme of that same axis, with the highest compute-instruction share (61.36\%) and the lowest branch (0.013 MPKI) and instruction-side pressure of any group measured.
The agentic AI suite sits at the other end, with effectively no vector work and a compute-instruction share (34.81\%) below every group except \SPECO INT Rate.
As expected, \SPEC is not primarily a vector-throughput stress test; its main value lies in frontend, cache, control-flow, and translation behavior studies. \par

\noindent{\textit{Frontend and translation pressure:}}
The largest differences between \SPECO and \SPEC appear in instruction-cache and translation behavior.
For L1I\$ MPKI, \SPEC rises by $2.11\times$ over \SPECO in INT Rate and $2.69\times$ in FP Rate, substantially increasing frontend pressure while remaining well below DCPerf (17.41 MPKI).
Still, \SPEC bridges the DCPerf gap more closely than \SPECO: 709.cactus\_r reaches L1I\$ stress comparable to some DCPerf workloads, while remaining distinct on service-style dimensions (\S~\ref{section_03_02}).
L1 iTLB pressure grows faster, from 2.51 to 7.99 MPMI in INT Rate ($3.18\times$) and from 0.99 to 1.98 in FP Rate ($2.00\times$), and L1 dTLB pressure rises by $1.42\times$ in INT Rate and $2.32\times$ in FP Rate.
The agentic AI suite lands between \SPEC and DCPerf on the instruction side (L1I\$ 4.02 MPKI vs.\ 2.26 for \SPEC INT Rate and 17.41 for DCPerf; L1 iTLB 14.69 MPMI vs.\ 7.99 and 76.26), while its data-side pressure is the lowest measured (L1D\$ 1.52 MPKI, L2\$ 0.50 MPKI) -- a Python/HTTP orchestration profile rather than a datacenter-service one~\cite{raj2025cpu, hendrycks2021apps, zhuo2024bigcodebench, vu2024freshllms, khot2020qasc}.
\SPEC thus expands frontend and translation stress relative to \SPECO, though DCPerf's instruction-side stresses (L1I\$ and iTLB) remain far larger. \par

\noindent{\textit{Cache and branch behavior:}}
Data-cache changes are more moderate.
L1D\$ MPKI increases by $1.45\times$ in INT Rate but is essentially unchanged in FP Rate (19.83 vs.\ 18.31), while L2\$ MPKI rises modestly, from 1.20 to 1.32 in INT Rate and from 1.22 to 1.66 in FP Rate.
Both \textit{SPEC} suites exhibit far lower L2\$ pressure than DCPerf (8.38) and MLPerf (4.75) at the suite level.
GEMM, despite its minimal instruction-side footprint, matches MLPerf's data-cache intensity (L1D\$ 22.48 vs.\ 22.29 MPKI; L2\$ 4.60 vs.\ 4.75): dense blocked math streams data while exercising little else.
In contrast, branch pressure decreases: \SPEC INT reduces branch MPKI from 3.12 to 1.43 ($0.46\times$ of \SPECO), and FP Rate also decreases slightly. \par

\noindent\underline{\textbf{Takeaway:}}
(1) Compared with \SPECO, \SPEC broadens general-purpose CPU coverage primarily by increasing instruction-cache, frontend, and translation pressure while substantially reducing branch-miss pressure.
(2) \SPEC remains complementary to domain-specific suites: \SPEC captures some DCPerf-like characteristics without reproducing DCPerf's extreme hyperscale instruction-cache and translation demands, and remains distinct from MLPerf's highly parallel inference behavior.
\begin{figure*}[t]
    \centerline{\includegraphics[width=2.1\columnwidth, trim = 2mm 2mm 2mm 2mm, page=1, clip=true]{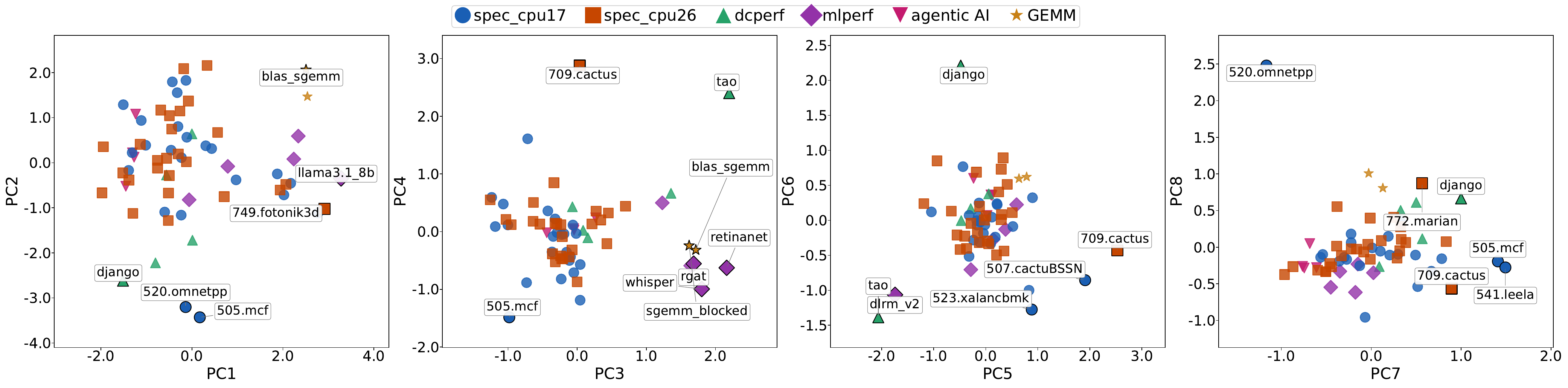}}
	\caption{
    \small{Principal Component (PC) comparison across six benchmark suites. The top 8 PCs capture 80.9\% of total variability; outlier and unique workloads plot far from the central cluster.
    Notably, 505.mcf\_r is an outlier in \SPECO and is not included in \SPEC.
    DCPerf (tao, django) exhibits unique signatures on PC3/PC4 and PC6; MLPerf workloads are primarily distinguished by PC3, with the GEMM kernels alongside them toward the far right of PC3.
    }}
	\label{fig_overview_pc}
\end{figure*}

\subsection{Similarity Analysis}
\label{section_03_02}
\noindent\textbf{{Similarity analysis methodology.}}
The per-metric comparison treats each dimension independently; we next quantify holistic workload similarity using clustering and principal component analysis~\cite{phansalkar2007analysis, panda2018wait, prokopec2019renaissance}.
First, we normalize all performance metrics so differences reflect workload behavior rather than metric scale or machine-specific magnitude.
Second, we apply Principal Component Analysis (PCA) to the cross-platform feature matrix, where each of the 19 metrics on each of the nine platforms is a distinct feature (171 dimensions before reduction).
PCA removes correlated dimensions while preserving dominant behavioral variation~\cite{phansalkar2007analysis, panda2018wait, prokopec2019renaissance}.
Finally, we apply hierarchical clustering in the PCA-reduced space and use linkage distance to quantify workload similarity and guide representative-subset selection. \par

Clustering the \SPEC Rate, \SPECO Rate, DCPerf, MLPerf, agentic AI, and GEMM workload features yields the dendrogram in Figure~\ref{fig_dendrogram_all}.
DCPerf workloads have linkage distances of 1-2 to several \SPEC workloads, indicating that \SPEC captures some datacenter-relevant microarchitectural properties within its broader general-purpose scope, while DCPerf still represents service-specific extremes.
In contrast, MLPerf workloads remain far from both \SPEC and \SPECO (linkage distance $\ge 3$, up to 8), reflecting a different execution regime dominated by vector-heavy, data-parallel inference kernels.
The MLPerf linkage distance is not a coverage failure; it indicates \SPEC's scope as a general-purpose CPU suite rather than a specialized accelerator-oriented inference suite. 
The agentic AI workloads fall inside \SPEC's cluster, joining 710.omnetpp\_r, 735.gem5\_r, 714.cpython\_r, and 753.ns3\_r at linkage 2.0; its distance to the nearest \SPEC workload (0.9-1.6) lies within \SPEC's internal nearest-neighbor spread (median 1.0), so \SPEC already covers this behavior. 
Mixed \textit{CPU17/CPU26} clusters reflect overlap in general-purpose behavior, whereas isolated locations indicate domain-specific or outlier characteristics. 
Overall, \SPEC approaches some DCPerf workload behaviors while remaining distinct from the MLPerf suite.\par

The GEMM kernel probes another boundary: would adding a dense matrix multiply -- similar to a matrix benchmark \textit{SPEC89}~\cite{dixit1991spec} carried until 1991 -- bring \SPEC closer to MLPerf?
Only partly.
The kernel itself is MLPerf-like -- in Figure~\ref{fig_dendrogram_all} it merges into the dense-inference branch, nearer to \textit{retinanet} (2.38) than to any \SPEC workload (3.72) -- so including it cuts the suite's distance to MLPerf's dense kernels from 3.58 to 3.09.
The recommendation and graph workloads (\textit{dlrm\_v2}, \textit{rgat}) -- already \SPEC's nearest MLPerf neighbors -- are untouched, so the augmented suite closes only 9\% of the overall gap: a GEMM buys the dense-kernel corner, not MLPerf. \par

\begin{table}[t]
\caption{\small{Dominant metrics per PC (top 4 by $\vert$mean loading$\vert$).}}
\label{table_loadings_metrics_only}
\centering
\footnotesize
\setlength{\tabcolsep}{1pt}
\renewcommand{\arraystretch}{1.1}
\begin{tabular}{|| m{0.3cm} | m{8cm} ||}
\hline
PC & Top 4 metrics (value in parens = mean of loadings over machines) \\
\hline
1 & BranchIns(-0.12), L3\$\_MPKI(0.10), FLOPs(0.10), VectorIns(0.10) \\
\hline
2 & ipc(0.15), L2\$\_MPKI(-0.11), L3\$\_MPKI(-0.09), BranchIns(-0.09) \\
\hline
3 & VectorIns(0.12), FLOPs(0.10), Mem\_Access(-0.09), KernelIns(0.08) \\
\hline
4 & L1I\$\_MPKI(0.16), BranchIns(-0.14), Branch\_MPKI(-0.11), L1D\$\_MPKI(0.08) \\
\hline
5 & LoadIns(0.14), L1D\$\_MPKI(0.14), KernelIns(-0.10), UserIns(0.10) \\
\hline
6 & L1\_iTLB\_MPMI(0.13), L2\$\_MPKI(0.10), Frontend(0.09), L1\_dTLB\_MPMI(-0.09) \\
\hline
7 & Branch\_MPKI(0.21), StoreIns(-0.11), BranchIns(-0.10), L1\_dTLB\_MPMI(-0.06) \\
\hline
8 & L2\_TLB\_MPMI(0.16), L1\_dTLB\_MPMI(0.13), VectorIns(0.08), BranchIns(-0.08) \\
\hline
\end{tabular}
\end{table}

Figure~\ref{fig_overview_pc} and Table~\ref{table_loadings_metrics_only} further illustrate the top eight principal components (in our PCA based clustering analysis).
Figure~\ref{fig_overview_pc} shows that \SPECO had outlier workloads (505.mcf\_r and 520.omnetpp\_r), while \SPEC has fewer outliers and more stable suite-level behavior.
DCPerf programs \textit{tao} and \textit{django} show uniqueness in many principal components. 
MLPerf is unique primarily along vector and memory-intensity dimensions (PC3), whereas DCPerf separates through service-style frontend behavior (e.g., \textit{tao} along PC3/PC4 and \textit{django} along PC6)~\cite{su2025dcperf}. 
The GEMM kernels sit near the MLPerf programs toward the far right of PC3, while the agentic workloads plot inside the \SPEC cloud.

\noindent\underline{\textbf{Takeaway:}}
\SPEC contains fewer extreme outliers than \SPECO, yielding more stable suite-level behavior.
Meanwhile, \SPEC moves closer to datacenter-like behavior represented by DCPerf while remaining clearly separated from highly vector-centric MLPerf, defining its modern general-purpose scope.
The two probes bound this scope from both sides: agentic AI is largely covered by \SPEC's existing high-footprint workloads, whereas a dense GEMM reaches only MLPerf's dense-kernel corner and leaves its recommendation and graph workloads as far away as before.

\section{Use Cases of \SPEC}
\label{section_case}
Beyond suite-level characterization, \SPEC exposes practical architectural and system-level sensitivities through case studies on system configurations (\S~\ref{section_04_01}), architectural choices (\S~\ref{section_04_02}), and proxy workload construction (\S~\ref{section_04_03}).

\begin{figure*}[t]
    \begin{minipage}[c]{0.5\linewidth}
        \centering
        \includegraphics[width=\linewidth, trim = 2mm 2mm 2mm 2mm, clip=true, page=1]{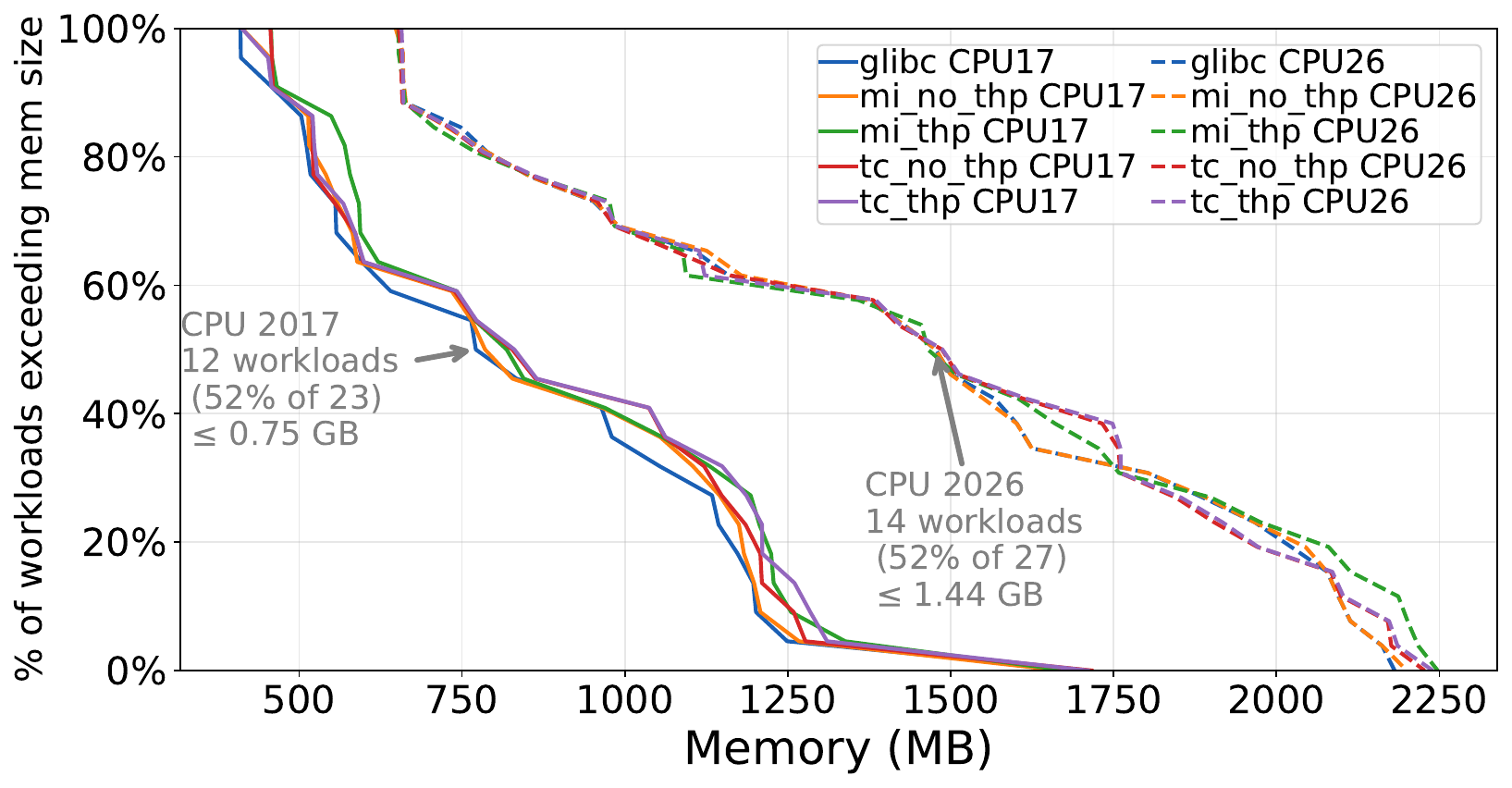}
        \subcaption{Complementary CDF of all Rate workloads.}
        \label{fig_cpu2017_cpu2026_rate_rss}
    \end{minipage}\hfill
    \begin{minipage}[c]{0.5\linewidth}
        \centering
        \includegraphics[width=0.95\linewidth, trim = 2mm 2mm 2mm 2mm, clip=true, page=1]{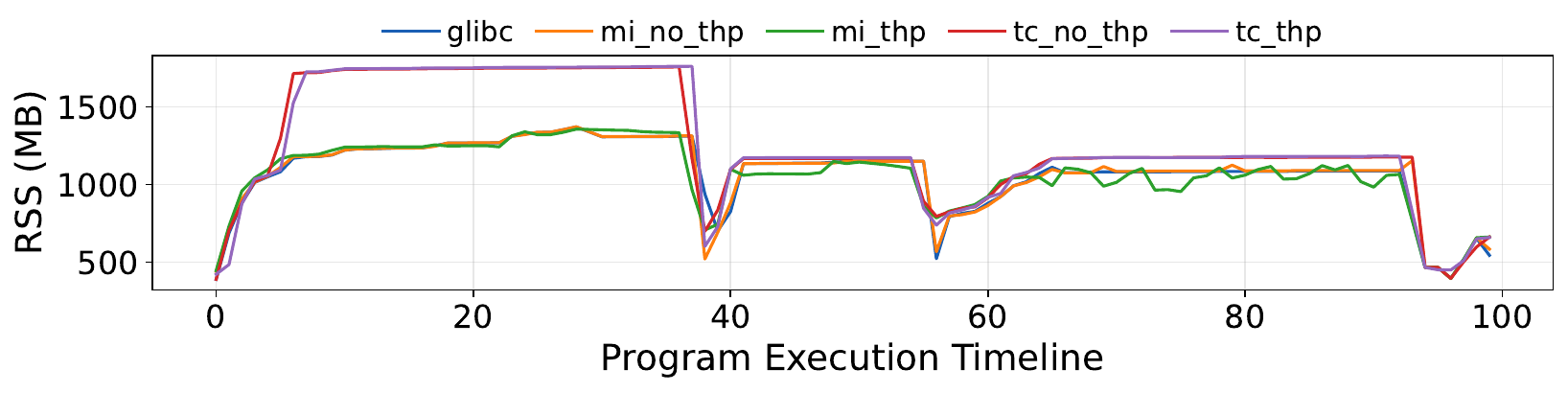}
        \subcaption{RSS over time for 721.gcc\_r.}
        \label{fig_rss_gcc}
        \includegraphics[width=0.95\linewidth, trim = 2mm 2mm 2mm 2mm, clip=true, page=1]{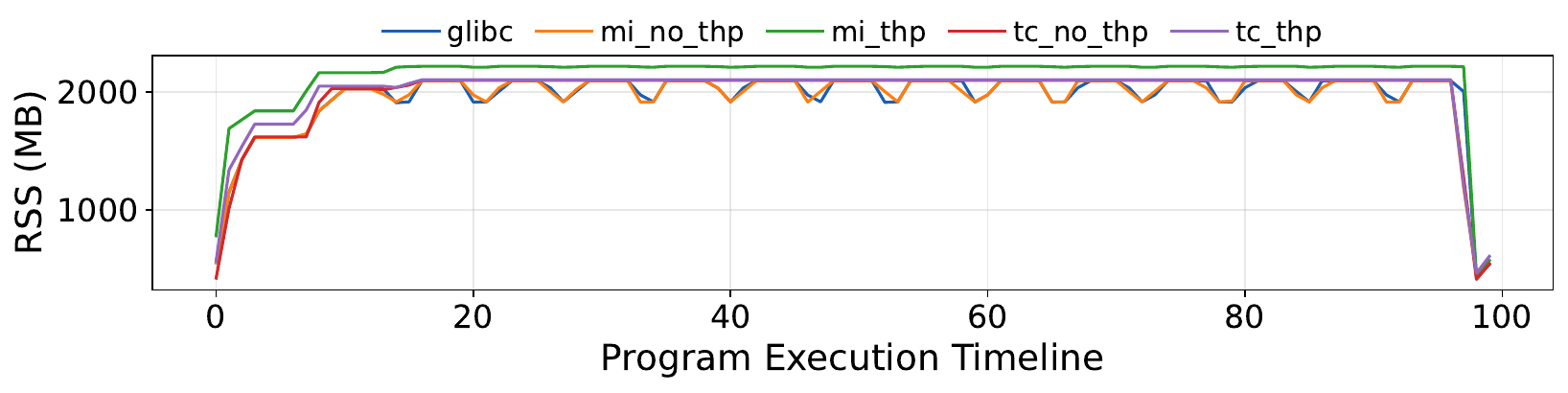}
        \subcaption{RSS over time for 709.cactus\_r.}
        \label{fig_rss_cactus}
    \end{minipage}
    
    \caption{\small{Memory Resident Set Size (RSS) for \SPEC/\SPECO Rate on CPU-C. (a) \SPEC RSS is larger than \SPECO (per-workload data in Appendix Table~\ref{tab_rss_detailed_cpu2017} and Table~\ref{tab_rss_detailed_cpu2026}). (b) gcc has varying RSS, also heavily dependent on THP. (c) cactus has varying RSS, also heavily dependent on THP.}}
    \label{fig_rss}
\end{figure*}

\begin{figure*}[t]
    \centerline{\includegraphics[width=2.1\columnwidth, trim = 2mm 7mm 2mm 2mm, page=1, clip=true]{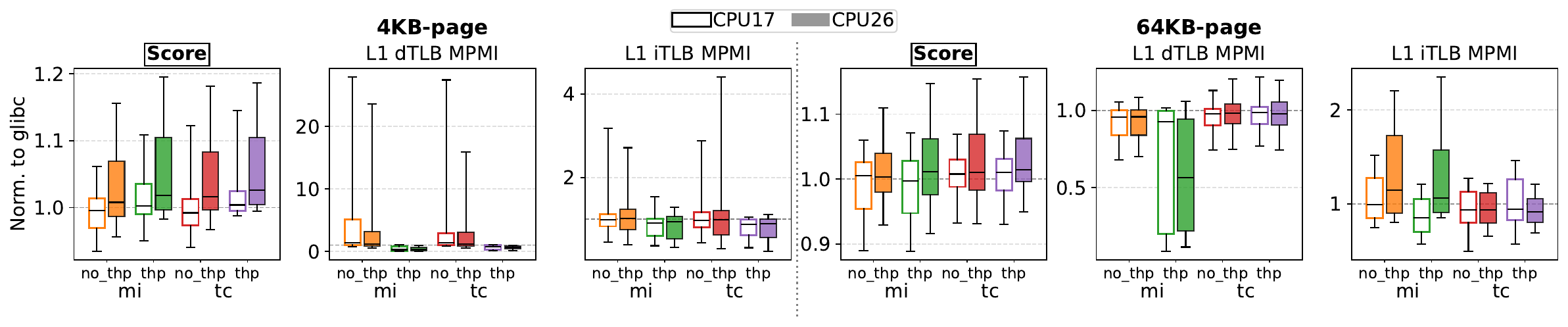}}
    \caption{\small{
    Performance of \texttt{mimalloc} (mi) and \texttt{tcmalloc} (tc), with and without THP, across \SPEC and \SPECO Rate workloads, normalized to the corresponding \texttt{glibc} baseline (per-workload data in Appendix Figure~\ref{fig_thp_appendix_cpu_c} and Figure~\ref{fig_thp_appendix_cpu_i}, the most 2 representative platforms CPU-C and CPU-I). 
    With 4 KB pages, advanced allocators and THP benefit \SPEC more than \SPECO. With 64 KB pages, both techniques are less effective. 
    The performance gains primarily come from reductions in both L1 dTLB and L1 iTLB misses, with THP producing larger reductions in L1 dTLB misses.
    }}
	\label{fig_thp}
\end{figure*}

\subsection{Evaluating System Configurations}
\label{section_04_01}

We show how \SPEC can evaluate key system-configuration choices that improve performance without changing the underlying silicon~\cite{sriraman2019softsku}, including memory requirements (\S~\ref{section_04_01_01}) and page sizes (\S~\ref{section_04_01_02}), hardware prefetcher (\S~\ref{section_04_01_03}), and compiler (\S~\ref{section_04_01_04}).

\subsubsection{Memory Requirements}
\label{section_04_01_01}
Working Set Size (WSS) is an important attribute that determines memory usage and system requirements; the concept dates back to Denning's early work~\cite{denning1968working}.
We use the resident set size (RSS, \texttt{rss} in the Unix utility \texttt{ps}), which measures how much memory an application actually uses, as a practical proxy for WSS~\cite{gove2007cpu2006} to compare the memory footprints of the Rate workloads.
Because memory usage is heavily affected by libraries, allocators, and optimizations, we study three allocators: \texttt{glibc}, \texttt{tcmalloc}, and \texttt{mimalloc} (more details in \S~\ref{section_04_01_02}).
\SPEC RSS reaches 2.2 GB, compared to 1.7 GB for \SPECO. \par

Figure~\ref{fig_cpu2017_cpu2026_rate_rss} shows the results on CPU-C; other platforms show similar trends. \SPEC has a larger memory footprint than \SPECO, with a median RSS of approximately $1.44$ GB versus $0.75$ GB.
A few \SPEC workloads share names with predecessors, but their codebases and inputs are entirely new, so we treat cross-suite comparisons as indicative rather than direct continuations.
Within \SPEC, six workloads exceed 2.0 GB (Table~\ref{table_rss}), whereas no \SPECO Rate workload reaches that threshold.
The shift is therefore driven primarily by newly introduced workloads with natively high memory demands and by enlarged inputs on same-named scientific benchmarks, not by moderate datacenter-oriented additions (708.sqlite\_r, 714.cpython\_r, 723.llvm\_r, 753.ns3\_r). \par

\noindent{\textbf{How allocators affect RSS?}}
In general, the choice of allocator does not have a strong effect on the RSS at the suite-level (Figure~\ref{fig_cpu2017_cpu2026_rate_rss}).
Beyond suite averages, temporal RSS traces reveal strong allocator sensitivities (Figure~\ref{fig_rss_gcc} and Figure~\ref{fig_rss_cactus}).
For 721.gcc\_r, peak RSS differs by up to $\sim$$0.4$ GB across allocators, while transient variations exceed 1 GB, indicating strong allocator sensitivity beyond the steady-state working-set size (similar difference exists in 709.cactus\_r as well).
\par

\noindent{\underline{\textbf{Takeaway:}}}
The median memory consumption of \SPEC (1.44 GB) is almost twice that of \SPECO (0.75 GB), which amplifies its sensitivity to page size and translation overheads for memory-intensive workloads. \par

\begin{table}[t]
\caption{\small{Memory resident set size (RSS) groupings of \SPEC Rate workloads (max RSS across allocators on CPU-C).}}
\label{table_rss}
\centering
\footnotesize
\setlength{\tabcolsep}{1pt}
\renewcommand{\arraystretch}{1.1}
\begin{tabular}{|| m{2.8cm} | m{2.8cm} | m{2.8cm} ||}
\hline
\textbf{Low RSS ($<$1 GB)} & \textbf{Medium RSS (1--2 GB)} & \textbf{High RSS ($\geq$2 GB)} \\
\hline
707.ntest, 727.cppcheck, 731.astcenc, 748.flightdm, 750.sealcrypto, 753.ns3, 766.femflow, 777.zstd & 708.sqlite, 710.omnetpp, 714.cpython, 721.gcc, 723.llvm, 729.abc, 735.gem5, 736.ocio, 737.gmsh, 765.roms, 767.nest, 772.marian & 706.stockfish, 709.cactus, 722.palm, 734.vpr, 749.fotonik3d, 782.lbm \\
\hline
\end{tabular}
\end{table}

\begin{figure*}[t]
    \begin{minipage}{2.1\columnwidth}
	\centering
	\includegraphics[width=\columnwidth, trim = 0mm 0mm 0mm 0mm, clip=true, page=1]{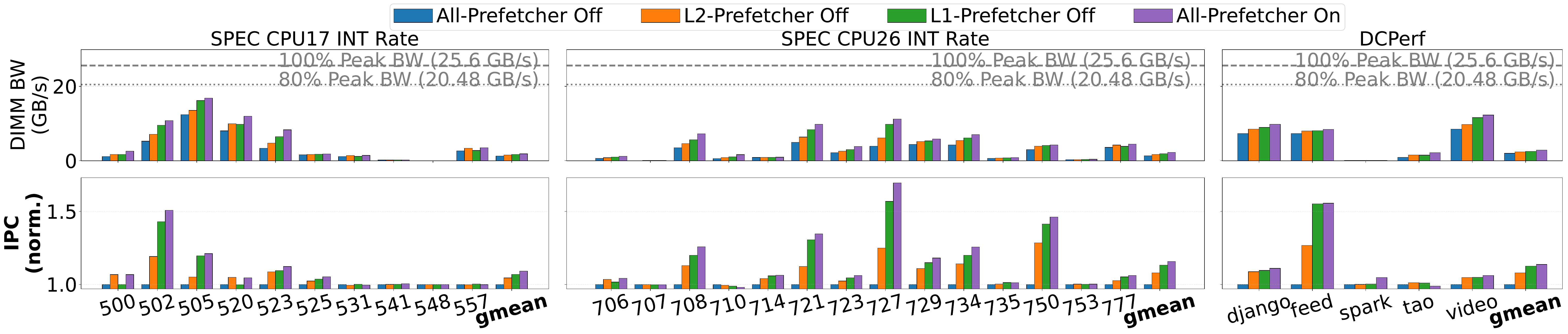}
	\subcaption{\small{Results on CPU-B. Higher IPC comes from better DIMM bandwidth utilization.}}
	\label{fig_prefetch_cpub}
	\end{minipage} 
    \begin{minipage}{2.1\columnwidth}
	\centering
	\includegraphics[width=\columnwidth, trim = 0mm 0mm 0mm 0mm, clip=true, page=1]{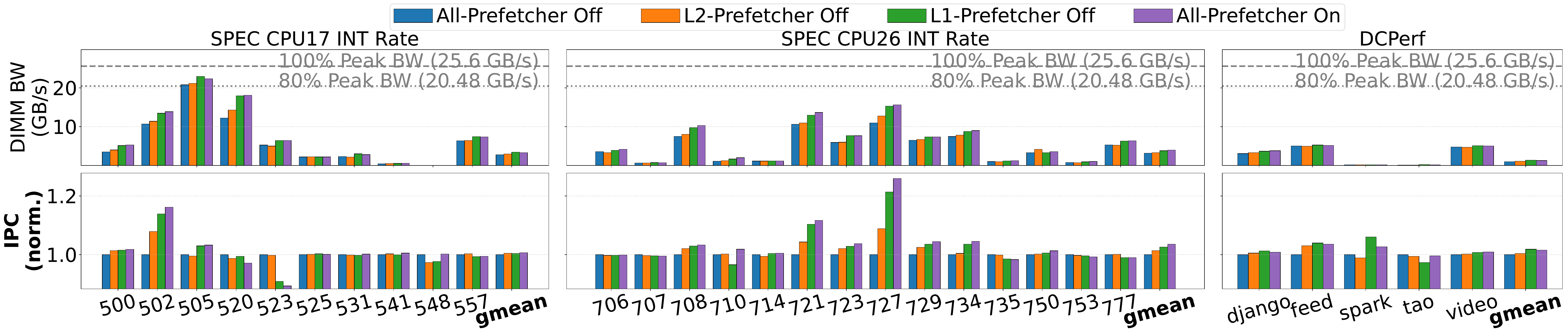}
	\subcaption{\small{Prefetcher efficiency drops on CPU-E compared with CPU-B as more cores (64 vs. 40) share the same DIMM bandwidth. }}
	\label{fig_prefetch_cpue}
	\end{minipage} 
    \caption{\small{
    IPC and DIMM bandwidth across prefetcher configurations on CPU-B and CPU-E. 
    IPC is normalized to the all-prefetchers-off configuration (4th bar in each group shows the overall prefetching benefit). 
    On CPU-B, 6 of 14 \SPEC INT Rate workloads improve by at least $10\%$, and \SPEC shows sensitivity comparable to DCPerf ($1.15\times$ vs.\ $1.13\times$) and greater than \SPECO ($1.06\times$). 
    On the bandwidth-limited CPU-E, \SPEC remains the most sensitive suite ($1.04\times$ vs.\ $1.02\times$ for DCPerf and $1.01\times$ for \SPECO).
    }}
 \label{fig_prefetch_per_workload}
\end{figure*}

\subsubsection{Page Size}
\label{section_04_01_02}
Page size is a common system-level tuning knob. Larger pages can reduce translation overhead and page-walk traffic, but may increase fragmentation and alter address translation behavior~\cite{cascaval2005multiple, guvenilir2020tailored, cox2017efficient, ausavarungnirun2018mosaic, zhou2023impact, lee2023memtis}. 
To evaluate how \SPEC and \SPECO respond to page size and transparent huge pages (THP), we analyze seven platforms: five \texttt{x86\_64} servers with \emph{4KB base pages} and two \texttt{AArch64} servers with \emph{64KB base pages}. We exclude the oldest (CPU-A) and newest (CPU-G) \texttt{x86\_64} systems to better align hardware generations with the \texttt{AArch64} nodes.
The microarchitectural impact of page size can be obscured by allocator behavior, since the default \texttt{glibc} allocator does not consistently support allocations with THPs~\cite{zhou2024characterizing}.
To reduce this bias, we use two additional allocators, \texttt{mimalloc} and \texttt{tcmalloc}, which explicitly support and manage huge pages~\cite{leijen2019mimalloc, hunter2021beyond}.
Figure~\ref{fig_thp} plots per-workload score distributions for \SPEC and \SPECO for 4KB and 64KB page sizes. \par

On 4KB-page systems, exposing proper THP behavior via these allocators reveals noticeable suite-level sensitivity to translation overheads. Relative to the baseline \texttt{glibc}, migrating to THP-aware allocators improves the geometric-mean \SPEC score by 3.1-6.7\% and \SPECO by 1.1-5.7\% (averaged across the 4KB platforms), with individual microarchitectures seeing \SPEC gains up to 12.7\%.
These gains confirm that 4KB-page systems suffer from translation bottlenecks that proper huge-page support can alleviate.
Specifically, \texttt{tcmalloc} with THP reduces L1 dTLB and L1 iTLB MPMI by $\sim$$52\%$ and $\sim$$28\%$, respectively, on average, reaching up to 65.3\% and 56.2\% reductions on the most TLB-sensitive architectures. \par

The suite-level gains aggregate widely varying per-workload responses.
THP speedups couple only loosely to baseline TLB pressure: 737.gmsh\_r, the most translation-bound \SPEC Rate workload (Figure~\ref{fig_overview_uarch}), gains about as much (4-8\%) as the far less translation-bound 777.zstd\_r and 706.stockfish\_r (5-7\%).
The extremes are instead set by allocator -- THP interactions: under \texttt{mimalloc}+THP, 507.cactuBSSN\_r (\SPECO) and 709.cactus\_r (\SPEC) range from a $2\times$ slowdown to a $1.26\times$ speedup across the 4KB platforms.
Page-size tuning therefore requires end-to-end validation across translation-bound, allocator-sensitive, and fragmentation-prone workloads -- coverage the Rate suites provide in one run. \par

In contrast, systems with native 64KB pages largely avoid these translation bottlenecks, leaving little room for THP to improve performance. 
Across the 64,KB-page \texttt{AArch64} platforms, enabling THP-aware allocators changes suite-level \SPEC scores by only -0.1\% to 3.0\%. 
The downside risk remains, however: with \texttt{mimalloc}, THP degrades both 507.cactuBSSN\_r and 709.cactus\_r, as it does on the 4KB-page platforms. 
Huge-page fragmentation and metadata costs therefore persist, while the larger base page provides less translation overhead for THP to eliminate.
\par

\noindent{\underline{\textbf{Takeaway:}}}
Native 64KB base pages effectively mitigate translation overheads.
On 4KB-page systems, THP successfully reduces translation overheads, but trades that efficiency for software fragmentation risks.
\begin{table}[t]
\caption{\small{Prefetch friendliness of \SPEC INT Rate workloads (IPC gain from enabling all prefetchers on CPU-B).}}
\label{table_prefetch}
\centering
\footnotesize
\setlength{\tabcolsep}{1pt}
\renewcommand{\arraystretch}{1.1}
\begin{tabular}{|| m{2.8cm} | m{2.8cm} | m{2.8cm} ||}
\hline
\textbf{Low ($\leq$2\%)} & \textbf{Moderate (4--6\%)} & \textbf{High ($\geq$10\%)} \\
\hline
707.ntest, 710.omnetpp, 735.gem5, 753.ns3 & 706.stockfish, 714.cpython, 723.llvm, 777.zstd & 708.sqlite, 721.gcc, 727.cppcheck, 729.abc, 734.vpr, 750.sealcrypto \\
\hline
\end{tabular}
\end{table}

\begin{figure*}[t]
    \centerline{\includegraphics[width=2.1\columnwidth, trim = 2mm 2mm 2mm 2mm, page=1, clip=true]{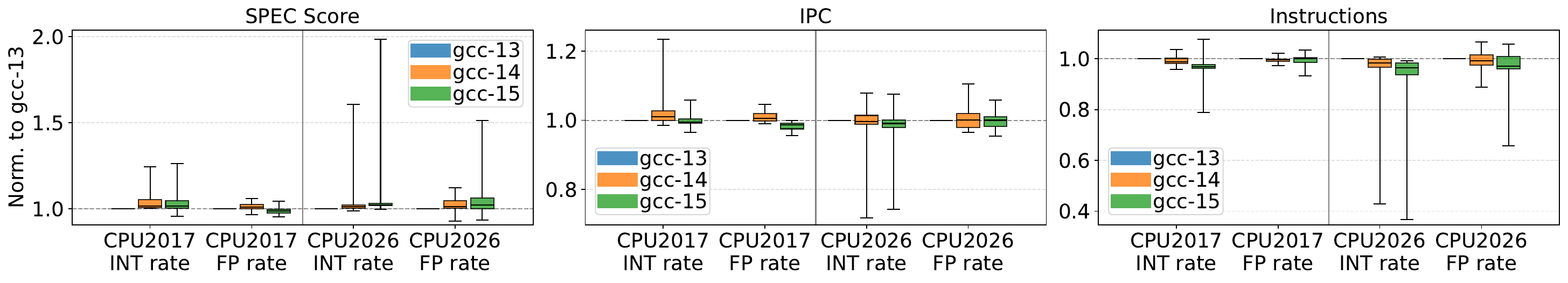}}
    \caption{\small{Performance distributions across gcc-13, gcc-14, and gcc-15 for \SPEC and \SPECO Rate workloads (per-workload data in Appendix Figure~\ref{fig_compiler_appendix_int} and Figure~\ref{fig_compiler_appendix_fp}). \SPEC exhibits significantly greater performance sensitivity to compiler evolution, primarily driven by fluctuations in dynamic instruction count. 
    }}
    \label{fig_compiler}
\end{figure*}

\subsubsection{Hardware Prefetchers}
\label{section_04_01_03}
Modern CPUs rely on hardware prefetchers to hide memory latency, often at the cost of increased memory bandwidth~\cite{lee2012prefetching,bakhshalipour2019evaluation,navarro2019memory,panda2023clip,jain2024limoncello}. We evaluate CPU-B (Intel Ice Lake) and CPU-E (AMD Milan) to avoid conclusions tied to a single vendor's implementation; the remaining platforms were excluded because we lacked root access to configure their prefetchers. We measure prefetcher effects using \SPECO, \SPEC, and DCPerf. \par

Figure~\ref{fig_prefetch_per_workload} compares IPC and per-channel memory bandwidth across four prefetcher configurations: \textit{All-Prefetcher Off}, \textit{L2-Prefetcher Off}, \textit{L1-Prefetcher Off}, and \textit{All-Prefetcher On}.
On CPU-B, enabling all prefetchers improves the geometric-mean IPC by $1.06\times$ for \SPECO, $1.15\times$ for \SPEC, and $1.13\times$ for DCPerf. The L2 prefetcher boosts IPC more than the L1 prefetcher (e.g., $1.12\times$ vs.\ $1.07\times$ for \SPEC).
In contrast, on CPU-E, the geometric-mean IPC gains are heavily muted: $1.01\times$ for \SPECO, $1.04\times$ for \SPEC, and $1.02\times$ for DCPerf.
One reason is the higher core count on CPU-E (64 vs.\ 40 cores) sharing the same 3200 MT/s per-channel memory bandwidth, leaving less bandwidth headroom for speculative prefetch requests. \par

Despite these vendor and architectural differences, both platforms exhibit the same qualitative trend: prefetching yields substantial improvements only when sufficient memory bandwidth headroom exists. Furthermore, these gains remain highly workload-dependent; for instance, \SPEC exhibits a significant upside in 727.cppcheck\_r, where prefetching improves IPC by $1.69\times$ on CPU-B and $1.26\times$ on CPU-E. \par

The six \emph{prefetch-friendly} workloads (Table~\ref{table_prefetch}) gain $1.18$-$1.69\times$ IPC on CPU-B. 
In every case, the L2 prefetcher contributes more than the L1 prefetcher; for 727.cppcheck\_r, the gains are $1.57\times$ and $1.25\times$, respectively. 
These improvements increase DRAM traffic by up to $2.9\times$ as latency-bound misses are converted into timely prefetches.
On CPU-E, the gains shrink to $1.01$-$1.26\times$ for two reasons. 
727.cppcheck\_r, 721.gcc\_r, 708.sqlite\_r, and 734.vpr\_r already consume 9--16 GB/s per channel, leaving little bandwidth headroom. 
In contrast, 750.sealcrypto\_r gains only 1.3\% despite using just $\sim$$3.5$ GB/s, because CPU-E already achieves high IPC with prefetchers disabled.
Prefetching can also cause performance degradations. 
On CPU-B, 710.omnetpp\_r loses 2\% IPC while its DRAM traffic nearly triples ($0.56$$\rightarrow$$1.65$ GB/s), indicating that inaccurate prefetches displace useful data. 
Overall, CPU-B's $1.15\times$ geometric-mean IPC gain comes at the cost of an 80\% increase in DRAM traffic ($2.34$$\rightarrow$$4.20$ GB/s per channel).
\par

\SPEC also excludes certain ``stress-test'' workloads present in \SPECO, such as 505.mcf\_r, which exert substantially higher pressure on the memory subsystem than typical datacenter applications~\cite{su2025dcperf}. 
These workloads can reach nearly 80\% of the system's peak bandwidth -- approaching practical saturation~\cite{esmaili2024mess} -- while such extreme cases are often better represented by dedicated memory subsystem benchmarks such as \textit{Mess}~\cite{esmaili2024mess, bostanci2026cleaningmessreevaluatingrealsystem}. \par

\noindent\underline{\textbf{Takeaway:}}
\SPEC successfully exposes prefetcher sensitivities without relying on extreme, unrealistic bandwidth stress-tests (e.g., 505.mcf\_r in \SPECO). 
Furthermore, cross-vendor analysis confirms that absolute prefetcher efficacy depends heavily on the platform's core-to-bandwidth ratio rather than the workload alone.
\begin{figure*}[t]
    \centerline{\includegraphics[width=2.1\columnwidth, trim = 2mm 2mm 2mm 2mm, page=1, clip=true]{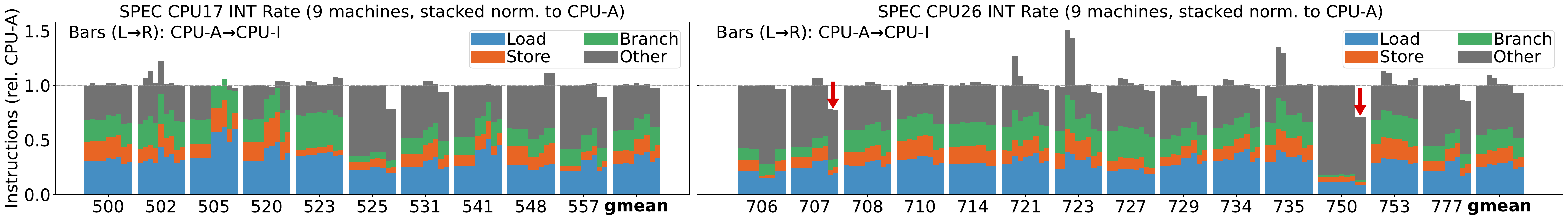}}
    \caption{\small{
    Per-copy retired-instruction distributions for \SPECO and \SPEC INT Rate across nine \texttt{x86\_64} and \texttt{AArch64} systems. For each benchmark, 9 bars are pasted in each cluster for CPU-A to CPU-I.
    As shown by the last two bars for workloads 707 and 750, these \SPEC workloads retire fewer instructions on \texttt{AArch64} than on \texttt{x86\_64}. 
    Overall, \SPEC exhibits greater cross-platform variation than \SPECO.
    }}
	\label{fig_isa}
\end{figure*}

\subsubsection{Compiler}
\label{section_04_01_04}

\SPEC can be used to evaluate the effectiveness of compilers, extending \SPECO's long-standing role~\cite{schmitt2020energy, sr2019battle, schmitt2020performance}.
Comparing compiler generations across both suites quantifies whether score changes reflect improved code generation and how much optimization headroom \SPEC restores. \par

We evaluate compiler sensitivity using \texttt{gcc-13, gcc-14}, and \texttt{gcc-15}.
As shown in Figure~\ref{fig_compiler} (distribution includes results of all nine platforms), gcc-15 improves INT Rate performance over gcc-13 for both suites, with geometric-mean speedups of $1.04\times$ on \SPECO and $1.08\times$ on \SPEC.
For FP Rate, gcc-15 slightly regresses on \SPECO ($0.99\times$) but improves \SPEC ($1.05\times$).
IPC and instruction counts reveal a clear contrast between the two suites. For \SPEC, the performance gains come primarily from a 17.1\% reduction in dynamic instruction count relative to gcc-13, despite a 3.5\% decrease in IPC. 
In comparison, \SPECO is much less sensitive, with instruction count decreasing by only 4.8\% and IPC changing by just 0.1\%.
This may reflect diminishing returns from a decade of compiler co-evolution with \SPECO; the fresh workloads in \SPEC offer renewed optimization headroom. 
At the per-workload level, certain \SPEC programs respond robustly to modern optimizations; for example, gcc-15 reduces the instruction count of 706.stockfish\_r by up to $3\times$.
This single workload forms the deep lower whisker of the \SPEC INT rate instruction-count distribution in Figure~\ref{fig_compiler} ($0.37\times$); the corresponding \SPECO extreme, 525.x264\_r, shrinks by only 21\% ($0.79\times$). \par

\noindent\underline{\textbf{Takeaway:}}
\SPEC restores the compiler headroom that \SPECO exhausted, which came mainly from eliminating dynamic instructions (e.g., $3\times$ fewer instructions for 706.stockfish\_r).
Compiler choice is therefore a first-order experimental variable for \SPEC: studies should report toolchain versions alongside scores and avoid frozen-binary simulation for compiler-sensitive workloads. \par

\subsection{Evaluating Architectural Designs}
\label{section_04_02}

We show how \SPEC can evaluate architectural design choices, including instruction-set comparisons (\texttt{x86\_64} vs.\ \texttt{AArch64}, \S~\ref{section_04_02_01}) and many-core SoC design (\S~\ref{section_04_02_02}).

\subsubsection{ISA}
\label{section_04_02_01}
We use \SPEC to study ISA-level trade-offs, focusing on \texttt{x86\_64} (CPU-A to CPU-G) versus \texttt{AArch64} (CPU-H and CPU-I) and using per-copy retired instruction count, which isolates ISA-level differences from platform-specific core and cache effects. \par

As shown in Figure~\ref{fig_isa}, \SPEC exhibits substantially higher cross-platform variation than \SPECO even for INT Rate. 
Across all platforms, \SPECO instruction counts span $0.78$-$1.22\times$ (standard deviation 5.9\%), whereas \SPEC expands the range to $0.72$-$1.51\times$ (standard deviation 10.1\%).
\SPEC also contains more workloads with clear ISA-separated behavior; for example, 750.sealcrypto\_r and 707.ntest\_r retire approximately 20-30\% fewer instructions on \texttt{AArch64} than on \texttt{x86\_64}, while \SPECO has fewer cases with gaps of similar magnitude. \par

\noindent\underline{\textbf{Takeaway:}}
\SPEC exhibits larger cross-ISA and cross-platform variation in instruction count than \SPECO, making it more effective for ISA comparisons.

\begin{figure*}[t]
    \centerline{\includegraphics[width=2.1\columnwidth, trim = 2mm 2mm 2mm 2mm, page=1, clip=true]{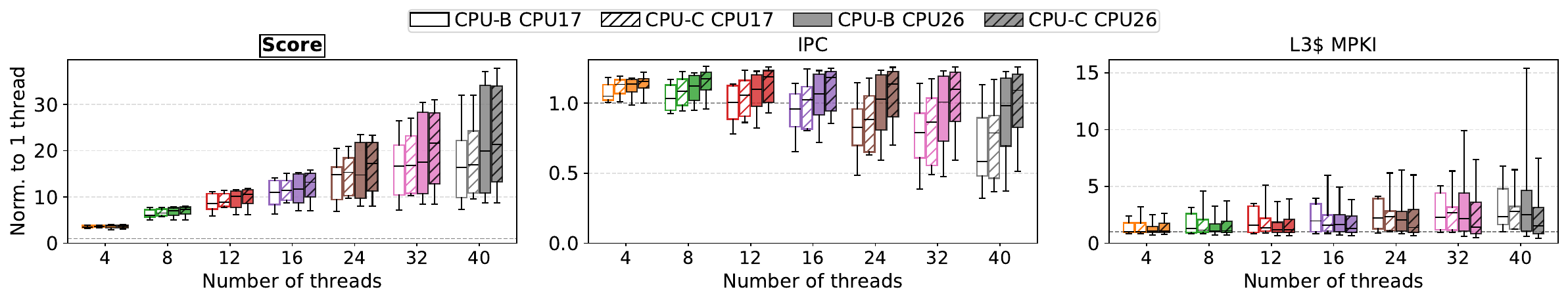}}
    \caption{\small{
    Scaling distributions from 1 to 40 threads for \SPEC and \SPECO Speed (per-workload data in Appendix Figure~\ref{fig_scale_appendix_cpu_b} and Figure~\ref{fig_scale_appendix_cpu_c}) on CPU-B (40-core monolithic) and CPU-C (4$\times$12-core chiplet). 
    \SPEC scales better overall, while CPU-C outperforms CPU-B due to its larger aggregate LLC. 
    \SPECO more often exposes communication-bound workloads that scale poorly and favor monolithic designs.
    }}
    \label{fig_speed_scale}
\end{figure*}

\begin{figure}[t]
    \centerline{\includegraphics[width=\columnwidth, trim = 2mm 2mm 2mm 2mm, page=1, clip=true]{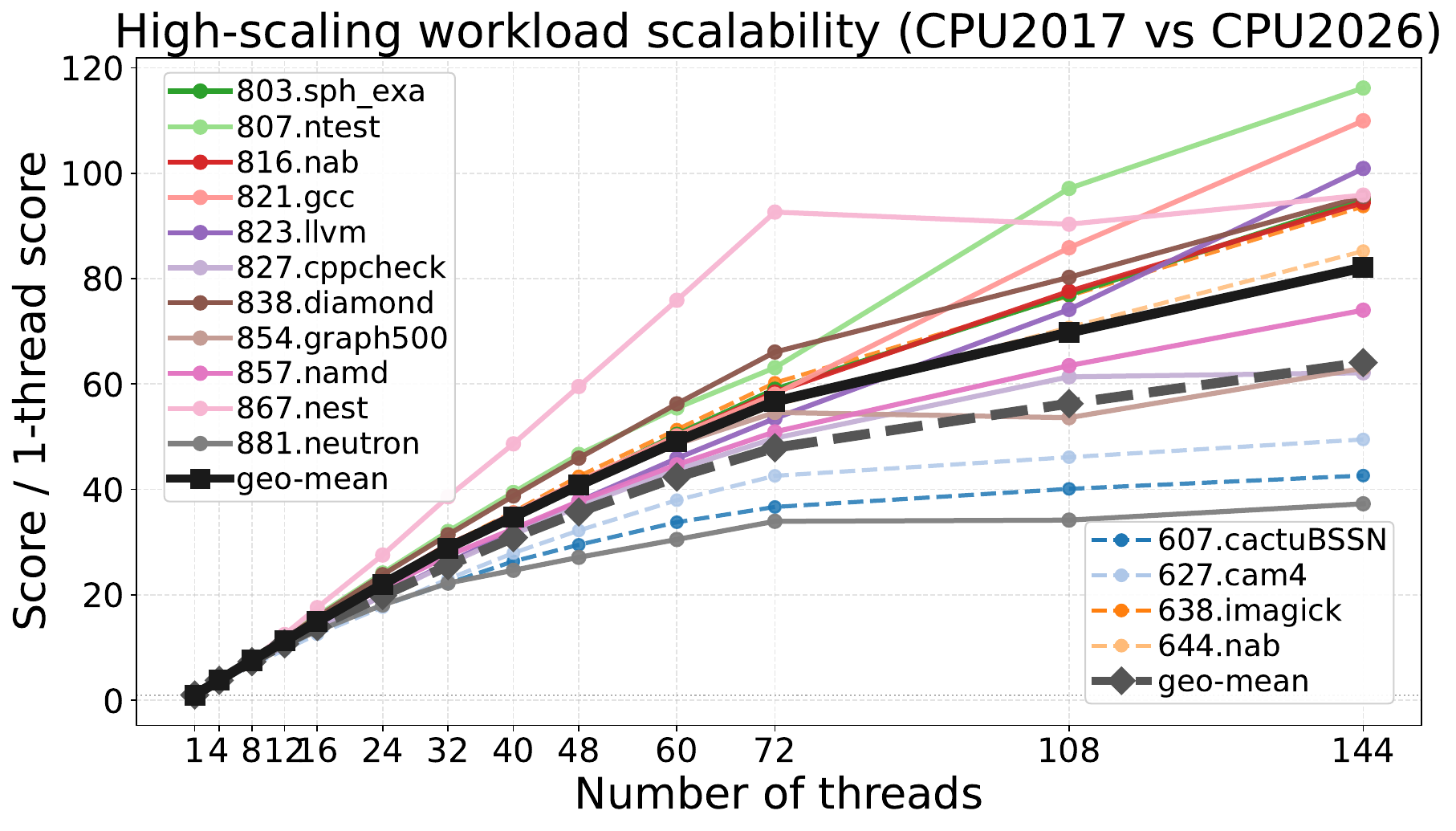}}
    \caption{\small{Scalability up to 144 threads (CPU-I). High-scalability workloads in \SPEC continue to scale significantly better than \SPECO past 40 threads into the ultra-many-core regime.}}
    \label{fig_gg_speed_scale}
\end{figure}

\begin{table}[t]
\caption{\small{Workload scalability classification by scaling (40-thread score $>20\times$ 1-thread score = High-scaling).}}
\label{table_workload_scaling}
\centering
\footnotesize
\setlength{\tabcolsep}{1pt}
\renewcommand{\arraystretch}{1.1}
\begin{tabular}{|| m{1cm} | m{3.8cm} | m{3.5cm} ||}
\hline
\textbf{Suite} & \textbf{Low-scaling} & \textbf{High-scaling} \\
\hline
\SPECO & 603.bwaves, 619.lbm, 621.wrf, 628.pop2, 649.fotonik3d, 654.roms, 657.xz & 607.cactuBSSN, 627.cam4, 638.imagick, 644.nab \\
\hline
\SPEC & 800.pot3d, 801.xz, 809.cactus, 811.tealeaf, 817.flac, 820.cloverleaf, 822.palm, 846.minizinc, 849.fotonik3d, 865.roms, 872.marian & 803.sph\_exa, 807.ntest, 816.nab, 821.gcc, 823.llvm, 827.cppcheck, 838.diamond, 854.graph500, 857.namd, 867.nest, 881.neutron \\
\hline
\end{tabular}
\end{table}

\subsubsection{Many-Core SoC}
\label{section_04_02_02}
As processor core counts have increased substantially over the past decade~\cite{papazian2020new, nassif2022sapphire, soltis2023next, evers2022amd, bhargava2024amd, singh2025zen, ampere, grace}, multi-threaded workloads have become increasingly important for evaluating many-core SoC design. 
Their scalability is shaped by shared on-chip resources and cross-core communication, e.g., LLC slices, coherence protocols, and memory-system traffic, all of which become even more critical with the widespread adoption of chiplet-based CPUs~\cite{naffziger2021pioneering, stojkovic2023mumanycore, smith2024realizing}.
Chiplets enable modular composition and improve performance scaling, but they can introduce non-uniform latencies or reduced bandwidth across chip boundaries~\cite{naffziger2021pioneering, velten2022memory, feng2022chiplet, stojkovic2023mumanycore, kwon2023mccore, smith2024realizing, zhang2023characterizing}.
We compare CPU-B (Intel Icelake, 40-core monolithic) and CPU-C (Intel Sapphire Rapids, $4\times12$-core chiplets). \par

Compared with \SPECO, which includes 11 multi-threaded Speed workloads, \SPEC includes 22, providing broader coverage of modern many-core systems.
To study many-core SoC design effects, we sweep thread counts from 1 to 40 and report Score, IPC, and L3\$ MPKI distributions in Figure~\ref{fig_speed_scale}.
At 40 threads, the geometric-mean normalized Score reaches $15.31\times/16.31\times$ (\SPECO, CPU-B/CPU-C) vs.\ $16.53\times/18.28\times$ (\SPEC, CPU-B/CPU-C).
Moreover, \SPEC continues to scale from 32 to 40 threads ($+9.5\%$/$+10.3\%$), while \SPECO shows earlier saturation ($+3.1\%$/$+5.0\%$).
\SPEC thus better stresses modern many-core systems. \par

\noindent{\textbf{Chiplet vs. Monolithic.}}
CPU-C achieves $1.13\times$ and $1.07\times$ higher absolute geometric-mean performance than CPU-B on \SPECO and \SPEC INT Rate (per-copy), respectively. However, our scaling results are normalized to a single-thread baseline, factoring out raw core improvements to isolate cross-core interactions.
In \SPECO, we observe distinct monolithic advantages for a small subset of low-scaling workloads (Table~\ref{table_workload_scaling}) that are highly sensitive to synchronization and serial cross-core communication overheads. 
For example, at 40 threads, 628.pop2\_s achieves $11.72\times$ on CPU-B but only $9.52\times$ on CPU-C (a $1.23\times$ monolithic advantage), while 621.wrf\_s reaches $15.64\times$ on CPU-B versus $12.35\times$ on CPU-C ($1.27\times$).
These gaps persist despite CPU-C's lower relative L3\$ MPKI increase (e.g., 628.pop2\_s L3\$ MPKI at 40 threads scales by $1.97\times$ on CPU-B vs.\ $1.34\times$ on CPU-C).
Non-uniform cross-chiplet coherence latencies can thus severely penalize older software patterns relying on tight synchronization, even with superior cache capacity. \par

In contrast, \SPEC exhibits a significantly more balanced scaling profile. CPU-B outscales CPU-C for far fewer workloads, and the monolithic advantages are considerably narrower. The sole prominent outlier is the low-scaling 846.minizinc\_s, which reaches $2.04\times$ on CPU-B versus $1.68\times$ on CPU-C at 40 threads (a $1.21\times$ advantage).
Beyond this single case, the performance deltas between the two topologies for the remaining low-scaling workloads fall within approximately 1-2\%. Consequently, while \SPECO highlights unscalable cases where old synchronization models favor monolithic designs, \SPEC indicates that modern code structures are largely robust against chiplet-induced latency penalties up to 40 threads, consistent with expanded aggregate L3 capacity absorbing much of the added cross-chiplet communication cost. \par

\noindent{\textbf{Scaling beyond 40 threads: Ultra-many-core behavior.}}
To assess whether \SPEC eventually succumbs to the same scalability plateaus as \SPECO on larger topologies, we extend our evaluation to 144 threads on CPU-I (Nvidia Grace Superchip, 144 cores across two sockets).
Figure~\ref{fig_gg_speed_scale} plots the high-scaling workloads identified in Table~\ref{table_workload_scaling}.
\SPEC continues to scale significantly better than \SPECO past 40 threads: the geometric-mean speedup of high-scaling workloads increases from $34.8\times$ (\SPEC) versus $30.9\times$ (\SPECO) at 40 threads, to $82.1\times$ versus $64.1\times$ at 144 threads, a relative scalability gain of $+136\%$ versus $+107\%$.
Consequently, \SPEC's higher high-scaling workload fraction provides a more rigorous and rugged vehicle for evaluating ultra-many-core fabrics. \par

\noindent\underline{\textbf{Takeaway:}}
(1) \SPEC is better suited to many-core SoC evaluation: it doubles the number of multi-threaded Speed workloads and maintains strong scalability into ultra-many-core (144-thread) regimes, where \SPECO begins to plateau.
(2) \SPEC also avoids extreme scalability anomalies. Its only apparent monolithic-topology advantage, 846.minizinc\_s, saturates by four threads on both systems, indicating limited parallelism rather than coherence sensitivity.
Thus, architects studying cross-chiplet coherence should retain \SPECO's 628.pop2\_s and 621.wrf\_s as targeted probes, while using \SPEC to evaluate the strengths of chiplet designs, including larger aggregate LLC capacity (Figure~\ref{fig_speed_scale}) and sustained high-concurrency scaling.

\subsection{Creating Workload Proxies Using RRR Mode}
\label{section_04_03}

Standard \SPEC Rate execution is homogeneous: all copies run the same benchmark before the suite advances.
\SPEC's new Rolling Round-Robin (RRR) mode instead staggers $N$ benchmarks across $M$ cores, each core executing the same sequence at a different phase~\cite{jacobvitz2015multi, llull2017cooper, patel2020clite, zhao2021understanding}, producing a time-rolled heterogeneous mix. \par

We use RRR as a composition primitive for creating proxy workloads.
Full-suite RRR closely reproduces homogeneous Rate execution~\cite{madhav2026spec}, because suite-wide averaging smooths workload-specific extremes.
Targeted RRR mixes instead combine complementary workloads into controlled stress points that bracket a target pressure point.
We evaluate this workflow in three steps:
\begin{itemize}
    \item Derive a workload pair and mix ratio that bracket a desired target pressure point (Alg.~\ref{alg_proxy}).
    \item Test whether RRR mixes follow predictions from weighted solo profiles.
    \item Quantify the contention omitted by time-rolled execution through true co-location experiments.
\end{itemize}
\par

\begin{algorithm}[t]
\footnotesize
\LinesNumbered
\SetAlgoNoEnd
\DontPrintSemicolon
\SetKwInOut{Input}{Input}
\SetKwInOut{Output}{Output}
\caption{\small{Targeted RRR proxy construction.}}
\label{alg_proxy}
\Input{target profile $t$; declared axes $K$; pool $\mathcal{P}$ of solo \SPEC profiles $v^{X}$ with instruction counts $I_X$; slot budget $S$}
\Output{pair $(A,B)$, slot weights $w_A{:}w_B$, unreachable axes $U$}
\BlankLine
\tcc{mixing $A$ and $B$ reaches only the segment $v^{AB}(\alpha)=\alpha v^{A}+(1{-}\alpha)v^{B}$}
$\mathcal{C}\leftarrow$ candidate pairs $(A,B)$ from $\mathcal{P}$ with $\min(v^{A}_{k},v^{B}_{k})\le t_k\le\max(v^{A}_{k},v^{B}_{k})$ $\forall k\in K$\label{line_bracket}\;
\ForEach{$(A,B)\in\mathcal{C}$\label{line_loop}}{
  $\alpha_{AB}\leftarrow\arg\min_{0\leq\alpha\leq1}\ \textstyle\sum_{k\in K}\big(v^{AB}_k(\alpha)/t_k-1\big)^{2}$\label{line_fit}\;
}
$(A,B)\leftarrow$ the best-fitting pair;\quad $\alpha^\ast\leftarrow\alpha_{AB}$\label{line_select}\;
$w_A{:}w_B\leftarrow$ integer slots, $w_A{+}w_B\leq S$, with instruction share nearest $\alpha^\ast$\label{line_slots}\;
$U\leftarrow$ axes outside $K$ on which $(A,B)$ does not bracket $t$\label{line_unreach}\;
Run RRR at $w_A{:}w_B$; report the measured residual and $U$\label{line_measure}\;
\end{algorithm}

\begin{figure}[t]
    \centerline{\includegraphics[width=\columnwidth, trim = 2mm 2mm 2mm 2mm, page=1, clip=true]{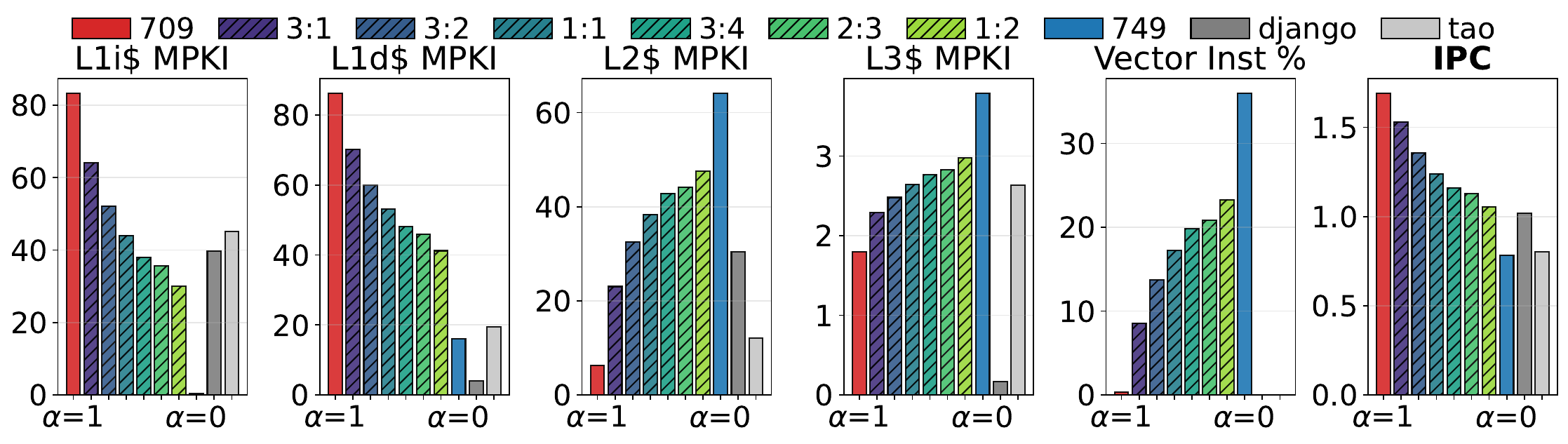}}
    \caption{\small{The RRR mix ratio is a tuning knob (CPU-C): hatched bars sweep the 709.cactus\_r:749.fotonik3d\_r \emph{slot} ratio from $3{:}1$ to $1{:}2$, lowering 709.cactus\_r's instruction share $\alpha$ from 0.77 to 0.36. The sweep brackets \textit{django}'s L1I\$ MPKI between the $1{:}1$ and $3{:}4$ mixes ($\alpha=0.53$ and 0.45; 44.0 and 38.0 MPKI vs.\ 39.7).}}
    \label{fig_rrr}
\end{figure}

\begin{figure}[t]
    \centerline{\includegraphics[width=\columnwidth, trim = 2mm 2mm 2mm 2mm, page=1, clip=true]{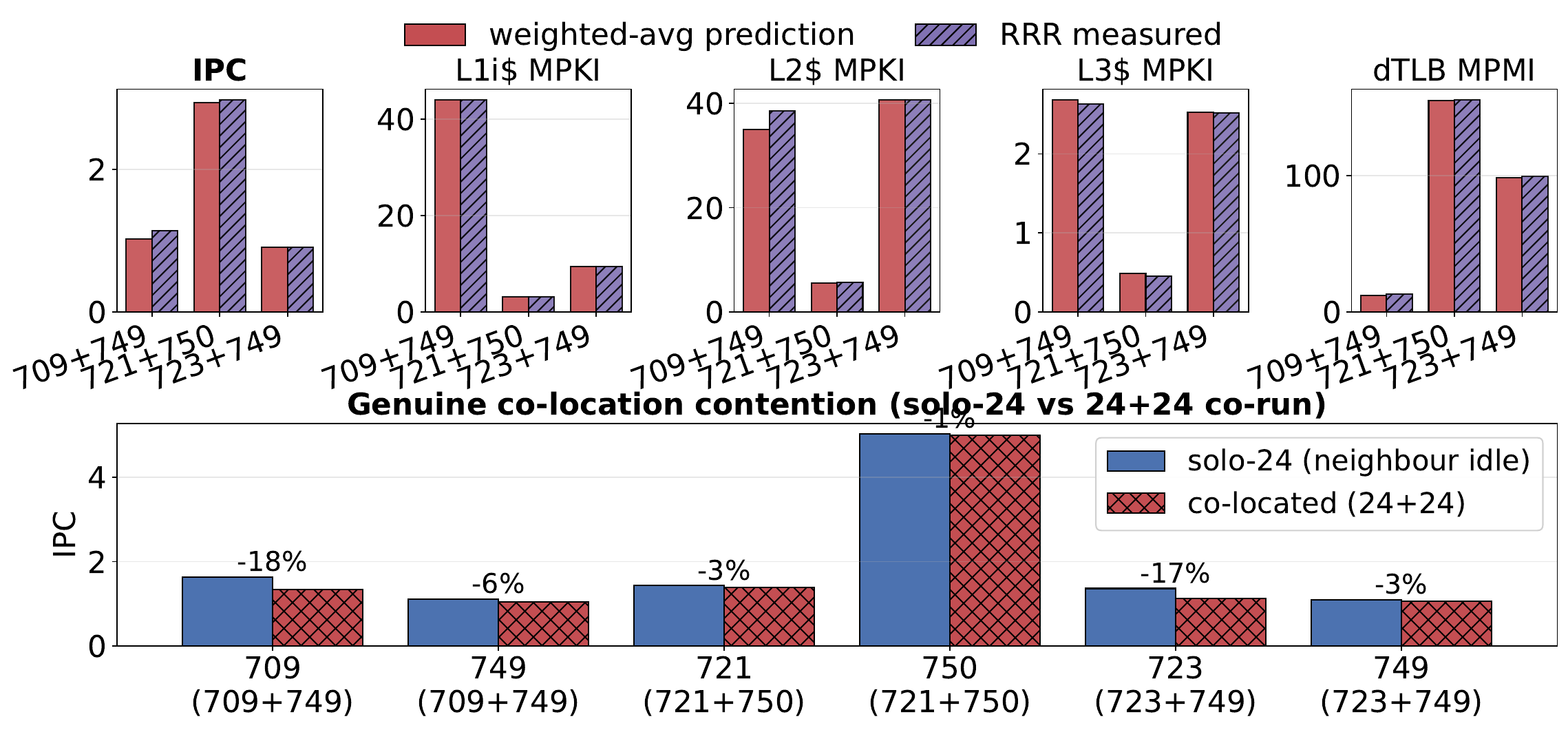}}
    \caption{\small{RRR mixes and true co-location on CPU-C. 
    Top: the measured RRR mix against the instruction-weighted prediction built from the two solo profiles, for pairs 709+749, 721+750, and 723+749. 
    Bottom: each workload's IPC with its partner running on the other 24 cores versus with that half idle (loss annotated per bar). 
    }}
    \label{fig_rrr_generalize}
\end{figure}

\noindent{\textbf{Bracketing a target profile.}}
Alg.~\ref{alg_proxy} derives both the workload pair and its mix ratio from a first-order composition model.
An RRR mix retires the full instruction stream of each constituent to within $0.02\%$, so every per-instruction metric is linear in the instruction share $\alpha=I_A/(I_A+I_B)$, and IPC composes as the weighted harmonic mean
$\mathrm{IPC}_{\mathrm{mix}}=(I_A{+}I_B)/(C_A{+}C_B)
=\big(\alpha/\mathrm{IPC}_A+(1{-}\alpha)/\mathrm{IPC}_B\big)^{-1}$.
Line~\ref{line_bracket} enumerates the pairs in the pool and keeps those whose solo values straddle the target on every declared axis $k\in K$.
Because the achievable set is the segment, a pair that fails to straddle $t_k$ cannot reach it at any ratio, so bracketing is a necessary condition and is cheap enough to test exhaustively at suite scale.
Lines~\ref{line_loop}-\ref{line_fit} fit each surviving pair, choosing the instruction share that minimizes the squared \emph{relative} deviation from the target, so that axes of very different magnitude count equally.
Entering IPC as its reciprocal keeps every axis linear in $\alpha$, leaving a one-dimensional quadratic that is solved in closed form and clipped to $[0,1]$.
Line~\ref{line_select} takes the pair with the smallest residual, and line~\ref{line_slots} converts its $\alpha^\ast$ into whole slots, since RRR schedules entire benchmark slots rather than fractional shares.
The search is over integer weights within the slot budget $S$ whose instruction share $w_AI_A/(w_AI_A{+}w_BI_B)$ is nearest $\alpha^\ast$, so the rounding tracks instructions rather than slot counts.
Line~\ref{line_unreach} records the axes outside $K$ that the chosen pair cannot span, the proxy's fidelity bound, and line~\ref{line_measure} measures the composed mix, because the weighted model is a prediction rather than a guarantee.
\par

We apply this procedure to 26 \SPEC Rate benchmarks profiled at 48 copies, targeting DCPerf \textit{django}'s L1I\$ MPKI and IPC (Figure~\ref{fig_overview_uarch}).
The bracketing requirement sharply narrows the candidates.
709.cactus\_r is the only benchmark whose L1I\$ MPKI (82.3) exceeds \textit{django}'s 39.7, while only 721.gcc\_r and 749.fotonik3d\_r run more slowly than \textit{django} (IPC 1.02).
The fit selects 749.fotonik3d\_r (IPC 0.79), which has high L2 pressure and vector intensity (64.1 L2\$ MPKI and 36\% vector instructions) but negligible L1I\$ pressure (0.3 MPKI).
The resulting 709.cactus\_r + 749.fotonik3d\_r proxy, run at stock RRR's default $1{:}1$ slot ratio, lies between its constituents on the measured metrics (Figure~\ref{fig_rrr}, CPU-C): L1I\$ MPKI 43.9, L2\$/L3\$ MPKI 38.5/2.64, and IPC 1.16.
It lands between \textit{django}'s and \textit{tao}'s L1I\$ MPKI (39.7 and 45.1) at an IPC comparable to \textit{django}'s 1.02.
It lands between \textit{django}'s and \textit{tao}'s L1I\$ MPKI (39.7 and 45.1), while its IPC overshoots \textit{django}'s 1.02 by 14\%.
\par

he mix ratio is likewise derived, not assumed.
The default $1{:}1$ mix overshoots \textit{django}'s L1I\$ MPKI (44.0 vs.\ 39.7) and $3{:}4$ undershoots it (38.0), so the ratio matching \textit{django} on both declared axes lies between them -- exactly where the model places its optimum, at $\alpha=0.476$ against the 0.53 and 0.45 of those two mixes (Figure~\ref{fig_rrr}).
Because stock RRR assigns one slot per benchmark, we added a per-benchmark slot-weight option for profiling only; it is not used for reportable \SPEC scores.
The procedure also bounds its own fidelity by reporting unreachable axes: both candidate pairs bottom out at L1D\$ MPKI 10.7 and 16.1 (721.gcc\_r and 749.fotonik3d\_r), far above \textit{django}'s 4.1.
\par

\noindent{\textbf{Generalization across pairs.}}
We repeat the analysis with two additional pairs, 723.llvm\_r + 749.fotonik3d\_r and 721.gcc\_r + 750.sealcrypto\_r.
The measured RRR profiles generally follow the instruction-weighted solo predictions (Figure~\ref{fig_rrr_generalize}, top).
Weighting by instructions matters for asymmetric pairs: 721.gcc\_r + 750.sealcrypto\_r measures 156 dTLB MPMI and is predicted at 155, versus an unweighted midpoint of 279.
Only four of the fifteen plotted quantities depart by more than 5\%: IPC, L2\$ MPKI, and dTLB MPMI for 709.cactus\_r + 749.fotonik3d\_r ($+12\%$, $+10\%$, and $+7\%$), and L3\$ MPKI for 721.gcc\_r + 750.sealcrypto\_r ($-7\%$).
Weighted solo profiles thus provide useful first-order estimates, but the constructed mix must still be measured.
\par

\noindent{\textbf{Time-rolled RRR vs.\ true co-location.}}
What makes RRR predictable also limits it: because RRR mixes workloads over time rather than running them together, a mix tracks the instruction-weighted average of its constituents and does not capture the contention that simultaneous tenants impose on one another~\cite{jiang2020characterizing, gan2019open}.
The 709.cactus\_r + 749.fotonik3d\_r mix, in fact, exceeds its instruction-weighted prediction by $12\%$ in IPC (Figure~\ref{fig_rrr_generalize}, top): only part of the machine runs the memory-streaming 749.fotonik3d\_r at any instant, so mixing eases contention rather than adding it.
To measure the omitted contention, we partition CPU-C's 48 cores into two disjoint 24-core halves and run one workload of a pair on each half at the same time, 24 copies per half.
The baseline reruns each workload on its own half with the other half idle (Figure~\ref{fig_rrr_generalize}, bottom), so copy count and core count are unchanged and the difference isolates shared-cache and memory interference.
\par

The interference is strongly asymmetric.
Streaming 749.fotonik3d\_r costs its partners 18\% (709.cactus\_r) and 17\% (723.llvm\_r) of their IPC while losing only 3-6\% itself; compute-bound 750.sealcrypto\_r is almost immune (-1\%) and costs 721.gcc\_r only 3\%.
A proxy's disruptiveness is therefore set by its memory demand rather than the share of the machine it occupies, and it must be reported per tenant rather than as a single number for the pair. \par

\noindent\underline{\textbf{Takeaway:}}
RRR can be used to construct desired microarchitectural proxies by combining distinct \SPEC workloads.
A target profile and declared axes determine the initial workload pair and mix ratio, and measured mixes can be tuned to achieve desired goals.
However, because RRR time-rolls its constituents instead of running them simultaneously, it generates proxies that may increase or decrease contention; true co-location effects may require separate measurement.
\par

\section{Representative Workloads in \SPEC}
\label{section_subset}
\begin{figure*}[t]
    \begin{minipage}{\columnwidth}
	\centering
	\includegraphics[width=\columnwidth, trim = 0mm 0mm 0mm 0mm, clip=true, page=1]{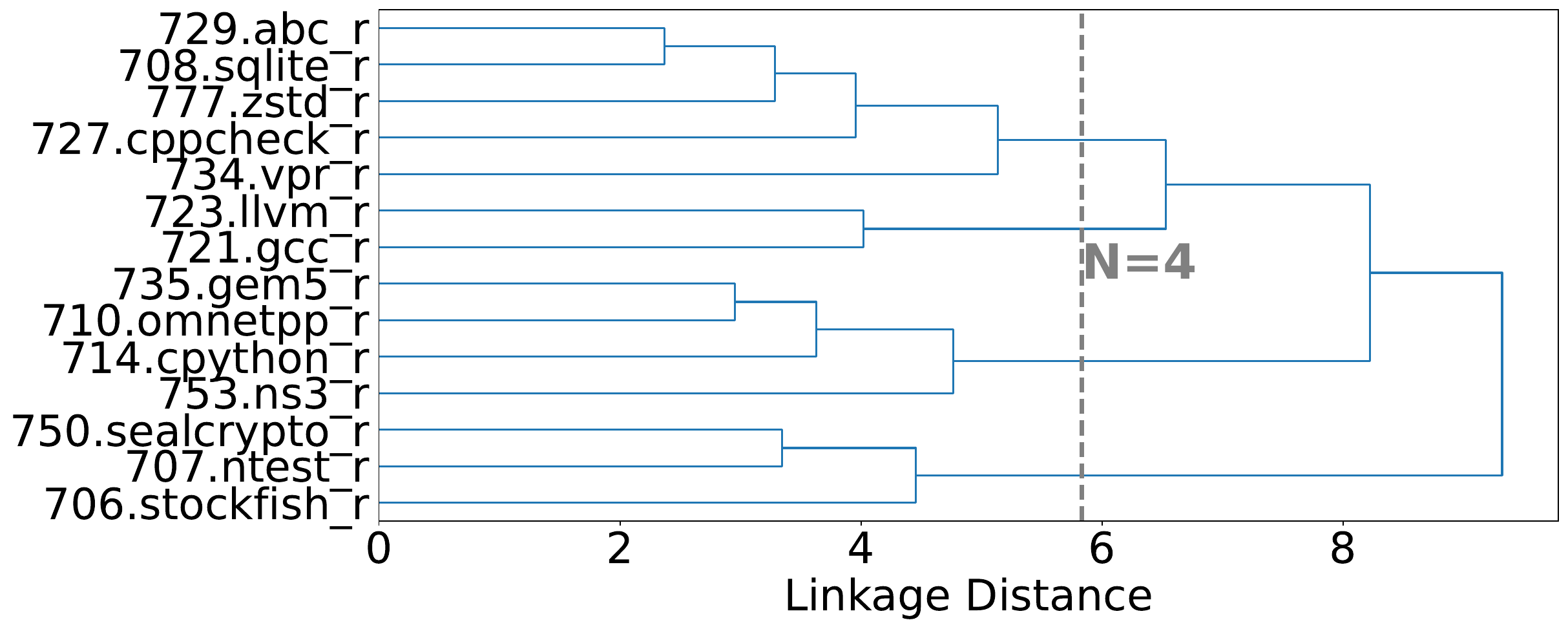}
	\subcaption{\SPEC INT Rate. }
	\label{fig_dendrogram_cpu2026_intrate}
	\end{minipage} 
    \begin{minipage}{\columnwidth}
	\centering
	\includegraphics[width=\columnwidth, trim = 0mm 0mm 0mm 0mm, clip=true, page=1]{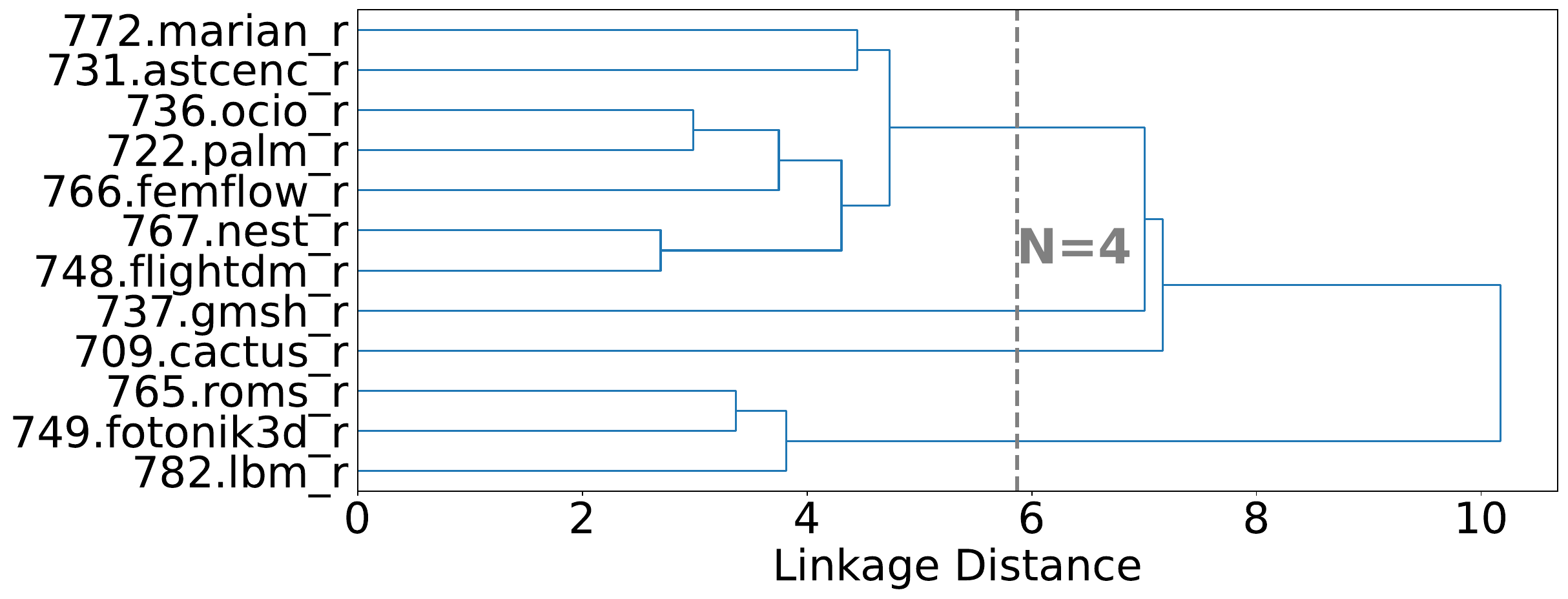}
	\subcaption{\SPEC FP Rate. }
	\label{fig_dendrogram_cpu2026_fprate}
	\end{minipage} 
    \begin{minipage}{\columnwidth}
	\centering
	\includegraphics[width=\columnwidth, trim = 0mm 0mm 0mm 0mm, clip=true, page=1]{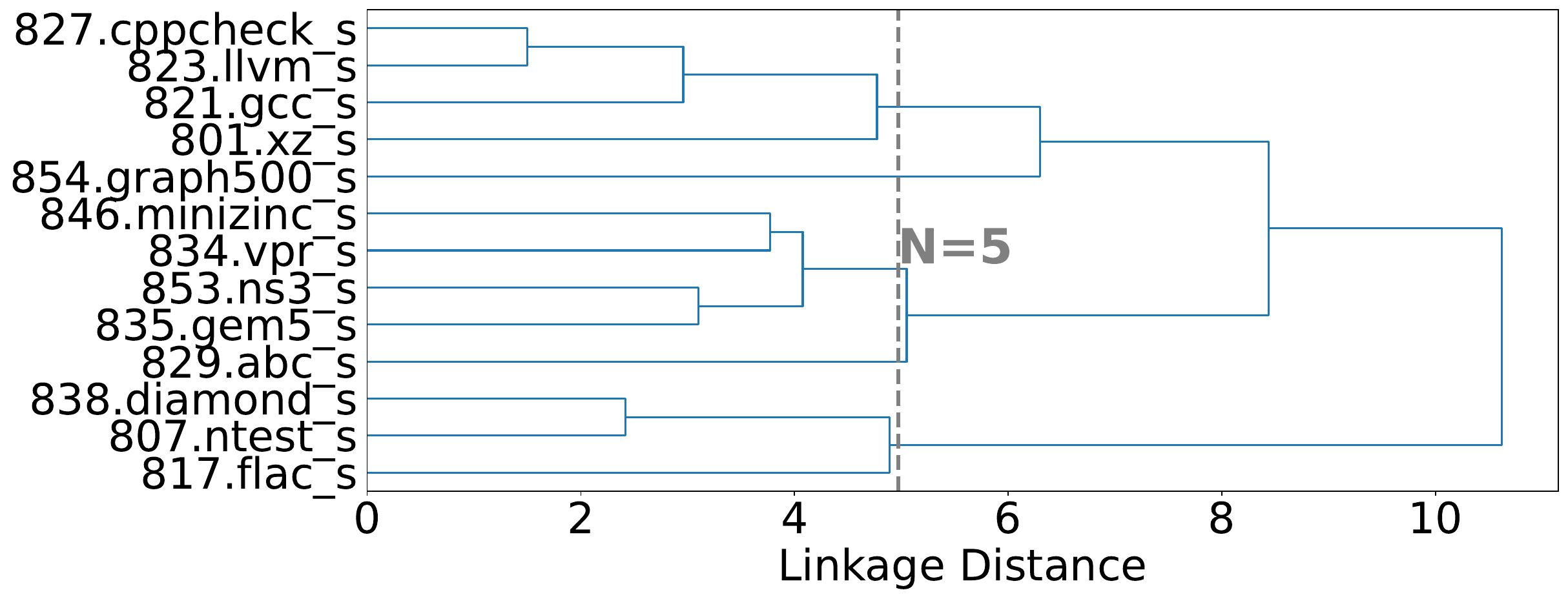}
	\subcaption{\SPEC INT Speed. }
	\label{fig_dendrogram_cpu2026_intspeed}
	\end{minipage}
    \begin{minipage}{\columnwidth}
	\centering
	\includegraphics[width=\columnwidth, trim = 0mm 0mm 0mm 0mm, clip=true, page=1]{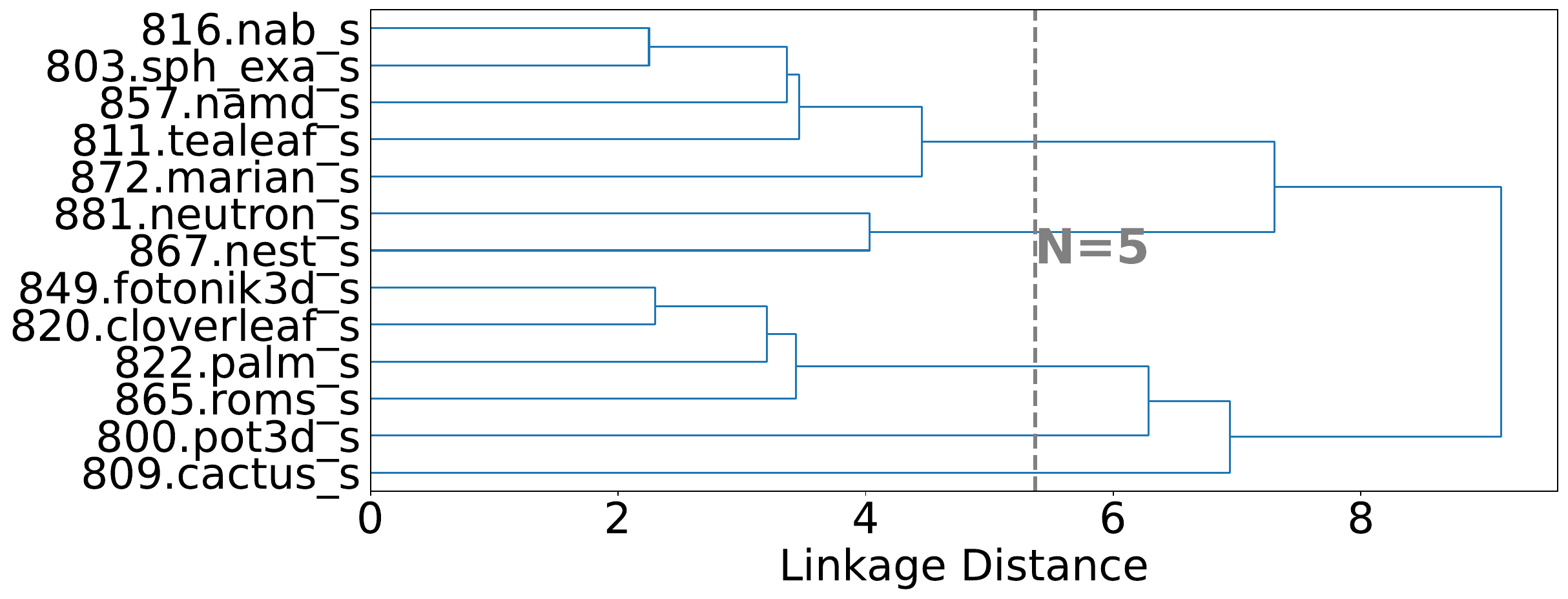}
	\subcaption{\SPEC FP Speed. }
	\label{fig_dendrogram_cpu2026_fpspeed}
	\end{minipage}
 \caption{Dendrogram showing similarity between \SPEC workloads; shorter linkage distance indicates higher similarity.}
 \label{fig_dendrogram_cpu2026}
\end{figure*}

This section applies the hierarchical clustering method of \S~\ref{section_03_02} to identify representative subsets that preserve the full \SPEC suite's behavioral diversity.
We cut each dendrogram at a fixed linkage distance to form behaviorally similar groups; Figure~\ref {fig_dendrogram_cpu2026} illustrates this process for the INT/FP Rate/Speed suites in \SPEC.
Within each group, we select the medoid, i.e., the workload with minimum average distance to other group members in PCA space.
To quantify subset fidelity, we define subset \emph{accuracy} as $1-\mathrm{err}$, where $\mathrm{err}=|\mathrm{GM}(\text{subset})-\mathrm{GM}(\text{suite})|/\mathrm{GM}(\text{suite})$, and $\mathrm{GM}(\cdot)$ is the geometric-mean score of \SPEC.
Table~\ref{table_subset_accuracy} summarizes the subsets for all four \SPEC groups.

\noindent{\underline{\textbf{Takeaway:}}} Compact subsets of 4-5 workloads per group achieve 96.4-99.9\% accuracy, substantially reducing evaluation cost while preserving representative behavior.

\begin{table}[t]
\caption{\small{Representative subset per group and accuracy.}} 
\label{table_subset_accuracy}
\centering
\footnotesize
\setlength{\tabcolsep}{1pt}
\renewcommand{\arraystretch}{1.1}
\begin{tabular}{|| >{\arraybackslash}p{1.1cm} | >{\arraybackslash}p{6cm} | >{\centering\arraybackslash}p{1.1cm} ||}
\hline
\rowcolor{blue!20}
Group & Subset workloads & Accuracy \\
\hline
int\_rate & 750.sealcrypto\_r, 735.gem5\_r, 721.gcc\_r, 708.sqlite\_r & 99.90\% \\
\hline
fp\_rate & 749.fotonik3d\_r, 709.cactus\_r, 737.gmsh\_r, 767.nest\_r & 99.07\% \\
\hline
int\_speed & 807.ntest\_s, 854.graph500\_s, 823.llvm\_s, 829.abc\_s, 835.gem5\_s & 96.61\% \\
\hline
fp\_speed & 867.nest\_s, 816.nab\_s, 809.cactus\_s, 800.pot3d\_s, 849.fotonik3d\_s & 96.40\% \\
\hline
\end{tabular}
\end{table}

\section{Conclusion}
\label{section_conclusion}
This paper presents a comprehensive characterization of the just-released \SPEC benchmark suite.
Across nine platforms, \SPEC broadens coverage over \SPECO through larger instruction volumes and memory footprints and greater pressure on CPU bottlenecks, especially the instruction cache.
Comparisons with \SPECO, DCPerf, and MLPerf show that \SPEC remains a general-purpose suite with proximity to foundational datacenter workloads while staying distinct from vector-centric MLPerf inference; an agentic AI pipeline falls within that coverage, while adding a dense GEMM kernel reaches only MLPerf's dense-kernel corner rather than closing the gap to MLPerf.
Our case studies demonstrate value for many explorations, including page size tradeoffs, prefetching, compiler sensitivity, and scalability, making \SPEC a forward-looking complement to domain-specific suites.

\section*{Acknowledgment}
The authors would like to thank the SPEC CPU Subcommittee for the opportunity to collaborate on the clustering analysis for the SPEC CPU2026 program selection. 
We are grateful to the member companies -- specifically AMD, Ampere, IBM, and Intel -- for providing the necessary data. 
Special thanks go to John Henning, Mahesh Madhav, and Frédérique Silber-Chaussumier for their invaluable technical support and for facilitating the analysis. 
Additionally, we thank the Texas Advanced Computing Center (TACC) at the University of Texas at Austin for providing the computing resources that supported this work.

\bibliographystyle{ACM-Reference-Format}
\bibliography{refs}

@inproceedings{panda2018wait,
  title={Wait of a decade: Did spec cpu 2017 broaden the performance horizon?},
  author={Panda, Reena and Song, Shuang and Dean, Joseph and John, Lizy K},
  booktitle={2018 IEEE International Symposium on High Performance Computer Architecture (HPCA)},
  pages={271--282},
  year={2018},
  organization={IEEE}
}

@inproceedings{limaye2018workload,
  title={A workload characterization of the spec cpu2017 benchmark suite},
  author={Limaye, Ankur and Adegbija, Tosiron},
  booktitle={2018 IEEE International Symposium on Performance Analysis of Systems and Software (ISPASS)},
  pages={149--158},
  year={2018},
  organization={IEEE}
}

@inproceedings{phansalkar2007analysis,
  title={Analysis of redundancy and application balance in the SPEC CPU2006 benchmark suite},
  author={Phansalkar, Aashish and Joshi, Ajay and John, Lizy K},
  booktitle={Proceedings of the 34th annual International Symposium on Computer architecture},
  pages={412--423},
  year={2007}
}

@article{spradling2007spec,
  title={SPEC CPU2006 benchmark tools},
  author={Spradling, Cloyce D},
  journal={ACM SIGARCH Computer Architecture News},
  volume={35},
  number={1},
  pages={130--134},
  year={2007},
  publisher={ACM New York, NY, USA}
}

@article{henning2006spec,
  title={SPEC CPU2006 benchmark descriptions},
  author={Henning, John L},
  journal={ACM SIGARCH Computer Architecture News},
  volume={34},
  number={4},
  pages={1--17},
  year={2006},
  publisher={ACM New York, NY, USA}
}

@inproceedings{ye2006performance,
  title={Performance characterization of SPEC CPU2006 integer benchmarks on x86-64 architecture},
  author={Ye, Dong and Ray, Joydeep and Harle, Christophe and Kaeli, David},
  booktitle={2006 IEEE International Symposium on Workload Characterization},
  pages={120--127},
  year={2006},
  organization={IEEE}
}

@inproceedings{su2025dcperf,
  title={DCPerf: An Open-Source, Battle-Tested Performance Benchmark Suite for Datacenter Workloads},
  author = {Su, Wei and Dhanotia, Abhishek and Torres, Carlos and Gandhi, Jayneel and Gholkar, Neha and Kanaujia, Shobhit and Naumov, Maxim and Subramanian, Kalyan and Andrei, Valentin and Yuan, Yifan and Tang, Chunqiang},
  booktitle={Proceedings of the 52nd Annual International Symposium on Computer Architecture},
  pages={1717--1730},
  year={2025}
}

@inproceedings{rasley2020deepspeed,
  title={Deepspeed: System optimizations enable training deep learning models with over 100 billion parameters},
  author={Rasley, Jeff and Rajbhandari, Samyam and Ruwase, Olatunji and He, Yuxiong},
  booktitle={Proceedings of the 26th ACM SIGKDD international conference on knowledge discovery \& data mining},
  pages={3505--3506},
  year={2020}
}

@inproceedings{aminabadi2022deepspeed,
  title={Deepspeed-inference: enabling efficient inference of transformer models at unprecedented scale},
  author = {Aminabadi, Reza Yazdani and Rajbhandari, Samyam and Awan, Ammar Ahmad and Li, Cheng and Li, Du and Zheng, Elton and Ruwase, Olatunji and Smith, Shaden and Zhang, Minjia and Rasley, Jeff and He, Yuxiong},
  booktitle={SC22: International Conference for High Performance Computing, Networking, Storage and Analysis},
  pages={1--15},
  year={2022},
  organization={IEEE}
}

@inproceedings{kwon2023efficient,
  title={Efficient memory management for large language model serving with pagedattention},
  author={Kwon, Woosuk and Li, Zhuohan and Zhuang, Siyuan and Sheng, Ying and Zheng, Lianmin and Yu, Cody Hao and Gonzalez, Joseph and Zhang, Hao and Stoica, Ion},
  booktitle={Proceedings of the 29th symposium on operating systems principles},
  pages={611--626},
  year={2023}
}

@inproceedings{patel2024splitwise,
  title={Splitwise: Efficient generative llm inference using phase splitting},
  author={Patel, Pratyush and Choukse, Esha and Zhang, Chaojie and Shah, Aashaka and Goiri, {\'I}{\~n}igo and Maleki, Saeed and Bianchini, Ricardo},
  booktitle={2024 ACM/IEEE 51st Annual International Symposium on Computer Architecture (ISCA)},
  pages={118--132},
  year={2024},
  organization={IEEE}
}

@inproceedings{zhao2025insights,
  title={Insights into deepseek-v3: Scaling challenges and reflections on hardware for ai architectures},
  author = {Zhao, Chenggang and Deng, Chengqi and Ruan, Chong and Dai, Damai and Gao, Huazuo and Li, Jiashi and Zhang, Liyue and Huang, Panpan and Zhou, Shangyan and Ma, Shirong and Liang, Wenfeng and He, Ying and Wang, Yuqing and Liu, Yuxuan and Wei, Y.X.},
  booktitle={Proceedings of the 52nd Annual International Symposium on Computer Architecture},
  pages={1731--1745},
  year={2025}
}

@inproceedings{chu2025scaling,
  title={Scaling Llama 3 Training with Efficient Parallelism Strategies},
  author = {Chu, Weiwei and Xie, Xinfeng and Yu, Jiecao and Wang, Jie and Phanishayee, Amar and Tang, Chunqiang and Hao, Yuchen and Huang, Jianyu and Ozdal, Mustafa and Wang, Jun and Goswami, Vedanuj and Goyal, Naman and Kadian, Abhishek and Gu, Andrew and Cai, Chris and Tian, Feng and Wang, Xiaodong and Si, Min and Balaji, Pavan and Chu, Ching-Hsiang and Park, Jongsoo},
  booktitle={Proceedings of the 52nd Annual International Symposium on Computer Architecture},
  pages={1703--1716},
  year={2025}
}

@inproceedings{jouppi2017datacenter,
  title={In-datacenter performance analysis of a tensor processing unit},
  author = {Jouppi, Norman P. and Young, Cliff and Patil, Nishant and Patterson, David and Agrawal, Gaurav and Bajwa, Raminder and Bates, Sarah and Bhatia, Suresh and Boden, Nan and Borchers, Al and Boyle, Rick and Cantin, Pierre-luc and Chao, Clifford and Clark, Chris and Coriell, Jeremy and Daley, Mike and Dau, Matt and Dean, Jeffrey and Gelb, Ben and Ghaemmaghami, Tara Vazir and Gottipati, Rajendra and Gulland, William and Hagmann, Robert and Ho, C. Richard and Hogberg, Doug and Hu, John and Hundt, Robert and Hurt, Dan and Ibarz, Julian and Jaffey, Aaron and Jaworski, Alek and Kaplan, Alexander and Khaitan, Harshit and Killebrew, Daniel and Koch, Andy and Kumar, Naveen and Lacy, Steve and Laudon, James and Law, James and Le, Diemthu and Leary, Chris and Liu, Zhuyuan and Lucke, Kyle and Lundin, Alan and MacKean, Gordon and Maggiore, Adriana and Mahony, Maire and Miller, Kieran and Nagarajan, Rahul and Narayanaswami, Ravi and Ni, Ray and Nix, Kathy and Norrie, Thomas and Omernick, Mark and Penukonda, Narayana and Phelps, Andy and Ross, Jonathan and Ross, Matt and Salek, Amir and Samadiani, Emad and Severn, Chris and Sizikov, Gregory and Snelham, Matthew and Souter, Jed and Steinberg, Dan and Swing, Andy and Tan, Mercedes and Thorson, Gregory and Tian, Bo and Toma, Horia and Tuttle, Erick and Vasudevan, Vijay and Walter, Richard and Wang, Walter and Wilcox, Eric and Yoon, Doe Hyun},
  booktitle={Proceedings of the 44th annual international symposium on computer architecture},
  pages={1--12},
  year={2017}
}

@inproceedings{jouppi2023tpu,
  title={Tpu v4: An optically reconfigurable supercomputer for machine learning with hardware support for embeddings},
  author = {Jouppi, Norm and Kurian, George and Li, Sheng and Ma, Peter and Nagarajan, Rahul and Nai, Lifeng and Patil, Nishant and Subramanian, Suvinay and Swing, Andy and Towles, Brian and Young, Clifford and Zhou, Xiang and Zhou, Zongwei and Patterson, David A},
  booktitle={Proceedings of the 50th annual international symposium on computer architecture},
  pages={1--14},
  year={2023}
}

@inproceedings{firoozshahian2023mtia,
  title={Mtia: First generation silicon targeting meta's recommendation systems},
  author = {Firoozshahian, Amin and Coburn, Joel and Levenstein, Roman and Nattoji, Rakesh and Kamath, Ashwin and Wu, Olivia and Grewal, Gurdeepak and Aepala, Harish and Jakka, Bhasker and Dreyer, Bob and Hutchin, Adam and Diril, Utku and Nair, Krishnakumar and Aredestani, Ehsan K. and Schatz, Martin and Hao, Yuchen and Komuravelli, Rakesh and Ho, Kunming and Abu Asal, Sameer and Shajrawi, Joe and Quinn, Kevin and Sreedhara, Nagesh and Kansal, Pankaj and Wei, Willie and Jayaraman, Dheepak and Cheng, Linda and Chopda, Pritam and Wang, Eric and Bikumandla, Ajay and Karthik Sengottuvel, Arun and Thottempudi, Krishna and Narasimha, Ashwin and Dodds, Brian and Gao, Cao and Zhang, Jiyuan and Al-Sanabani, Mohammed and Zehtabioskuie, Ana and Fix, Jordan and Yu, Hangchen and Li, Richard and Gondkar, Kaustubh and Montgomery, Jack and Tsai, Mike and Dwarakapuram, Saritha and Desai, Sanjay and Avidan, Nili and Ramani, Poorvaja and Narayanan, Karthik and Mathews, Ajit and Gopal, Sethu and Naumov, Maxim and Rao, Vijay and Noru, Krishna and Reddy, Harikrishna and Venkatapuram, Prahlad and Bjorlin, Alexis},
  booktitle={Proceedings of the 50th Annual International Symposium on Computer Architecture},
  pages={1--13},
  year={2023}
}

@inproceedings{coburn2025meta,
  title={Meta's Second Generation AI Chip: Model-Chip Co-Design and Productionization Experiences},
  author = {Coburn, Joel and Tang, Chunqiang and Asal, Sameer Abu and Agrawal, Neeraj and Chinta, Raviteja and Dixit, Harish and Dodds, Brian and Dwarakapuram, Saritha and Firoozshahian, Amin and Gao, Cao and Gondkar, Kaustubh and Graf, Tyler and Hu, Junhan and Huang, Jian and Hughes, Sterling and Hutchin, Adam and Jakka, Bhasker and Chen, Guoqiang Jerry and Kalyanaraman, Indu and Kamath, Ashwin and Kansal, Pankaj and Kazi, Erum and Levenstein, Roman and Maddury, Mahesh and Mastro, Alex and Medaiyese, Siji and Modi, Pritesh and Montgomery, Jack and Nadathur, Satish and Nagpal, Amit and Narasimha, Ashwin and Naumov, Maxim and Ozer, Eleanor and Park, Jongsoo and Ramani, Poorvaja and Reddy, Harikrishna and Reiss, David and Roy, Deboleena and Sekar, Sathish and Sharma, Arushi and Shetty, Pavan and Sukumaran-Rajam, Aravind and Tal, Eran and Tsai, Mike and Varshini, Shreya and Wareing, Richard and Wu, Olivia and Xie, Xiaolong and Yang, Jinghan and Yu, Hangchen and Zargar, Tanmay and Zeng, Zitong and Zhang, Feixiong and Matthews, Ajit and Jiao, Xun and Zhang, Jiyuan and Menage, Emmanuel and Stokke, Truls Edvard and Sourouri, Mohammed},
  booktitle={Proceedings of the 52nd Annual International Symposium on Computer Architecture},
  pages={1689--1702},
  year={2025}
}

@inproceedings{imani2019floatpim,
  title={Floatpim: In-memory acceleration of deep neural network training with high precision},
  author={Imani, Mohsen and Gupta, Saransh and Kim, Yeseong and Rosing, Tajana},
  booktitle={Proceedings of the 46th International Symposium on Computer Architecture},
  pages={802--815},
  year={2019}
}

@article{chen2016eyeriss,
  title={Eyeriss: An energy-efficient reconfigurable accelerator for deep convolutional neural networks},
  author={Chen, Yu-Hsin and Krishna, Tushar and Emer, Joel S and Sze, Vivienne},
  journal={IEEE journal of solid-state circuits},
  volume={52},
  number={1},
  pages={127--138},
  year={2016},
  publisher={IEEE}
}

@inproceedings{xu2025wsc,
  title={WSC-LLM: Efficient LLM Service and Architecture Co-exploration for Wafer-scale Chips},
  author={Xu, Zheng and Kong, Dehao and Liu, Jiaxin and Li, Jinxi and Hou, Jingxiang and Dai, Xu and Li, Chao and Wei, Shaojun and Hu, Yang and Yin, Shouyi},
  booktitle={Proceedings of the 52nd Annual International Symposium on Computer Architecture},
  pages={1--17},
  year={2025}
}

@inproceedings{ishida2020supernpu,
  title={SuperNPU: An extremely fast neural processing unit using superconducting logic devices},
  author={Ishida, Koki and Byun, Ilkwon and Nagaoka, Ikki and Fukumitsu, Kosuke and Tanaka, Masamitsu and Kawakami, Satoshi and Tanimoto, Teruo and Ono, Takatsugu and Kim, Jangwoo and Inoue, Koji},
  booktitle={2020 53rd Annual IEEE/ACM International Symposium on Microarchitecture (MICRO)},
  pages={58--72},
  year={2020},
  organization={IEEE}
}

@inproceedings{fowers2018configurable,
  title={A configurable cloud-scale DNN processor for real-time AI},
  author={Fowers, Jeremy and Ovtcharov, Kalin and Papamichael, Michael and Massengill, Todd and Liu, Ming and Lo, Daniel and Alkalay, Shlomi and Haselman, Michael and Adams, Logan and Ghandi, Mahdi and Heil, Stephen and Patel, Prerak and Sapek, Adam and Weisz, Gabriel and Woods, Lisa and Lanka, Sitaram and Reinhardt, Steven K. and Caulfield, Adrian M. and Chung, Eric S. and Burger, Doug},
  booktitle={2018 ACM/IEEE 45th Annual International Symposium on Computer Architecture (ISCA)},
  pages={1--14},
  year={2018},
  organization={IEEE}
}

@inproceedings{zhang2015optimizing,
  title={Optimizing FPGA-based accelerator design for deep convolutional neural networks},
  author={Zhang, Chen and Li, Peng and Sun, Guangyu and Guan, Yijin and Xiao, Bingjun and Cong, Jason},
  booktitle={Proceedings of the 2015 ACM/SIGDA international symposium on field-programmable gate arrays},
  pages={161--170},
  year={2015}
}

@inproceedings{alwani2016fused,
  title={Fused-layer CNN accelerators},
  author={Alwani, Manoj and Chen, Han and Ferdman, Michael and Milder, Peter},
  booktitle={2016 49th Annual IEEE/ACM International Symposium on Microarchitecture (MICRO)},
  pages={1--12},
  year={2016},
  organization={IEEE}
}

@article{shen2017maximizing,
  title={Maximizing CNN accelerator efficiency through resource partitioning},
  author={Shen, Yongming and Ferdman, Michael and Milder, Peter},
  journal={ACM SIGARCH Computer Architecture News},
  volume={45},
  number={2},
  pages={535--547},
  year={2017},
  publisher={ACM New York, NY, USA}
}

@article{chen2024understanding,
  title={Understanding the potential of fpga-based spatial acceleration for large language model inference},
  author = {Chen, Hongzheng and Zhang, Jiahao and Du, Yixiao and Xiang, Shaojie and Yue, Zichao and Zhang, Niansong and Cai, Yaohui and Zhang, Zhiru},
  journal={ACM Transactions on Reconfigurable Technology and Systems},
  volume={18},
  number={1},
  pages={1--29},
  year={2024},
  publisher={ACM New York, NY}
}

@article{mattson2020mlperf,
  title={Mlperf training benchmark},
  author={Peter Mattson and Christine Cheng and Cody Coleman and Greg Diamos and Paulius Micikevicius and David Patterson and Hanlin Tang and Gu-Yeon Wei and Peter Bailis and Victor Bittorf and David Brooks and Dehao Chen and Debojyoti Dutta and Udit Gupta and Kim Hazelwood and Andrew Hock and Xinyuan Huang and Atsushi Ike and Bill Jia and Daniel Kang and David Kanter and Naveen Kumar and Jeffery Liao and Guokai Ma and Deepak Narayanan and Tayo Oguntebi and Gennady Pekhimenko and Lillian Pentecost and Vijay Janapa Reddi and Taylor Robie and Tom St. John and Tsuguchika Tabaru and Carole-Jean Wu and Lingjie Xu and Masafumi Yamazaki and Cliff Young and Matei Zaharia},
  journal={Proceedings of Machine Learning and Systems},
  volume={2},
  pages={336--349},
  year={2020}
}

@inproceedings{reddi2020mlperf,
  title={Mlperf inference benchmark},
  author={Vijay Janapa Reddi and Christine Cheng and David Kanter and Peter Mattson and Guenther Schmuelling and Carole-Jean Wu and Brian Anderson and Maximilien Breughe and Mark Charlebois and William Chou and Ramesh Chukka and Cody Coleman and Sam Davis and Pan Deng and Greg Diamos and Jared Duke and Dave Fick and J. Scott Gardner and Itay Hubara and Sachin Idgunji and Thomas B. Jablin and Jeff Jiao and Tom St. John and Pankaj Kanwar and David Lee and Jeffery Liao and Anton Lokhmotov and Francisco Massa and Peng Meng and Paulius Micikevicius and Colin Osborne and Gennady Pekhimenko and Arun Tejusve Raghunath Rajan and Dilip Sequeira and Ashish Sirasao and Fei Sun and Hanlin Tang and Michael Thomson and Frank Wei and Ephrem Wu and Lingjie Xu and Koichi Yamada and Bing Yu and George Yuan and Aaron Zhong and Peizhao Zhang and Yuchen Zhou},
  booktitle={2020 ACM/IEEE 47th Annual International Symposium on Computer Architecture (ISCA)},
  pages={446--459},
  year={2020},
  organization={IEEE}
}

@article{janapa2022mlperf,
  title={MLPerf mobile inference benchmark: An industry-standard open-source machine learning benchmark for on-device AI},
  author={Vijay Janapa Reddi and David Kanter and Peter Mattson and Jared Duke and Thai Nguyen and Ramesh Chukka and Ken Shiring and Koan-Sin Tan and Mark Charlebois and William Chou and Mostafa El-Khamy and Jungwook Hong and Tom St. John and Cindy Trinh and Michael Buch and Mark Mazumder and Relia Markovic and Thomas Atta and Fatih Cakir and Masoud Charkhabi and Xiaodong Chen and Cheng-Ming Chiang and Dave Dexter and Terry Heo and Gunther Schmuelling and Maryam Shabani and Dylan Zika},
  journal={Proceedings of Machine Learning and Systems},
  volume={4},
  pages={352--369},
  year={2022}
}

@inproceedings{tschand2025mlperf,
  title={MLPerf Power: Benchmarking the Energy Efficiency of Machine Learning Systems from $\mu$Watts to MWatts for Sustainable AI},
  author={Tschand, Arya and Rajan, Arun Tejusve Raghunath and Idgunji, Sachin and Ghosh, Anirban and Holleman, Jeremy and Kiraly, Csaba and Ambalkar, Pawan and Borkar, Ritika and Chukka, Ramesh and Cockrell, Trevor and Curtis, Oliver and Fursin, Grigori and Hodak, Miro and Kassa, Hiwot and Lokhmotov, Anton and Miskovic, Dejan and Pan, Yuechao and Manmathan, Manu Prasad and Raymond, Liz and John, Tom St. and Suresh, Arjun and Taubitz, Rowan and Zhan, Sean and Wasson, Scott and Kanter, David and Reddi, Vijay Janapa},
  booktitle={2025 IEEE International Symposium on High Performance Computer Architecture (HPCA)},
  pages={1201--1216},
  year={2025},
  organization={IEEE}
}

@inproceedings{zhao2022understanding,
  title={Understanding data storage and ingestion for large-scale deep recommendation model training: Industrial product},
  author = {Zhao, Mark and Agarwal, Niket and Basant, Aarti and Gedik, Bu\u{g}ra and Pan, Satadru and Ozdal, Mustafa and Komuravelli, Rakesh and Pan, Jerry and Bao, Tianshu and Lu, Haowei and Narayanan, Sundaram and Langman, Jack and Wilfong, Kevin and Rastogi, Harsha and Wu, Carole-Jean and Kozyrakis, Christos and Pol, Parik},
  booktitle={Proceedings of the 49th annual international symposium on computer architecture},
  pages={1042--1057},
  year={2022}
}

@inproceedings{sriraman2019softsku,
  title={Softsku: Optimizing server architectures for microservice diversity@ scale},
  author={Sriraman, Akshitha and Dhanotia, Abhishek and Wenisch, Thomas F},
  booktitle={Proceedings of the 46th International Symposium on Computer Architecture},
  pages={513--526},
  year={2019}
}

@inproceedings{kanev2015profiling,
  title={Profiling a warehouse-scale computer},
  author={Kanev, Svilen and Darago, Juan Pablo and Hazelwood, Kim and Ranganathan, Parthasarathy and Moseley, Tipp and Wei, Gu-Yeon and Brooks, David},
  booktitle={Proceedings of the 42nd annual international symposium on computer architecture},
  pages={158--169},
  year={2015}
}

@inproceedings{zhao2023contiguitas,
  title={Contiguitas: The pursuit of physical memory contiguity in datacenters},
  author = {Zhao, Kaiyang and Xue, Kaiwen and Wang, Ziqi and Schatzberg, Dan and Yang, Leon and Manousis, Antonis and Weiner, Johannes and Van Riel, Rik and Sharma, Bikash and Tang, Chunqiang and Skarlatos, Dimitrios},
  booktitle={Proceedings of the 50th Annual International Symposium on Computer Architecture},
  pages={1--15},
  year={2023}
}

@inproceedings{gonzalez2023profiling,
  title={Profiling hyperscale big data processing},
  author={Gonzalez, Abraham and Kolli, Aasheesh and Khan, Samira and Liu, Sihang and Dadu, Vidushi and Karandikar, Sagar and Chang, Jichuan and Asanovic, Krste and Ranganathan, Parthasarathy},
  booktitle={Proceedings of the 50th Annual International Symposium on Computer Architecture},
  pages={1--16},
  year={2023}
}

@inproceedings{nasr2025concorde,
  title={Concorde: Fast and Accurate CPU Performance Modeling with Compositional Analytical-ML Fusion},
  author = {Nasr-Esfahany, Arash and Alizadeh, Mohammad and Lee, Victor and Alam, Hanna and Coon, Brett W. and Culler, David and Dadu, Vidushi and Dixon, Martin and Levy, Henry M. and Pandey, Santosh and Ranganathan, Parthasarathy and Yazdanbakhsh, Amir},
  booktitle={Proceedings of the 52nd Annual International Symposium on Computer Architecture},
  pages={1480--1494},
  year={2025}
}

@article{um2023fastflow,
  title={Fastflow: Accelerating deep learning model training with smart offloading of input data pipeline},
  author={Um, Taegeon and Oh, Byungsoo and Seo, Byeongchan and Kweun, Minhyeok and Kim, Goeun and Lee, Woo-Yeon},
  journal={Proceedings of the VLDB Endowment},
  volume={16},
  number={5},
  pages={1086--1099},
  year={2023},
  publisher={VLDB Endowment}
}

@article{jiang2025neo,
  title={Neo: Saving gpu memory crisis with cpu offloading for online llm inference},
  author={Jiang, Xuanlin and Zhou, Yang and Cao, Shiyi and Stoica, Ion and Yu, Minlan},
  journal={Proceedings of Machine Learning and Systems},
  volume={7},
  year={2025}
}

@article{na2025flexinfer,
  title={Flexinfer: Flexible llm inference with cpu computations},
  author={Na, Seonjin and Jeong, Geonhwa and Ahn, Byung H and Jezghani, Aaron and Young, Jeffrey and Hughes, Christopher J and Krishna, Tushar and Kim, Hyesoon},
  journal={Proceedings of Machine Learning and Systems},
  volume={7},
  year={2025}
}

@inproceedings{jain2023optimizing,
  title={Optimizing cpu performance for recommendation systems at-scale},
  author = {Jain, Rishabh and Cheng, Scott and Kalagi, Vishwas and Sanghavi, Vrushabh and Kaul, Samvit and Arunachalam, Meena and Maeng, Kiwan and Jog, Adwait and Sivasubramaniam, Anand and Kandemir, Mahmut Taylan and Das, Chita R.},
  booktitle={Proceedings of the 50th Annual International Symposium on Computer Architecture},
  pages={1--15},
  year={2023}
}

@inproceedings{yahya2022agilewatts,
  title={Agilewatts: An energy-efficient cpu core idle-state architecture for latency-sensitive server applications},
  author={Yahya, Jawad Haj and Volos, Haris and Bartolini, Davide B. and Antoniou, Georgia and Kim, Jeremie S. and Wang, Zhe and Kalaitzidis, Kleovoulos and Rollet, Tom and Chen, Zhirui and Geng, Ye and Mutlu, Onur and Sazeides, Yiannakis},
  booktitle={2022 55th IEEE/ACM International Symposium on Microarchitecture (MICRO)},
  pages={835--850},
  year={2022},
  organization={IEEE}
}

@article{wang2024atrec,
  title={AtRec: Accelerating recommendation model training on CPUs},
  author={Wang, Siqi and Feng, Tianyu and Yang, Hailong and You, Xin and Chen, Bangduo and Liu, Tongxuan and Luan, Zhongzhi and Qian, Depei},
  journal={IEEE Transactions on Parallel and Distributed Systems},
  volume={35},
  number={6},
  pages={905--918},
  year={2024},
  publisher={IEEE}
}

@article{antoniou2024agile,
  title={Agile C-states: a core C-state architecture for latency critical applications optimizing both transition and cold-start latency},
  author = {Antoniou, Georgia and Bartolini, Davide and Volos, Haris and Kleanthous, Marios and Wang, Zhe and Kalaitzidis, Kleovoulos and Rollet, Tom and Li, Ziwei and Mutlu, Onur and Sazeides, Yiannakis and Haj Yahya, Jawad},
  journal={ACM Transactions on Architecture and Code Optimization},
  volume={21},
  number={4},
  pages={1--26},
  year={2024},
  publisher={ACM New York, NY}
}

@inproceedings{park2025ecocore,
  title={EcoCore: Dynamic Core Management for Improving Energy Efficiency in Latency-Critical Applications},
  author={Park, Gyeongseo and Kim, Minho and Kang, Ki-Dong and Jeon, Yunhyeong and Kim, Seulki and Kim, Daehoon},
  booktitle={Proceedings of the 58th IEEE/ACM International Symposium on Microarchitecture{\textregistered}},
  pages={1132--1146},
  year={2025}
}

@book{zahran2019heterogeneous,
  title={Heterogeneous computing: Hardware and software perspectives},
  author={Zahran, Mohamed},
  year={2019},
  publisher={Morgan \& Claypool}
}

@inproceedings{yu2024twinpilots,
  title={Twinpilots: A new computing paradigm for gpu-cpu parallel llm inference},
  author={Yu, Chengye and Wang, Tianyu and Shao, Zili and Zhu, Linjie and Zhou, Xu and Jiang, Song},
  booktitle={Proceedings of the 17th ACM International Systems and Storage Conference},
  pages={91--103},
  year={2024}
}

@inproceedings{na2024understanding,
  title={Understanding performance implications of llm inference on cpus},
  author={Na, Seonjin and Jeong, Geonhwa and Ahn, Byung Hoon and Young, Jeffrey and Krishna, Tushar and Kim, Hyesoon},
  booktitle={2024 IEEE International Symposium on Workload Characterization (IISWC)},
  pages={169--180},
  year={2024},
  organization={IEEE}
}

@article{henning2002spec,
  title={SPEC CPU2000: Measuring CPU performance in the new millennium},
  author={Henning, John L},
  journal={Computer},
  volume={33},
  number={7},
  pages={28--35},
  year={2002},
  publisher={IEEE}
}

@inproceedings{phansalkar2005measuring,
  title={Measuring program similarity: Experiments with SPEC CPU benchmark suites},
  author={Phansalkar, Aashish and Joshi, Ajay and Eeckhout, Lieven and John, Lizy Kurian},
  booktitle={IEEE International Symposium on Performance Analysis of Systems and Software, 2005. ISPASS 2005.},
  pages={10--20},
  year={2005},
  organization={IEEE}
}

@misc{perf,
  author = {{Linux Kernel Community}},
  title = {Linux perf tool},
  year = {2026},
  howpublished = {\url{https://perf.wiki.kernel.org/index.php/Main_Page}}
}

@misc{cpunews,
  author = {{Jowi Morales}},
  title = {Are we staring down the barrel of an AI-driven CPU shortage?},
  year = {2026},
  howpublished = {\url{https://www.tomshardware.com/pc-components/cpus/cpus-are-cool-again-intel-and-amd-reporting-spikes-in-cpu-demand-due-to-agentic-ai-shortages-lisa-su-says-business-exceeded-expectations-while-intel-is-looking-at-long-term-agreements-with-potential-customers}}
}

@article{binkert2011gem5,
  title={The gem5 simulator},
  author = {Binkert, Nathan and Beckmann, Bradford and Black, Gabriel and Reinhardt, Steven K. and Saidi, Ali and Basu, Arkaprava and Hestness, Joel and Hower, Derek R. and Krishna, Tushar and Sardashti, Somayeh and Sen, Rathijit and Sewell, Korey and Shoaib, Muhammad and Vaish, Nilay and Hill, Mark D. and Wood, David A.},
  journal={ACM SIGARCH computer architecture news},
  volume={39},
  number={2},
  pages={1--7},
  year={2011},
  publisher={ACM New York, NY, USA}
}

@article{perelman2003using,
  title={Using simpoint for accurate and efficient simulation},
  author={Perelman, Erez and Hamerly, Greg and Van Biesbrouck, Michael and Sherwood, Timothy and Calder, Brad},
  journal={ACM SIGMETRICS Performance Evaluation Review},
  volume={31},
  number={1},
  pages={318--319},
  year={2003},
  publisher={ACM New York, NY, USA}
}

@article{sherwood2002automatically,
  title={Automatically characterizing large scale program behavior},
  author={Sherwood, Timothy and Perelman, Erez and Hamerly, Greg and Calder, Brad},
  journal={ACM SIGPLAN Notices},
  volume={37},
  number={10},
  pages={45--57},
  year={2002},
  publisher={ACM New York, NY, USA}
}

@inproceedings{gottschall2023balancing,
  title={Balancing accuracy and evaluation overhead in simulation point selection},
  author={Gottschall, Bj{\"o}rn and de Santana, Silvio Campelo and Jahre, Magnus},
  booktitle={2023 IEEE International Symposium on Workload Characterization (IISWC)},
  pages={43--53},
  year={2023},
  organization={IEEE}
}

@inproceedings{karandikar2018firesim,
  title={FireSim: FPGA-accelerated cycle-exact scale-out system simulation in the public cloud},
  author = {Karandikar, Sagar and Mao, Howard and Kim, Donggyu and Biancolin, David and Amid, Alon and Lee, Dayeol and Pemberton, Nathan and Amaro, Emmanuel and Schmidt, Colin and Chopra, Aditya and Huang, Qijing and Kovacs, Kyle and Nikolic, Borivoje and Katz, Randy and Bachrach, Jonathan and Asanovi\'{c}, Krste},
  booktitle={2018 ACM/IEEE 45th Annual International Symposium on Computer Architecture (ISCA)},
  pages={29--42},
  year={2018},
  organization={IEEE}
}

@inproceedings{prokopec2019renaissance,
  title={Renaissance: Benchmarking suite for parallel applications on the jvm},
  author = {Prokopec, Aleksandar and Ros\`{a}, Andrea and Leopoldseder, David and Duboscq, Gilles and T\r{u}ma, Petr and Studener, Martin and Bulej, Lubom\'{\i}r and Zheng, Yudi and Villaz\'{o}n, Alex and Simon, Doug and W\"{u}rthinger, Thomas and Binder, Walter},
  booktitle={Proceedings of the 40th ACM SIGPLAN Conference on Programming Language Design and Implementation},
  pages={31--47},
  year={2019}
}

@article{gove2007cpu2006,
  title={CPU2006 working set size},
  author={Gove, Darryl},
  journal={ACM SIGARCH Computer Architecture News},
  volume={35},
  number={1},
  pages={90--96},
  year={2007},
  publisher={ACM New York, NY, USA}
}

@inproceedings{singh2019memory,
  title={Memory centric characterization and analysis of spec cpu2017 suite},
  author={Singh, Sarabjeet and Awasthi, Manu},
  booktitle={Proceedings of the 2019 ACM/SPEC International Conference on Performance Engineering},
  pages={285--292},
  year={2019}
}

@inproceedings{sr2019battle,
  title={Battle of compilers: An experimental evaluation using spec cpu2017},
  author={SR, Ranjan Hebbar and Ponugoti, Mounika and Milenkovi{\'c}, Aleksandar},
  booktitle={2019 SoutheastCon},
  pages={1--8},
  year={2019},
  organization={IEEE}
}

@inproceedings{schmitt2020energy,
  title={Energy efficiency analysis of compiler optimizations on the spec cpu 2017 benchmark suite},
  author={Schmitt, Norbert and Bucek, James and Lange, Klaus-Dieter and Kounev, Samuel},
  booktitle={Companion of the ACM/SPEC International Conference on Performance Engineering},
  pages={38--41},
  year={2020}
}

@inproceedings{schmitt2020performance,
  title={Performance, power, and energy-efficiency impact analysis of compiler optimizations on the spec cpu 2017 benchmark suite},
  author={Schmitt, Norbert and Bucek, James and Beckett, John and Cragin, Aaron and Lange, Klaus-Dieter and Kounev, Samuel},
  booktitle={2020 IEEE/ACM 13th International Conference on Utility and Cloud Computing (UCC)},
  pages={292--301},
  year={2020},
  organization={IEEE}
}

@inproceedings{naffziger2021pioneering,
  title={Pioneering chiplet technology and design for the amd epyc™ and ryzen™ processor families: Industrial product},
  author={Naffziger, Samuel and Beck, Noah and Burd, Thomas and Lepak, Kevin and Loh, Gabriel H and Subramony, Mahesh and White, Sean},
  booktitle={2021 ACM/IEEE 48th Annual International Symposium on Computer Architecture (ISCA)},
  pages={57--70},
  year={2021},
  organization={IEEE}
}

@inproceedings{velten2022memory,
  title={Memory performance of AMD EPYC Rome and Intel Cascade Lake SP server processors},
  author={Velten, Markus and Sch{\"o}ne, Robert and Ilsche, Thomas and Hackenberg, Daniel},
  booktitle={Proceedings of the 2022 ACM/SPEC on International Conference on Performance Engineering},
  pages={165--175},
  year={2022}
}

@inproceedings{feng2022chiplet,
  title={Chiplet actuary: A quantitative cost model and multi-chiplet architecture exploration},
  author={Feng, Yinxiao and Ma, Kaisheng},
  booktitle={Proceedings of the 59th ACM/IEEE Design Automation Conference},
  pages={121--126},
  year={2022}
}

@inproceedings{smith2024realizing,
  title={Realizing the AMD Exascale Heterogeneous Processor Vision: Industry Product},
  author={Smith, Alan and Loh, Gabriel H. and Schulte, Michael J. and Ignatowski, Mike and Naffziger, Samuel and Mantor, Mike and Kalyanasundharam, Mark Fowler Nathan and Alla, Vamsi and Malaya, Nicholas and Greathouse, Joseph L. and Chapman, Eric and Swaminathan, Raja},
  booktitle={2024 ACM/IEEE 51st Annual International Symposium on Computer Architecture (ISCA)},
  pages={876--889},
  year={2024},
  organization={IEEE}
}

@inproceedings{stojkovic2023mumanycore,
  title={$\mu$Manycore: A Cloud-Native CPU for Tail at Scale},
  author={Stojkovic, Jovan and Liu, Chunao and Shahbaz, Muhammad and Torrellas, Josep},
  booktitle={Proceedings of the 50th Annual International Symposium on Computer Architecture},
  pages={1--15},
  year={2023}
}

@inproceedings{kwon2023mccore,
  title={McCore: A Holistic Management of High-Performance Heterogeneous Multicores},
  author={Kwon, Jaewon and Lee, Yongju and Kal, Hongju and Kim, Minjae and Kim, Youngsok and Ro, Won Woo},
  booktitle={Proceedings of the 56th Annual IEEE/ACM International Symposium on Microarchitecture},
  pages={1044--1058},
  year={2023}
}

@inproceedings{jacobvitz2015multi,
  title={Multi-program benchmark definition},
  author={Jacobvitz, Adam N and Hilton, Andrew D and Sorin, Daniel J},
  booktitle={2015 IEEE international symposium on performance analysis of systems and software (ISPASS)},
  pages={72--82},
  year={2015},
  organization={IEEE}
}

@inproceedings{llull2017cooper,
  title={Cooper: Task colocation with cooperative games},
  author={Llull, Qiuyun and Fan, Songchun and Zahedi, Seyed Majid and Lee, Benjamin C},
  booktitle={2017 IEEE International Symposium on High Performance Computer Architecture (HPCA)},
  pages={421--432},
  year={2017},
  organization={IEEE}
}

@inproceedings{patel2020clite,
  title={Clite: Efficient and qos-aware co-location of multiple latency-critical jobs for warehouse scale computers},
  author={Patel, Tirthak and Tiwari, Devesh},
  booktitle={2020 IEEE International Symposium on High Performance Computer Architecture (HPCA)},
  pages={193--206},
  year={2020},
  organization={IEEE}
}

@inproceedings{zhao2021understanding,
  title={Understanding, predicting and scheduling serverless workloads under partial interference},
  author={Zhao, Laiping and Yang, Yanan and Li, Yiming and Zhou, Xian and Li, Keqiu},
  booktitle={Proceedings of the International conference for high performance computing, networking, storage and analysis},
  pages={1--15},
  year={2021}
}

@inproceedings{hunter2021beyond,
  title={Beyond malloc efficiency to fleet efficiency: a hugepage-aware memory allocator},
  author={Hunter, Andrew Hamilton and Kennelly, Chris and Turner, Paul and Gove, Darryl and Moseley, Tipp and Ranganathan, Parthasarathy},
  booktitle={15th $\{$USENIX$\}$ Symposium on Operating Systems Design and Implementation ($\{$OSDI$\}$ 21)},
  pages={257--273},
  year={2021}
}

@inproceedings{zhou2024characterizing,
  title={Characterizing a memory allocator at warehouse scale},
  author={Zhou, Zhuangzhuang and Gogte, Vaibhav and Vaish, Nilay and Kennelly, Chris and Xia, Patrick and Kanev, Svilen and Moseley, Tipp and Delimitrou, Christina and Ranganathan, Parthasarathy},
  booktitle={Proceedings of the 29th ACM International Conference on Architectural Support for Programming Languages and Operating Systems, Volume 3},
  pages={192--206},
  year={2024}
}

@inproceedings{cascaval2005multiple,
  title={Multiple page size modeling and optimization},
  author={Cascaval, Calin and Duesterwald, Evelyn and Sweeney, Peter F and Wisniewski, Robert W},
  booktitle={14th International Conference on Parallel Architectures and Compilation Techniques (PACT'05)},
  pages={339--349},
  year={2005},
  organization={IEEE}
}

@inproceedings{guvenilir2020tailored,
  title={Tailored page sizes},
  author={Guvenilir, Faruk and Patt, Yale N},
  booktitle={2020 ACM/IEEE 47th Annual International Symposium on Computer Architecture (ISCA)},
  pages={900--912},
  year={2020},
  organization={IEEE}
}

@article{cox2017efficient,
  title={Efficient address translation for architectures with multiple page sizes},
  author={Cox, Guilherme and Bhattacharjee, Abhishek},
  journal={ACM SIGPLAN Notices},
  volume={52},
  number={4},
  pages={435--448},
  year={2017},
  publisher={ACM New York, NY, USA}
}

@inproceedings{lee2023memtis,
  title={Memtis: Efficient memory tiering with dynamic page classification and page size determination},
  author={Lee, Taehyung and Monga, Sumit Kumar and Min, Changwoo and Eom, Young Ik},
  booktitle={Proceedings of the 29th Symposium on Operating Systems Principles},
  pages={17--34},
  year={2023}
}

@article{zhou2023impact,
  title={The Impact of Page Size and Microarchitecture on Instruction Address Translation Overhead},
  author={Zhou, Yufeng and Cox, Alan L and Dwarkadas, Sandhya and Dong, Xiaowan},
  journal={ACM Transactions on Architecture and Code Optimization},
  volume={20},
  number={3},
  pages={1--25},
  year={2023},
  publisher={ACM New York, NY}
}

@article{ausavarungnirun2018mosaic,
  title={Mosaic: Enabling application-transparent support for multiple page sizes in throughput processors},
  author={Ausavarungnirun, Rachata and Landgraf, Joshua and Miller, Vance and Ghose, Saugata and Gandhi, Jayneel and Rossbach, Christopher J and Mutlu, Onur},
  journal={ACM SIGOPS Operating Systems Review},
  volume={52},
  number={1},
  pages={27--44},
  year={2018},
  publisher={ACM New York, NY, USA}
}

@article{lee2012prefetching,
  title={When prefetching works, when it doesn’t, and why},
  author={Lee, Jaekyu and Kim, Hyesoon and Vuduc, Richard},
  journal={ACM Transactions on Architecture and Code Optimization (TACO)},
  volume={9},
  number={1},
  pages={1--29},
  year={2012},
  publisher={ACM New York, NY, USA}
}

@article{bakhshalipour2019evaluation,
  title={Evaluation of hardware data prefetchers on server processors},
  author={Bakhshalipour, Mohammad and Tabaeiaghdaei, Seyedali and Lotfi-Kamran, Pejman and Sarbazi-Azad, Hamid},
  journal={ACM Computing Surveys (CSUR)},
  volume={52},
  number={3},
  pages={1--29},
  year={2019},
  publisher={ACM New York, NY, USA}
}

@article{navarro2019memory,
  title={Memory hierarchy characterization of SPEC CPU2006 and SPEC CPU2017 on the Intel Xeon Skylake-SP},
  author={Navarro-Torres, Agust{\'\i}n and Alastruey-Bened{\'e}, Jes{\'u}s and Ib{\'a}{\~n}ez-Mar{\'\i}n, Pablo and Vi{\~n}als-Y{\'u}fera, V{\'\i}ctor},
  journal={Plos one},
  volume={14},
  number={8},
  pages={e0220135},
  year={2019},
  publisher={Public Library of Science San Francisco, CA USA}
}

@inproceedings{panda2023clip,
  title={Clip: Load criticality based data prefetching for bandwidth-constrained many-core systems},
  author={Panda, Biswabandan},
  booktitle={Proceedings of the 56th Annual IEEE/ACM International Symposium on Microarchitecture},
  pages={714--727},
  year={2023}
}

@inproceedings{jain2024limoncello,
  title={Limoncello: Prefetchers for scale},
  author={Jain, Akanksha and Lin, Hannah and Villavieja, Carlos and Kasikci, Baris and Kennelly, Chris and Hashemi, Milad and Ranganathan, Parthasarathy},
  booktitle={Proceedings of the 29th ACM International Conference on Architectural Support for Programming Languages and Operating Systems, Volume 3},
  pages={577--590},
  year={2024}
}

@inproceedings{leijen2019mimalloc,
  title={Mimalloc: Free list sharding in action},
  author={Leijen, Daan and Zorn, Benjamin and De Moura, Leonardo},
  booktitle={Asian Symposium on Programming Languages and Systems},
  pages={244--265},
  year={2019},
  organization={Springer}
}

@inproceedings{papazian2020new,
  title={New 3rd Gen Intel{\textregistered} Xeon{\textregistered} Scalable Processor (Codename: Ice Lake-SP).},
  author={Papazian, Irma Esmer},
  booktitle={Hot Chips Symposium},
  pages={1--22},
  year={2020}
}

@INPROCEEDINGS{nassif2022sapphire,
  author={Nassif, Nevine and Munch, Ashley O. and Molnar, Carleton L. and Pasdast, Gerald and Lyer, Sitaraman V. and Yang, Zibing and Mendoza, Oscar and Huddart, Mark and Venkataraman, Srikrishnan and Kandula, Sireesha and Marom, Rafi and Kern, Alexandra M. and Bowhill, Bill and Mulvihill, David R. and Nimmagadda, Srikanth and Kalidindi, Varma and Krause, Jonathan and Haq, Mohammad M. and Sharma, Roopali and Duda, Kevin},
  booktitle={2022 IEEE International Solid-State Circuits Conference (ISSCC)}, 
  title={Sapphire Rapids: The Next-Generation Intel Xeon Scalable Processor}, 
  year={2022},
  volume={65},
  number={},
  pages={44-46},
  doi={10.1109/ISSCC42614.2022.9731107}}

@inproceedings{soltis2023next,
  title={The Next Generation of High Performance, Energy-Efficient Computing: Intel{\textregistered} Xeon{\textregistered} Processors Built on Efficient-Core.},
  author={Soltis, Don and Robinson, Stephen},
  booktitle={HCS},
  pages={1--16},
  year={2023}
}

@article{evers2022amd,
  title={The AMD next-generation “Zen 3” core},
  author={Evers, Mark and Barnes, Leslie and Clark, Mike},
  journal={IEEE Micro},
  volume={42},
  number={3},
  pages={7--12},
  year={2022},
  publisher={IEEE}
}

@article{bhargava2024amd,
  title={AMD next-generation “Zen 4” core and 4th gen AMD EPYC server CPUs},
  author={Bhargava, Ravi and Troester, Kai},
  journal={IEEE Micro},
  volume={44},
  number={3},
  pages={8--17},
  year={2024},
  publisher={IEEE}
}

@INPROCEEDINGS{singh2025zen,
  author={Singh, Teja and Oliver, Spence and Rangarajan, Sundar and Southard, Shane and Henrion, Carson and Schaefer, Alex and Johnson, Brett and Tower, Sarah Bartaszewicz and Hoover, Kathy and John, Deepesh and Antoniadis, Ted and Lakshman, Shravan and Mittal, Vibhor and Kasprzyk, Brian and McCoy, Ross and Mohlman, Kurt and Mohan, Anitha and Wong, Hon-Hin and Lieu, Daryl and Schreiber, Russell and Singh, Sahilpreet and Lance, Nick and Prudich, Darryl and Coppin, Justin and Jackson, Tim and Karegar, Anita and Miller, Ryan and Balagangadharan, Sabeesh and Pistole, James and Li, Wilson and McCabe, Michael},
  booktitle={2025 IEEE International Solid-State Circuits Conference (ISSCC)}, 
  title={“Zen 5”: The AMD High-Performance 4nm x86-64 Microprocessor Core}, 
  year={2025},
  volume={68},
  number={},
  pages={1-3},
  doi={10.1109/ISSCC49661.2025.10904529}}

@misc{grace,
  author = {{NVIDIA}},
  title = {NVIDIA Grace CPU Superchip Whitepaper},
  year = {2026},
  howpublished = {\url{https://resources.nvidia.com/en-us-grace-cpu/nvidia-grace-cpu-superchip}}
}

@misc{ampere,
  author = {{Ampere}},
  title = {Ampere Processor Platforms},
  year = {2026},
  howpublished = {\url{https://amperecomputing.com/products/processors}}
}

@inproceedings{esmaili2024mess,
  title={A mess of memory system benchmarking, simulation and application profiling},
  author={Esmaili-Dokht, Pouya and Sgherzi, Francesco and Girelli, Valéria Soldera and Boixaderas, Isaac and Carmin, Mariana and Monemi, Alireza and Armejach, Adrià and Mercadal, Estanislao and Llort, Germán and Radojković, Petar and Moreto, Miquel and Giménez, Judit and Martorell, Xavier and Ayguadé, Eduard and Labarta, Jesus and Confalonieri, Emanuele and Dubey, Rishabh and Adlard, Jason},
  booktitle={2024 57th IEEE/ACM International Symposium on Microarchitecture (MICRO)}, 
  pages={136--152},
  year={2024},
  organization={IEEE}
}

@misc{romstad2008stockfish,
  title={Stockfish: A strong open source chess engine},
  author={Romstad, Tord and Costalba, Marco and Kiiski, Joona and Linscott, G},
  year={2008},
  howpublished = {\url{https://stockfishchess.org}}
}

@misc{NTest,
  title={NTest othello program},
  author={Welty, Chris},
  year={2026},
  howpublished = {\url{https://github.com/weltyc/ntest}}
}

@article{gaffney2022sqlite,
  title={SQLite: past, present, and future},
  author={Gaffney, Kevin P and Prammer, Martin and Brasfield, Larry and Hipp, D Richard and Kennedy, Dan and Patel, Jignesh M},
  journal={Proceedings of the VLDB Endowment},
  volume={15},
  number={12},
  year={2022}
}

@inproceedings{varga2008overview,
  title={An overview of the OMNeT++ simulation environment},
  author={Varga, Andr{\'a}s and Hornig, Rudolf},
  booktitle={Proceedings of the 1st international conference on Simulation tools and techniques for communications, networks and systems \& workshops},
  pages={1--10},
  year={2008}
}

@book{van1995python,
  title={Python reference manual},
  author={Van Rossum, Guido},
  volume={111},
  year={1995},
  publisher={Centrum voor Wiskunde en Informatica Amsterdam}
}

@article{stallman2002gnu,
  title={GNU compiler collection internals},
  author={Stallman, Richard M},
  journal={Free Software Foundation},
  volume={46},
  pages={765},
  year={2002}
}

@inproceedings{lattner2004llvm,
  title={LLVM: A compilation framework for lifelong program analysis \& transformation},
  author={Lattner, Chris and Adve, Vikram},
  booktitle={International symposium on code generation and optimization, 2004. CGO 2004.},
  pages={75--86},
  year={2004},
  organization={IEEE}
}

@article{marjamaki2013cppcheck,
  title={Cppcheck: a tool for static c/c++ code analysis},
  author={Marjam{\"a}ki, Daniel},
  journal={URL: https://cppcheck. sourceforge. io},
  year={2013}
}

@inproceedings{brayton2010abc,
  title={ABC: An academic industrial-strength verification tool},
  author={Brayton, Robert and Mishchenko, Alan},
  booktitle={International Conference on Computer Aided Verification},
  pages={24--40},
  year={2010},
  organization={Springer}
}

@inproceedings{betz1997vpr,
  title={VPR: A new packing, placement and routing tool for FPGA research},
  author={Betz, Vaughn and Rose, Jonathan},
  booktitle={International Workshop on Field Programmable Logic and Applications},
  pages={213--222},
  year={1997},
  organization={Springer}
}

@misc{xzutils,
  title={XZ Utils},
  year={2026},
  author={The Tukaani Project},
  howpublished = {\url{https://github.com/tukaani-project/xz}}
}

@article{buchfink2021sensitive,
  title={Sensitive protein alignments at tree-of-life scale using DIAMOND},
  author={Buchfink, Benjamin and Reuter, Klaus and Drost, Hajk-Georg},
  journal={Nature methods},
  volume={18},
  number={4},
  pages={366--368},
  year={2021},
  publisher={Nature Publishing Group US New York}
}

@inproceedings{nethercote2007minizinc,
  title={MiniZinc: Towards a standard CP modelling language},
  author={Nethercote, Nicholas and Stuckey, Peter J and Becket, Ralph and Brand, Sebastian and Duck, Gregory J and Tack, Guido},
  booktitle={International conference on principles and practice of constraint programming},
  pages={529--543},
  year={2007},
  organization={Springer}
}

@inproceedings{chen2017simple,
  title={Simple encrypted arithmetic library-SEAL v2. 1},
  author={Chen, Hao and Laine, Kim and Player, Rachel},
  booktitle={International conference on financial cryptography and data security},
  pages={3--18},
  year={2017},
  organization={Springer}
}

@incollection{riley2010ns,
  title={The ns-3 network simulator},
  author={Riley, George F and Henderson, Thomas R},
  booktitle={Modeling and tools for network simulation},
  pages={15--34},
  year={2010},
  publisher={Springer}
}

@article{murphy2010introducing,
  title={Introducing the graph 500},
  author={Murphy, Richard C and Wheeler, Kyle B and Barrett, Brian W and Ang, James A},
  journal={Cray Users Group (CUG)},
  volume={19},
  number={45-74},
  pages={22},
  year={2010}
}

@misc{zstd,
  title={Zstandard - Fast real-time compression algorithm},
  year={2026},
  author={Facebook},
  howpublished = {\url{https://github.com/facebook/zstd}}
}

@article{downs2025near,
  title={A near-real-time data-assimilative model of the solar corona},
  author = {Cooper Downs  and Jon A. Linker  and Ronald M. Caplan  and Emily I. Mason  and Pete Riley  and Ryder Davidson  and Andres Reyes  and Erika Palmerio  and Roberto Lionello  and James Turtle  and Michal Ben-Nun  and Miko M. Stulajter  and Viacheslav S. Titov  and Tibor Török  and Lisa A. Upton  and Raphael Attie  and Bibhuti K. Jha  and Charles N. Arge  and Carl J. Henney  and Gherardo Valori  and Hanna Strecker  and Daniele Calchetti  and Dietmar Germerott  and Johann Hirzberger  and David Orozco Suárez  and Julian Blanco Rodríguez  and Sami K. Solanki  and Xin Cheng  and Sizhe Wu },
  journal={Science},
  volume={388},
  number={6753},
  pages={1306--1310},
  year={2025},
  publisher={American Association for the Advancement of Science}
}

@inproceedings{guerrera2018towards,
  title={Towards a mini-app for smoothed particle hydrodynamics at exascale},
  author={Guerrera, Danilo and Cabez{\'o}n, Rub{\'e}n M and Piccinali, Jean-Guillaume and Cavelan, Aur{\'e}lien and Ciorba, Florina M and Imbert, David and Mayer, Lucio and Reed, Darren},
  booktitle={2018 IEEE International Conference on Cluster Computing (CLUSTER)},
  pages={607--614},
  year={2018},
  organization={IEEE}
}

@inproceedings{goodale2002cactus,
  title={The cactus framework and toolkit: design and applications: invited talk},
  author={Goodale, Tom and Allen, Gabrielle and Lanfermann, Gerd and Mass{\'o}, Joan and Radke, Thomas and Seidel, Edward and Shalf, John},
  booktitle={International conference on high performance computing for computational science},
  pages={197--227},
  year={2002},
  organization={Springer}
}

@inproceedings{mcintosh2017tealeaf,
  title={Tealeaf: A mini-application to enable design-space explorations for iterative sparse linear solvers},
  author={McIntosh-Smith, Simon and Martineau, Matthew and Deakin, Tom and Pawelczak, Grzegorz and Gaudin, Wayne and Garrett, Paul and Liu, Wei and Smedley-Stevenson, Richard and Beckingsale, David},
  booktitle={2017 IEEE International Conference on Cluster Computing (CLUSTER)},
  pages={842--849},
  year={2017},
  organization={IEEE}
}

@incollection{macke1998modeling,
  title={Modeling unusual nucleic acid structures},
  author={Macke, Thomas J and Case, David A},
  year={1998},
  publisher={ACS Publications}
}

@misc{cloverleaf,
  title={Compressible Euler equations on a Cartesian grid},
  year={2026},
  author={CloverLeaf Mantevo Project },
  howpublished = {\url{https://github.com/Mantevo/CloverLeaf}}
}

@article{maronga2020overview,
  title={Overview of the PALM model system 6.0},
  author={Maronga, B. and Banzhaf, S. and Burmeister, C. and Esch, T. and Forkel, R. and Fr\"ohlich, D. and Fuka, V. and Gehrke, K. F. and Geleti\v{c}, J. and Giersch, S. and Gronemeier, T. and Gro{\ss}, G. and Heldens, W. and Hellsten, A. and Hoffmann, F. and Inagaki, A. and Kadasch, E. and Kanani-S\"uhring, F. and Ketelsen, K. and Khan, B. A. and Knigge, C. and Knoop, H. and Kr\v{c}, P. and Kurppa, M. and Maamari, H. and Matzarakis, A. and Mauder, M. and Pallasch, M. and Pavlik, D. and Pfafferott, J. and Resler, J. and Rissmann, S. and Russo, E. and Salim, M. and Schrempf, M. and Schwenkel, J. and Seckmeyer, G. and Schubert, S. and S\"uhring, M. and von Tils, R. and Vollmer, L. and Ward, S. and Witha, B. and Wurps, H. and Zeidler, J. and Raasch, S.},
  journal={Geoscientific Model Development},
  volume={13},
  number={3},
  pages={1335--1372},
  year={2020},
  publisher={Copernicus Publications G{\"o}ttingen, Germany}
}

@misc{astc,
  title={The Arm ASTC Encoder, a compressor for the Adaptive Scalable Texture Compression data format.},
  author={ARM},
  year={2026},
  howpublished = {\url{https://github.com/ARM-software/astc-encoder}}
}

@inproceedings{walker2020aswf,
  title={The ASWF takes OpenColorIO to the next level},
  author={Walker, Doug and Dolan, Michael and Hodoul, Patrick},
  booktitle={Proceedings of the 2020 Digital Production Symposium},
  pages={1--10},
  year={2020}
}

@article{geuzaine2009gmsh,
  title={Gmsh: A 3-D finite element mesh generator with built-in pre-and post-processing facilities},
  author={Geuzaine, Christophe and Remacle, Jean-Fran{\c{c}}ois},
  journal={International journal for numerical methods in engineering},
  volume={79},
  number={11},
  pages={1309--1331},
  year={2009},
  publisher={Wiley Online Library}
}

@inproceedings{berndt2004jsbsim,
  title={JSBSim: An open source flight dynamics model in C++},
  author={Berndt, Jon},
  booktitle={AIAA modeling and simulation technologies conference and exhibit},
  pages={4923},
  year={2004}
}

@article{andersson2006parallel,
  title={Parallel power computation for photonic crystal devices},
  author={Andersson, Ulf and Qiu, Min and Zhang, Ziyang},
  journal={Methods and applications of analysis},
  volume={13},
  number={2},
  pages={149--156},
  year={2006},
  publisher={International Press}
}

@article{phillips2005scalable,
  title={Scalable molecular dynamics with NAMD},
  author={Phillips, James C and Braun, Rosemary and Wang, Wei and Gumbart, James and Tajkhorshid, Emad and Villa, Elizabeth and Chipot, Christophe and Skeel, Robert D and Kale, Laxmikant and Schulten, Klaus},
  journal={Journal of computational chemistry},
  volume={26},
  number={16},
  pages={1781--1802},
  year={2005},
  publisher={Wiley Online Library}
}

@article{shchepetkin2005regional,
  title={The regional oceanic modeling system (ROMS): a split-explicit, free-surface, topography-following-coordinate oceanic model},
  author={Shchepetkin, Alexander F and McWilliams, James C},
  journal={Ocean modelling},
  volume={9},
  number={4},
  pages={347--404},
  year={2005},
  publisher={Elsevier}
}

@article{kronbichler2023enhancing,
  title={Enhancing data locality of the conjugate gradient method for high-order matrix-free finite-element implementations},
  author={Kronbichler, Martin and Sashko, Dmytro and Munch, Peter},
  journal={The International Journal of High Performance Computing Applications},
  volume={37},
  number={2},
  pages={61--81},
  year={2023},
  publisher={Sage Publications Sage UK: London, England}
}

@article{gewaltig2007nest,
  title={Nest (neural simulation tool)},
  author={Gewaltig, Marc-Oliver and Diesmann, Markus},
  journal={Scholarpedia},
  volume={2},
  number={4},
  pages={1430},
  year={2007}
}

@inproceedings{junczys2018marian,
  title={Marian: Fast neural machine translation in C++},
  author = "Junczys-Dowmunt, Marcin  and
      Grundkiewicz, Roman  and
      Dwojak, Tomasz  and
      Hoang, Hieu  and
      Heafield, Kenneth  and
      Neckermann, Tom  and
      Seide, Frank  and
      Germann, Ulrich  and
      Aji, Alham Fikri  and
      Bogoychev, Nikolay  and
      Martins, Andr{\'e} F. T.  and
      Birch, Alexandra",
  booktitle={Proceedings of ACL 2018, system demonstrations},
  pages={116--121},
  year={2018}
}

@article{pohl2003optimization,
  title={Optimization and profiling of the cache performance of parallel lattice Boltzmann codes},
  author={Pohl, Thomas and Kowarschik, Markus and Wilke, Jens and Iglberger, Klaus and R{\"u}de, Ulrich},
  journal={Parallel Processing Letters},
  volume={13},
  number={04},
  pages={549--560},
  year={2003},
  publisher={World Scientific}
}

@inproceedings{tramm2014performance,
  title={Performance analysis of a reduced data movement algorithm for neutron cross section data in monte carlo simulations},
  author={Tramm, John R and Siegel, Andrew R and Forget, Benoit and Josey, Colin},
  booktitle={International Conference on Exascale Applications and Software},
  pages={39--56},
  year={2014},
  organization={Springer}
}

@misc{armagi,
  title={The world's most efficient agentic CPU},
  author={ARM},
  year={2026},
  howpublished = {\url{https://www.arm.com/products/cloud-datacenter/arm-agi-cpu}}
}

@article{raj2025cpu,
  title={Towards Understanding, Analyzing, and Optimizing Agentic AI Execution: A CPU-Centric Perspective},
  author={Raj, Ritik and Kundu, Souvik and Vohra, Ishita and Wang, Hong and Krishna, Tushar},
  journal={arXiv preprint arXiv:2511.00739},
  year={2025}
}

@inproceedings{bucek2018spec,
  title={SPEC CPU2017: Next-generation compute benchmark},
  author={Bucek, James and Lange, Klaus-Dieter and v. Kistowski, J{\'o}akim},
  booktitle={Companion of the 2018 ACM/SPEC International Conference on Performance Engineering},
  pages={41--42},
  year={2018}
}

@inproceedings{hebbar2019spec,
  title={SPEC CPU2017: Performance, event, and energy characterization on the core i7-8700K},
  author={Hebbar SR, Ranjan and Milenkovi{\'c}, Aleksandar},
  booktitle={Proceedings of the 2019 ACM/SPEC International Conference on Performance Engineering},
  pages={111--118},
  year={2019}
}

@inproceedings{shan2024guser,
  title={Guser: A GPGPU power stressmark generator},
  author={Shan, Yalong and Yang, Yongkui and Qian, Xuehai and Yu, Zhibin},
  booktitle={2024 IEEE International Symposium on High-Performance Computer Architecture (HPCA)},
  pages={1111--1124},
  year={2024},
  organization={IEEE}
}

@article{zhang2023characterizing,
  title={Characterizing multi-chip GPU data sharing},
  author={Zhang, Shiqing and Naderan-Tahan, Mahmood and Jahre, Magnus and Eeckhout, Lieven},
  journal={ACM Transactions on Architecture and Code Optimization},
  volume={20},
  number={4},
  pages={1--24},
  year={2023},
  publisher={ACM New York, NY}
}

@misc{bostanci2026cleaningmessreevaluatingrealsystem,
      title={Cleaning up the Mess: Re-Evaluating the Real-System Modeling Accuracy of Ramulator 2.0}, 
      author={F. Nisa Bostanci and Haocong Luo and Ataberk Olgun and Maria Makeenkova and Geraldo F. Oliveira and A. Giray Yaglikci and Onur Mutlu},
      year={2026},
      eprint={2510.15744},
      archivePrefix={arXiv},
      primaryClass={cs.AR},
      url={https://arxiv.org/abs/2510.15744}, 
}

@inproceedings{stojkovic2026dorado,
  title={Dorado: Clustered Hardware Cache Coherence for 1,000+ Cores},
  author={Stojkovic, Jovan and Farrell, Abraham and Gerogiannis, Gerasimos and Gong, Zhangxiaowen and Hughes, C and Torrellas, Josep},
  booktitle={Proceedings of the 53rd Annual International Symposium on Computer Architecture (ISCA’26)},
  year={2026}
}

@article{kim2026phaseweave,
  title={PhaseWeave: Phase-Aware Execution on Heterogeneous Chiplet Architectures for Datacenters},
  author={Kim, Joshua and Zhang, Chaojie and Goiri, {\'I}nigo and Rossbach, Christopher J and Stojkovic, Jovan},
  booktitle={Proceedings of the 53rd Annual International Symposium on Computer Architecture (ISCA’26)},
  year={2026}
}

@inproceedings{stojkovic2026accelflow,
  title={AccelFlow: Orchestrating an On-Package Ensemble of Fine-Grained Accelerators for Microservices},
  author={Stojkovic, Jovan and Farrell, Abraham and Gong, Zhangxiaowen and Hughes, Christopher J and Torrellas, Josep},
  booktitle={2026 IEEE International Symposium on High Performance Computer Architecture (HPCA)},
  pages={1--17},
  year={2026},
  organization={IEEE}
}

@inproceedings{agarwal2026tina,
  title={TiNA: Tiered Network Buffer Architecture for Fast Networking in Chiplet-based CPUs},
  author={Agarwal, Siddharth and Wang, Tianchen and Huang, Jinghan and Agarwal, Saksham and Kim, Nam Sung},
  booktitle={Proceedings of the 31st ACM International Conference on Architectural Support for Programming Languages and Operating Systems, Volume 1},
  pages={298--313},
  year={2026}
}

@inproceedings{sun2023demystifying,
  title={Demystifying cxl memory with genuine cxl-ready systems and devices},
  author = {Sun, Yan and Yuan, Yifan and Yu, Zeduo and Kuper, Reese and Song, Chihun and Huang, Jinghan and Ji, Houxiang and Agarwal, Siddharth and Lou, Jiaqi and Jeong, Ipoom and Wang, Ren and Ahn, Jung Ho and Xu, Tianyin and Kim, Nam Sung},
  booktitle={Proceedings of the 56th Annual IEEE/ACM International Symposium on Microarchitecture},
  pages={105--121},
  year={2023}
}

@inproceedings{song2025hybridtier,
  title={Hybridtier: an adaptive and lightweight cxl-memory tiering system},
  author={Song, Kevin and Yang, Jiacheng and Wang, Zixuan and Zhao, Jishen and Liu, Sihang and Pekhimenko, Gennady},
  booktitle={Proceedings of the 30th ACM International Conference on Architectural Support for Programming Languages and Operating Systems, Volume 3},
  pages={112--128},
  year={2025}
}

@inproceedings{chen2019parties,
  title={Parties: Qos-aware resource partitioning for multiple interactive services},
  author={Chen, Shuang and Delimitrou, Christina and Mart{\'\i}nez, Jos{\'e} F},
  booktitle={Proceedings of the twenty-fourth international conference on architectural support for programming languages and operating systems},
  pages={107--120},
  year={2019}
}

@inproceedings{margaritov2019stretch,
  title={Stretch: Balancing qos and throughput for colocated server workloads on smt cores},
  author={Margaritov, Artemiy and Gupta, Siddharth and Gonzalez-Alberquilla, Rekai and Grot, Boris},
  booktitle={2019 IEEE International Symposium on High Performance Computer Architecture (HPCA)},
  pages={15--27},
  year={2019},
  organization={IEEE}
}

@inproceedings{zhang2022ocolos,
  title={Ocolos: Online code layout optimizations},
  author={Zhang, Yuxuan and Khan, Tanvir Ahmed and Pokam, Gilles and Kasikci, Baris and Litz, Heiner and Devietti, Joseph},
  booktitle={2022 55th IEEE/ACM International Symposium on Microarchitecture (MICRO)},
  pages={530--545},
  year={2022},
  organization={IEEE}
}

@inproceedings{bhuiyan2026wax,
  title={Wax: Optimizing Data Center Applications With Stale Profile},
  author={Bhuiyan, Tawhid and Hoque, Sumya and Moreira, Ang{\'e}lica Aparecida and Khan, Tanvir Ahmed},
  booktitle={Proceedings of the 31st ACM International Conference on Architectural Support for Programming Languages and Operating Systems, Volume 2},
  pages={2232--2248},
  year={2026}
}

@article{jiang2020characterizing,
  title={Characterizing co-located workloads in alibaba cloud datacenters},
  author={Jiang, Congfeng and Qiu, Yitao and Shi, Weisong and Ge, Zhefeng and Wang, Jiwei and Chen, Shenglei and C{\'e}rin, Christophe and Ren, Zujie and Xu, Guoyao and Lin, Jiangbin},
  journal={IEEE Transactions on Cloud Computing},
  volume={10},
  number={4},
  pages={2381--2397},
  year={2020},
  publisher={IEEE}
}

@inproceedings{gan2019open,
  title={An open-source benchmark suite for microservices and their hardware-software implications for cloud \& edge systems},
  author = {Gan, Yu and Zhang, Yanqi and Cheng, Dailun and Shetty, Ankitha and Rathi, Priyal and Katarki, Nayan and Bruno, Ariana and Hu, Justin and Ritchken, Brian and Jackson, Brendon and Hu, Kelvin and Pancholi, Meghna and He, Yuan and Clancy, Brett and Colen, Chris and Wen, Fukang and Leung, Catherine and Wang, Siyuan and Zaruvinsky, Leon and Espinosa, Mateo and Lin, Rick and Liu, Zhongling and Padilla, Jake and Delimitrou, Christina},
  booktitle={Proceedings of the twenty-fourth international conference on architectural support for programming languages and operating systems},
  pages={3--18},
  year={2019}
}

@Misc{spec, 
author = {SPEC org},
title = {SPEC CPU},
howpublished = {\url{https://www.spec.org/cpu/}}, 
year = {2026}
}

@inproceedings{madhav2026spec,
  title={SPEC CPU: The Next Generation},
  author={Mahesh Madhav and Allen Lee and Andres Mejia and Branden Moore and Charan Soppadandi and Chris Cambly and Christoph Müllner and Daniel Bowers and David Reiner and Denis Bakhvalov and Di Zhao and Duane Voth and Feng Xue and Frédérique Silber-Chaussumier and James Bucek and James Southern and Jiangning Liu and Jim Himer and John Henning and Kevin Smith and Kristen Yang and Kunal Kashyap and Mason Guy and Mat Colgrove and Michael Berg and Prasad Battini and Prasad Joshi and Rohit Prasad and Shayantika Bhattacharya and Sriyash Caculo and Stefan Reimbold and Sundar Iyengar and Van Smith and Zarko Todorovski},
  booktitle={2026 ACM/IEEE 53rd Annual International Symposium on Computer Architecture (ISCA)},
  pages={671-687},
  year={2026},
  organization={IEEE}
}

@article{hendrycks2021apps,
  title={Measuring coding challenge competence with APPS},
  author={Hendrycks, Dan and Basart, Steven and Kadavath, Saurav and Mazeika, Mantas and Arora, Akul and Guo, Ethan and Burns, Collin and Puranik, Samir and He, Horace and Song, Dawn and Steinhardt, Jacob},
  journal={Advances in Neural Information Processing Systems (NeurIPS)},
  year={2021}
}

@inproceedings{zhuo2024bigcodebench,
  title={{BigCodeBench}: Benchmarking Code Generation with Diverse Function Calls and Complex Instructions},
  author={Terry Yue Zhuo and Minh Chien Vu and Jenny Chim and Han Hu and Wenhao Yu and Ratnadira Widyasari and Imam Nur Bani Yusuf and Haolan Zhan and Junda He and Indraneil Paul and Simon Brunner and Chen Gong and Thong Hoang and Armel Randy Zebaze and Xiaoheng Hong and Wen-Ding Li and Jean Kaddour and Ming Xu and Zhihan Zhang and Prateek Yadav and Naman Jain and Alex Gu and Zhoujun Cheng and Jiawei Liu and Qian Liu and Zijian Wang and Binyuan Hui and Niklas Muennighoff and David Lo and Daniel Fried and Xiaoning Du and Harm de Vries and Leandro Von Werra},
  booktitle={International Conference on Learning Representations (ICLR)},
  year={2025}
}

@inproceedings{vu2024freshllms,
  title={Freshllms: Refreshing large language models with search engine augmentation},
  author={Vu, Tu and Iyyer, Mohit and Wang, Xuezhi and Constant, Noah and Wei, Jerry and Wei, Jason and Tar, Chris and Sung, Yun-Hsuan and Zhou, Denny and Le, Quoc and Luong, Thang},
  booktitle={Findings of the Association for Computational Linguistics: ACL 2024},
  pages={13697--13720},
  year={2024}
}

@inproceedings{khot2020qasc,
  title={Qasc: A dataset for question answering via sentence composition},
  author={Khot, Tushar and Clark, Peter and Guerquin, Michal and Jansen, Peter and Sabharwal, Ashish},
  booktitle={Proceedings of the AAAI Conference on Artificial Intelligence},
  volume={34},
  number={05},
  pages={8082--8090},
  year={2020}
}

@article{grattafiori2024llama,
  title={The {Llama} 3 Herd of Models},
  author={Aaron Grattafiori and Abhimanyu Dubey and Abhinav Jauhri and Abhinav Pandey and Abhishek Kadian and Ahmad Al-Dahle and Aiesha Letman and Akhil Mathur and Alan Schelten and Alex Vaughan and Amy Yang and Angela Fan and Anirudh Goyal and Anthony Hartshorn and Aobo Yang and Archi Mitra and Archie Sravankumar and Artem Korenev and Arthur Hinsvark and Arun Rao and Aston Zhang and Aurelien Rodriguez and Austen Gregerson and Ava Spataru and Baptiste Roziere and Bethany Biron and Binh Tang and Bobbie Chern and Charlotte Caucheteux and Chaya Nayak and Chloe Bi and Chris Marra and Chris McConnell and Christian Keller and Christophe Touret and Chunyang Wu and Corinne Wong and Cristian Canton Ferrer and Cyrus Nikolaidis and Damien Allonsius and Daniel Song and Danielle Pintz and Danny Livshits and Danny Wyatt and David Esiobu and Dhruv Choudhary and Dhruv Mahajan and Diego Garcia-Olano and Diego Perino and Dieuwke Hupkes and Egor Lakomkin and Ehab AlBadawy and Elina Lobanova and Emily Dinan and Eric Michael Smith and Filip Radenovic and Francisco Guzmán and Frank Zhang and Gabriel Synnaeve and Gabrielle Lee and Georgia Lewis Anderson and Govind Thattai and Graeme Nail and Gregoire Mialon and Guan Pang and Guillem Cucurell and Hailey Nguyen and Hannah Korevaar and Hu Xu and Hugo Touvron and Iliyan Zarov and Imanol Arrieta Ibarra and Isabel Kloumann and Ishan Misra and Ivan Evtimov and Jack Zhang and Jade Copet and Jaewon Lee and Jan Geffert and Jana Vranes and Jason Park and Jay Mahadeokar and Jeet Shah and Jelmer van der Linde and Jennifer Billock and Jenny Hong and Jenya Lee and Jeremy Fu and Jianfeng Chi and Jianyu Huang and Jiawen Liu and Jie Wang and Jiecao Yu and Joanna Bitton and Joe Spisak and Jongsoo Park and Joseph Rocca and Joshua Johnstun and Joshua Saxe and Junteng Jia and Kalyan Vasuden Alwala and Karthik Prasad and Kartikeya Upasani and Kate Plawiak and Ke Li and Kenneth Heafield and Kevin Stone and Khalid El-Arini and Krithika Iyer and Kshitiz Malik and Kuenley Chiu and Kunal Bhalla and Kushal Lakhotia and Lauren Rantala-Yeary and Laurens van der Maaten and Lawrence Chen and Liang Tan and Liz Jenkins and Louis Martin and Lovish Madaan and Lubo Malo and Lukas Blecher and Lukas Landzaat and Luke de Oliveira and Madeline Muzzi and Mahesh Pasupuleti and Mannat Singh and Manohar Paluri and Marcin Kardas and Maria Tsimpoukelli and Mathew Oldham and Mathieu Rita and Maya Pavlova and Melanie Kambadur and Mike Lewis and Min Si and Mitesh Kumar Singh and Mona Hassan and Naman Goyal and Narjes Torabi and Nikolay Bashlykov and Nikolay Bogoychev and Niladri Chatterji and Ning Zhang and Olivier Duchenne and Onur Çelebi and Patrick Alrassy and Pengchuan Zhang and Pengwei Li and Petar Vasic and Peter Weng and Prajjwal Bhargava and Pratik Dubal and Praveen Krishnan and Punit Singh Koura and Puxin Xu and Qing He and Qingxiao Dong and Ragavan Srinivasan and Raj Ganapathy and Ramon Calderer and Ricardo Silveira Cabral and Robert Stojnic and Roberta Raileanu and Rohan Maheswari and Rohit Girdhar and Rohit Patel and Romain Sauvestre and Ronnie Polidoro and Roshan Sumbaly and Ross Taylor and Ruan Silva and Rui Hou and Rui Wang and Saghar Hosseini and Sahana Chennabasappa and Sanjay Singh and Sean Bell and Seohyun Sonia Kim and Sergey Edunov and Shaoliang Nie and Sharan Narang and Sharath Raparthy and Sheng Shen and Shengye Wan and Shruti Bhosale and Shun Zhang and Simon Vandenhende and Soumya Batra and Spencer Whitman and Sten Sootla and Stephane Collot and Suchin Gururangan and Sydney Borodinsky and Tamar Herman and Tara Fowler and Tarek Sheasha and Thomas Georgiou and Thomas Scialom and Tobias Speckbacher and Todor Mihaylov and Tong Xiao and Ujjwal Karn and Vedanuj Goswami and Vibhor Gupta and Vignesh Ramanathan and Viktor Kerkez and Vincent Gonguet and Virginie Do and Vish Vogeti and Vítor Albiero and Vladan Petrovic and Weiwei Chu and Wenhan Xiong and Wenyin Fu and Whitney Meers and Xavier Martinet and Xiaodong Wang and Xiaofang Wang and Xiaoqing Ellen Tan and Xide Xia and Xinfeng Xie and Xuchao Jia and Xuewei Wang and Yaelle Goldschlag and Yashesh Gaur and Yasmine Babaei and Yi Wen and Yiwen Song and Yuchen Zhang and Yue Li and Yuning Mao and Zacharie Delpierre Coudert and Zheng Yan and Zhengxing Chen and Zoe Papakipos and Aaditya Singh and Aayushi Srivastava and Abha Jain and Adam Kelsey and Adam Shajnfeld and Adithya Gangidi and Adolfo Victoria and Ahuva Goldstand and Ajay Menon and Ajay Sharma and Alex Boesenberg and Alexei Baevski and Allie Feinstein and Amanda Kallet and Amit Sangani and Amos Teo and Anam Yunus and Andrei Lupu and Andres Alvarado and Andrew Caples and Andrew Gu and Andrew Ho and Andrew Poulton and Andrew Ryan and Ankit Ramchandani and Annie Dong and Annie Franco and Anuj Goyal and Aparajita Saraf and Arkabandhu Chowdhury and Ashley Gabriel and Ashwin Bharambe and Assaf Eisenman and Azadeh Yazdan and Beau James and Ben Maurer and Benjamin Leonhardi and Bernie Huang and Beth Loyd and Beto De Paola and Bhargavi Paranjape and Bing Liu and Bo Wu and Boyu Ni and Braden Hancock and Bram Wasti and Brandon Spence and Brani Stojkovic and Brian Gamido and Britt Montalvo and Carl Parker and Carly Burton and Catalina Mejia and Ce Liu and Changhan Wang and Changkyu Kim and Chao Zhou and Chester Hu and Ching-Hsiang Chu and Chris Cai and Chris Tindal and Christoph Feichtenhofer and Cynthia Gao and Damon Civin and Dana Beaty and Daniel Kreymer and Daniel Li and David Adkins and David Xu and Davide Testuggine and Delia David and Devi Parikh and Diana Liskovich and Didem Foss and Dingkang Wang and Duc Le and Dustin Holland and Edward Dowling and Eissa Jamil and Elaine Montgomery and Eleonora Presani and Emily Hahn and Emily Wood and Eric-Tuan Le and Erik Brinkman and Esteban Arcaute and Evan Dunbar and Evan Smothers and Fei Sun and Felix Kreuk and Feng Tian and Filippos Kokkinos and Firat Ozgenel and Francesco Caggioni and Frank Kanayet and Frank Seide and Gabriela Medina Florez and Gabriella Schwarz and Gada Badeer and Georgia Swee and Gil Halpern and Grant Herman and Grigory Sizov and Guangyi and Zhang and Guna Lakshminarayanan and Hakan Inan and Hamid Shojanazeri and Han Zou and Hannah Wang and Hanwen Zha and Haroun Habeeb and Harrison Rudolph and Helen Suk and Henry Aspegren and Hunter Goldman and Hongyuan Zhan and Ibrahim Damlaj and Igor Molybog and Igor Tufanov and Ilias Leontiadis and Irina-Elena Veliche and Itai Gat and Jake Weissman and James Geboski and James Kohli and Janice Lam and Japhet Asher and Jean-Baptiste Gaya and Jeff Marcus and Jeff Tang and Jennifer Chan and Jenny Zhen and Jeremy Reizenstein and Jeremy Teboul and Jessica Zhong and Jian Jin and Jingyi Yang and Joe Cummings and Jon Carvill and Jon Shepard and Jonathan McPhie and Jonathan Torres and Josh Ginsburg and Junjie Wang and Kai Wu and Kam Hou U and Karan Saxena and Kartikay Khandelwal and Katayoun Zand and Kathy Matosich and Kaushik Veeraraghavan and Kelly Michelena and Keqian Li and Kiran Jagadeesh and Kun Huang and Kunal Chawla and Kyle Huang and Lailin Chen and Lakshya Garg and Lavender A and Leandro Silva and Lee Bell and Lei Zhang and Liangpeng Guo and Licheng Yu and Liron Moshkovich and Luca Wehrstedt and Madian Khabsa and Manav Avalani and Manish Bhatt and Martynas Mankus and Matan Hasson and Matthew Lennie and Matthias Reso and Maxim Groshev and Maxim Naumov and Maya Lathi and Meghan Keneally and Miao Liu and Michael L. Seltzer and Michal Valko and Michelle Restrepo and Mihir Patel and Mik Vyatskov and Mikayel Samvelyan and Mike Clark and Mike Macey and Mike Wang and Miquel Jubert Hermoso and Mo Metanat and Mohammad Rastegari and Munish Bansal and Nandhini Santhanam and Natascha Parks and Natasha White and Navyata Bawa and Nayan Singhal and Nick Egebo and Nicolas Usunier and Nikhil Mehta and Nikolay Pavlovich Laptev and Ning Dong and Norman Cheng and Oleg Chernoguz and Olivia Hart and Omkar Salpekar and Ozlem Kalinli and Parkin Kent and Parth Parekh and Paul Saab and Pavan Balaji and Pedro Rittner and Philip Bontrager and Pierre Roux and Piotr Dollar and Polina Zvyagina and Prashant Ratanchandani and Pritish Yuvraj and Qian Liang and Rachad Alao and Rachel Rodriguez and Rafi Ayub and Raghotham Murthy and Raghu Nayani and Rahul Mitra and Rangaprabhu Parthasarathy and Raymond Li and Rebekkah Hogan and Robin Battey and Rocky Wang and Russ Howes and Ruty Rinott and Sachin Mehta and Sachin Siby and Sai Jayesh Bondu and Samyak Datta and Sara Chugh and Sara Hunt and Sargun Dhillon and Sasha Sidorov and Satadru Pan and Saurabh Mahajan and Saurabh Verma and Seiji Yamamoto and Sharadh Ramaswamy and Shaun Lindsay and Shaun Lindsay and Sheng Feng and Shenghao Lin and Shengxin Cindy Zha and Shishir Patil and Shiva Shankar and Shuqiang Zhang and Shuqiang Zhang and Sinong Wang and Sneha Agarwal and Soji Sajuyigbe and Soumith Chintala and Stephanie Max and Stephen Chen and Steve Kehoe and Steve Satterfield and Sudarshan Govindaprasad and Sumit Gupta and Summer Deng and Sungmin Cho and Sunny Virk and Suraj Subramanian and Sy Choudhury and Sydney Goldman and Tal Remez and Tamar Glaser and Tamara Best and Thilo Koehler and Thomas Robinson and Tianhe Li and Tianjun Zhang and Tim Matthews and Timothy Chou and Tzook Shaked and Varun Vontimitta and Victoria Ajayi and Victoria Montanez and Vijai Mohan and Vinay Satish Kumar and Vishal Mangla and Vlad Ionescu and Vlad Poenaru and Vlad Tiberiu Mihailescu and Vladimir Ivanov and Wei Li and Wenchen Wang and Wenwen Jiang and Wes Bouaziz and Will Constable and Xiaocheng Tang and Xiaojian Wu and Xiaolan Wang and Xilun Wu and Xinbo Gao and Yaniv Kleinman and Yanjun Chen and Ye Hu and Ye Jia and Ye Qi and Yenda Li and Yilin Zhang and Ying Zhang and Yossi Adi and Youngjin Nam and Yu and Wang and Yu Zhao and Yuchen Hao and Yundi Qian and Yunlu Li and Yuzi He and Zach Rait and Zachary DeVito and Zef Rosnbrick and Zhaoduo Wen and Zhenyu Yang and Zhiwei Zhao and Zhiyu Ma},
  journal={arXiv preprint arXiv:2407.21783},
  year={2024}
}

@article{goto2008anatomy,
  title={Anatomy of high-performance matrix multiplication},
  author={Goto, Kazushige and van de Geijn, Robert A},
  journal={ACM Transactions on Mathematical Software (TOMS)},
  volume={34},
  number={3},
  pages={1--25},
  year={2008}
}

@misc{openblas,
  title={{OpenBLAS}: An Optimized {BLAS} Library},
  author={Zhang, Xianyi and Kroeker, Martin and Saar, Werner and Qian, Wang and Chothia, Zaheer and Chen, Shaohu and Luo, Wen},
  howpublished={\url{https://www.openblas.net}},
  year={2026}
}

@article{dixit1991spec,
  title={The {SPEC} benchmarks},
  author={Dixit, Kaivalya M},
  journal={Parallel Computing},
  volume={17},
  number={10-11},
  pages={1195--1209},
  year={1991}
}

@article{denning1968working,
  title={The working set model for program behavior},
  author={Denning, Peter J},
  journal={Communications of the ACM},
  volume={11},
  number={5},
  pages={323--333},
  year={1968},
  publisher={ACM New York, NY, USA}
}

\clearpage

\appendix
\section{Appendix}
\label{section_appendix}
\begin{table}[htbp]
    \caption{Dynamic instruction count (billions), application domain, instruction mix, and IPC for each SPEC CPU26 benchmark measured on a representative Intel Sapphire Rapids Platinum platform (CPU-C in Table~\ref{table_machine_configurations}).}
    \label{table_spec_overview}
    \centering
    \footnotesize
    \setlength{\tabcolsep}{1pt}
    \renewcommand{\arraystretch}{1}
    \begin{tabular}{| >{\centering\arraybackslash}p{1.6cm} |
                       >{\centering\arraybackslash}p{2.2cm} |
                       >{\centering\arraybackslash}p{0.8cm} |
                       >{\centering\arraybackslash}p{0.8cm} |
                       >{\centering\arraybackslash}p{0.8cm} |
                       >{\centering\arraybackslash}p{0.9cm} |
                       >{\centering\arraybackslash}p{0.8cm} 
                       |}
        \hline
        Benchmark & Domain & Icount & Loads & Stores & Branches & IPC \\
        \hline
        \multicolumn{7}{|c|}{\SPEC Integer Rate -- 14 workloads} \\
        \hline
        706.stockfish\_r & Games (Chess)~\cite{romstad2008stockfish} & 6507 & 22.0 & 9.9 & 10.4 & 3.625 \\
        \hline
        707.ntest\_r & Games (Othello)~\cite{NTest} & 2507 & 25.0 & 9.6 & 9.2 & 3.268 \\
        \hline
        708.sqlite\_r & Database~\cite{gaffney2022sqlite} & 1716 & 26.9 & 11.9 & 20.9 & 2.228 \\
        \hline
        710.omnetpp\_r & Network Sim.~\cite{varga2008overview} & 1583 & 31.9 & 17.5 & 20.5 & 2.103 \\
        \hline
        714.cpython\_r & Lang. Runtime~\cite{van1995python} & 1475 & 27.9 & 15.8 & 21.4 & 2.843 \\
        \hline
        721.gcc\_r & Compiler~\cite{stallman2002gnu} & 1503 & 28.0 & 11.3 & 21.8 & 0.551 \\
        \hline
        723.llvm\_r & Compiler~\cite{lattner2004llvm} & 1534 & 26.0 & 13.7 & 20.8 & 1.484 \\
        \hline
        727.cppcheck\_r & Static Analysis~\cite{marjamaki2013cppcheck} & 1286 & 22.5 & 9.6 & 26.7 & 2.228 \\
        \hline
        729.abc\_r & EDA~\cite{brayton2010abc} & 1400 & 26.2 & 8.9 & 16.7 & 2.187 \\
        \hline
        734.vpr\_r & EDA (FPGA)~\cite{betz1997vpr} & 1367 & 30.9 & 11.2 & 19.2 & 2.097 \\
        \hline
        735.gem5\_r & Arch. Sim.~\cite{binkert2011gem5} & 1659 & 30.2 & 14.7 & 20.9 & 2.068 \\
        \hline
        750.sealcrypto\_r & Cryptography~\cite{chen2017simple} & 3087 & 12.0 & 4.7 & 1.9 & 4.961 \\
        \hline
        753.ns3\_r & Network Sim.~\cite{riley2010ns} & 1432 & 29.4 & 16.8 & 22.2 & 2.230 \\
        \hline
        777.zstd\_r & Compression~\cite{zstd} & 1817 & 22.2 & 9.0 & 13.3 & 1.911 \\
        \hline
        \multicolumn{7}{|c|}{\SPEC Integer Speed -- 13 workloads} \\
        \hline
        801.xz\_s & Compression~\cite{xzutils} & 17757 & 22.2 & 7.4 & 14.4 & 1.008 \\
        \hline
        807.ntest\_s & Games (Othello)~\cite{NTest} & 151005 & 20.8 & 7.8 & 6.7 & 3.460 \\
        \hline
        817.flac\_s & Compression & 90970 & 17.6 & 2.3 & 4.4 & 4.156 \\
        \hline
        821.gcc\_s & Compiler~\cite{stallman2002gnu} & 109486 & 26.8 & 12.5 & 21.7 & 2.016 \\
        \hline
        823.llvm\_s & Compiler~\cite{lattner2004llvm} & 103105 & 22.0 & 11.8 & 23.1 & 1.896 \\
        \hline
        827.cppcheck\_s & Static Analysis~\cite{marjamaki2013cppcheck} & 90423 & 23.0 & 11.2 & 26.5 & 2.375 \\
        \hline
        829.abc\_s & EDA~\cite{brayton2010abc} & 1433 & 25.1 & 11.7 & 18.1 & 0.858 \\
        \hline
        834.vpr\_s & EDA (FPGA)~\cite{betz1997vpr} & 3117 & 30.7 & 11.1 & 19.4 & 1.863 \\
        \hline
        835.gem5\_s & Arch. Sim.~\cite{binkert2011gem5} & 2858 & 29.4 & 13.5 & 17.6 & 1.805 \\
        \hline
        838.diamond\_s & Bioinformatics~\cite{buchfink2021sensitive} & 146966 & 20.1 & 6.8 & 5.5 & 3.203 \\
        \hline
        846.minizinc\_s & CSP Solver~\cite{nethercote2007minizinc} & 5062 & 26.5 & 18.2 & 15.9 & 1.228 \\
        \hline
        853.ns3\_s & Network Sim.~\cite{riley2010ns} & 11053 & 28.9 & 14.3 & 21.0 & 1.662 \\
        \hline
        854.graph500\_s & Graph Analytics~\cite{murphy2010introducing} & 37168 & 36.2 & 0.9 & 25.7 & 1.539 \\
        \hline
        \multicolumn{7}{|c|}{\SPEC Floating Point Rate -- 12 workloads} \\
        \hline
        709.cactus\_r & Physics (GR)~\cite{goodale2002cactus} & 1456 & 51.9 & 7.9 & 1.1 & 1.696 \\
        \hline
        722.palm\_r & Climate~\cite{maronga2020overview} & 3272 & 39.0 & 9.1 & 5.0 & 3.187 \\
        \hline
        731.astcenc\_r & Img. Compress.~\cite{astc} & 2615 & 28.3 & 6.5 & 8.7 & 2.718 \\
        \hline
        736.ocio\_r & Image Process.~\cite{walker2020aswf} & 2484 & 24.2 & 7.5 & 9.8 & 3.269 \\
        \hline
        737.gmsh\_r & Mesh Generation~\cite{geuzaine2009gmsh} & 1086 & 29.2 & 12.1 & 17.2 & 1.585 \\
        \hline
        748.flightdm\_r & Aerospace~\cite{berndt2004jsbsim} & 1721 & 29.5 & 14.2 & 18.8 & 3.071 \\
        \hline
        749.fotonik3d\_r & Photonics~\cite{andersson2006parallel} & 1291 & 36.8 & 13.7 & 1.8 & 0.785 \\
        \hline
        765.roms\_r & Ocean Modeling~\cite{shchepetkin2005regional} & 2738 & 34.8 & 8.4 & 7.3 & 1.830 \\
        \hline
        766.femflow\_r & Fluid Dynamics~\cite{kronbichler2023enhancing} & 5012 & 34.9 & 20.1 & 6.9 & 3.265 \\
        \hline
        767.nest\_r & Neuroscience~\cite{gewaltig2007nest} & 1848 & 33.5 & 12.3 & 14.0 & 2.844 \\
        \hline
        772.marian\_r & NLP~\cite{junczys2018marian} & 6389 & 8.7 & 1.3 & 3.0 & 3.953 \\
        \hline
        782.lbm\_r & Fluid Dynamics~\cite{tramm2014performance} & 2236 & 21.2 & 10.9 & 0.7 & 1.241 \\
        \hline
        \multicolumn{7}{|c|}{\SPEC Floating Point Speed -- 13 workloads} \\
        \hline
        800.pot3d\_s & Astrophysics~\cite{downs2025near} & 7603 & 34.8 & 8.1 & 9.5 & 0.754 \\
        \hline
        803.sph\_exa\_s & HPC (SPH)~\cite{guerrera2018towards} & 64626 & 24.8 & 3.4 & 11.2 & 2.465 \\
        \hline
        809.cactus\_s & Physics (GR)~\cite{goodale2002cactus} & 29190 & 51.9 & 8.1 & 1.6 & 1.338 \\
        \hline
        811.tealeaf\_s & Physics (Heat)~\cite{mcintosh2017tealeaf} & 40570 & 20.2 & 4.9 & 8.8 & 1.617 \\
        \hline
        816.nab\_s & Mol. Dynamics~\cite{macke1998modeling} & 67717 & 31.2 & 5.7 & 11.9 & 2.441 \\
        \hline
        820.cloverleaf\_s & Physics (Hydro)~\cite{cloverleaf} & 25781 & 33.2 & 4.7 & 5.9 & 1.349 \\
        \hline
        822.palm\_s & Climate~\cite{maronga2020overview} & 48883 & 38.2 & 8.9 & 6.4 & 1.920 \\
        \hline
        849.fotonik3d\_s & Photonics~\cite{andersson2006parallel} & 17777 & 56.0 & 9.8 & 2.7 & 0.955 \\
        \hline
        857.namd\_s & Mol. Dynamics~\cite{phillips2005scalable} & 168881 & 26.4 & 6.6 & 2.2 & 3.929 \\
        \hline
        865.roms\_s & Ocean Modeling~\cite{shchepetkin2005regional} & 28484 & 34.7 & 8.6 & 7.9 & 1.574 \\
        \hline
        867.nest\_s & Neuroscience~\cite{gewaltig2007nest} & 66774 & 30.1 & 9.3 & 14.7 & 1.790 \\
        \hline
        872.marian\_s & NLP~\cite{junczys2018marian} & 65980 & 10.8 & 2.8 & 3.8 & 3.258 \\
        \hline
        881.neutron\_s & Nuclear Physics~\cite{pohl2003optimization} & 33545 & 25.8 & 11.3 & 9.1 & 1.204 \\
        \hline
    \end{tabular}
\end{table}
\begin{table*}[t]
\caption{Detailed per-workload RSS of \SPECO (appendix for Figure~\ref{fig_cpu2017_cpu2026_rate_rss}), in GB.}
\label{tab_rss_detailed_cpu2017}
\centering
\scriptsize
\setlength{\tabcolsep}{2pt}
\renewcommand{\arraystretch}{1.05}
\resizebox{\textwidth}{!}{%
\begin{tabular}{l|cccccccccc}
\hline
Workload & C\_glibc & C\_mi\_no\_thp & C\_mi\_thp & C\_tc\_no\_thp & C\_tc\_thp & I\_glibc & I\_mi\_no\_thp & I\_mi\_thp & I\_tc\_no\_thp & I\_tc\_thp \\
\hline
500.perlbench\_r & 0.54 & 0.57 & 0.58 & 0.57 & 0.57 & 0.56 & 0.59 & 0.59 & 0.60 & 0.59 \\
502.gcc\_r & 1.66 & 1.63 & 1.64 & 1.68 & 1.67 & 1.67 & 1.66 & 1.64 & 1.70 & 1.69 \\
503.bwaves\_r & 1.15 & 1.15 & 1.18 & 1.16 & 1.16 & 1.16 & 1.16 & 1.16 & 1.17 & 1.17 \\
505.mcf\_r & 0.94 & 0.94 & 0.95 & 1.23 & 1.23 & 0.96 & 0.96 & 0.96 & 1.24 & 1.24 \\
507.cactuBSSN\_r & 1.12 & 1.12 & 1.20 & 1.12 & 1.12 & 1.14 & 1.14 & 1.14 & 1.15 & 1.15 \\
508.namd\_r & 0.51 & 0.50 & 0.56 & 0.51 & 0.51 & 0.53 & 0.54 & 0.54 & 0.55 & 0.55 \\
510.parest\_r & 0.75 & 0.77 & 0.80 & 0.81 & 0.81 & 0.77 & 0.81 & 0.81 & 0.84 & 0.85 \\
511.povray\_r & 0.45 & 0.44 & 0.45 & 0.45 & 0.45 & 0.42 & 0.43 & 0.46 & 0.46 & 0.48 \\
519.lbm\_r & 0.75 & 0.75 & 0.75 & 0.75 & 0.75 & 0.76 & 0.76 & 0.76 & 0.77 & 0.77 \\
520.omnetpp\_r & 0.58 & 0.58 & 0.58 & 0.58 & 0.59 & 0.60 & 0.59 & 0.59 & 0.60 & 0.60 \\
521.wrf\_r & 0.54 & 0.55 & 0.61 & 0.54 & 0.55 & 0.56 & 0.58 & 0.58 & 0.58 & 0.58 \\
523.xalancbmk\_r & 0.81 & 0.81 & 0.82 & 0.84 & 0.84 & 0.83 & 0.83 & 0.85 & 0.86 & 0.86 \\
525.x264\_r & 0.50 & 0.50 & 0.56 & 0.51 & 0.51 & 0.52 & 0.53 & 0.53 & 0.53 & 0.53 \\
526.blender\_r & 0.96 & 1.08 & 1.10 & 1.01 & 1.01 & 0.98 & 1.12 & 1.12 & 1.05 & 1.05 \\
527.cam4\_r & 1.22 & 1.24 & 1.31 & 1.25 & 1.25 & 1.23 & 1.26 & 1.26 & 1.27 & 1.27 \\
531.deepsjeng\_r & 1.03 & 1.03 & 1.03 & 1.04 & 1.04 & 1.04 & 1.04 & 1.04 & 1.05 & 1.05 \\
538.imagick\_r & 0.63 & 0.72 & 0.72 & 0.72 & 0.72 & 0.64 & 0.73 & 0.73 & 0.74 & 0.74 \\
541.leela\_r & 0.40 & 0.40 & 0.45 & 0.45 & 0.40 & 0.41 & 0.42 & 0.47 & 0.42 & 0.42 \\
544.nab\_r & 0.49 & 0.53 & 0.54 & 0.51 & 0.52 & 0.51 & 0.55 & 0.55 & 0.52 & 0.52 \\
548.exchange2\_r & 0.40 & 0.46 & 0.46 & 0.45 & 0.44 & 0.47 & 0.41 & 0.47 & 0.42 & 0.46 \\
549.fotonik3d\_r & 1.17 & 1.18 & 1.20 & 1.18 & 1.18 & 1.19 & 1.19 & 1.19 & 1.20 & 1.20 \\
554.roms\_r & 1.17 & 1.17 & 1.23 & 1.18 & 1.18 & 1.18 & 1.19 & 1.19 & 1.20 & 1.20 \\
557.xz\_r & 1.11 & 1.16 & 1.17 & 1.10 & 1.28 & 1.11 & 1.16 & 1.16 & 1.20 & 1.29 \\
\hline
\end{tabular}%
}
\end{table*}

\begin{table*}[t]
\caption{Detailed per-workload RSS of \SPEC (appendix for Figure~\ref{fig_cpu2017_cpu2026_rate_rss}), in GB.}
\label{tab_rss_detailed_cpu2026}
\centering
\scriptsize
\setlength{\tabcolsep}{2pt}
\renewcommand{\arraystretch}{1.05}
\resizebox{\textwidth}{!}{%
\begin{tabular}{l|cccccccccc}
\hline
Workload & C\_glibc & C\_mi\_no\_thp & C\_mi\_thp & C\_tc\_no\_thp & C\_tc\_thp & I\_glibc & I\_mi\_no\_thp & I\_mi\_thp & I\_tc\_no\_thp & I\_tc\_thp \\
\hline
706.stockfish\_r & 2.06 & 2.06 & 2.14 & 2.13 & 2.13 & 2.08 & 2.15 & 2.15 & 2.15 & 2.15 \\
707.ntest\_r & 0.64 & 0.63 & 0.64 & 0.64 & 0.64 & 0.66 & 0.67 & 0.67 & 0.67 & 0.66 \\
708.sqlite\_r & 1.59 & 1.59 & 1.43 & 1.35 & 1.35 & 1.61 & 1.43 & 1.45 & 1.35 & 1.35 \\
709.cactus\_r & 2.05 & 2.05 & 2.17 & 2.05 & 2.05 & 2.08 & 2.09 & 2.09 & 2.09 & 2.09 \\
710.omnetpp\_r & 1.08 & 1.15 & 1.07 & 1.13 & 1.09 & 1.16 & 1.14 & 1.15 & 1.10 & 1.08 \\
714.cpython\_r & 1.53 & 1.51 & 1.57 & 1.69 & 1.71 & 1.53 & 1.57 & 1.58 & 1.69 & 1.70 \\
721.gcc\_r & 1.34 & 1.34 & 1.33 & 1.72 & 1.72 & 1.38 & 1.36 & 1.35 & 1.76 & 1.75 \\
722.palm\_r & 2.13 & 2.15 & 2.19 & 2.18 & 2.19 & 2.14 & 2.19 & 2.19 & 2.21 & 2.21 \\
723.llvm\_r & 1.76 & 1.76 & 1.69 & 1.72 & 1.72 & 1.79 & 1.70 & 1.70 & 1.75 & 1.75 \\
727.cppcheck\_r & 0.97 & 0.97 & 0.95 & 0.96 & 0.96 & 0.99 & 0.97 & 0.97 & 0.98 & 0.98 \\
729.abc\_r & 1.84 & 1.85 & 1.86 & 1.86 & 1.87 & 1.87 & 1.86 & 1.86 & 1.89 & 1.89 \\
731.astcenc\_r & 0.65 & 0.64 & 0.64 & 0.64 & 0.64 & 0.67 & 0.67 & 0.67 & 0.67 & 0.67 \\
734.vpr\_r & 1.98 & 2.00 & 2.06 & 1.80 & 1.81 & 2.03 & 2.02 & 2.12 & 1.85 & 1.85 \\
735.gem5\_r & 1.13 & 1.10 & 1.06 & 1.05 & 1.10 & 1.16 & 1.07 & 1.07 & 1.08 & 1.08 \\
736.ocio\_r & 1.92 & 1.92 & 1.93 & 1.92 & 1.93 & 1.94 & 1.94 & 1.94 & 1.95 & 1.95 \\
737.gmsh\_r & 1.39 & 1.39 & 1.47 & 1.45 & 1.45 & 1.42 & 1.49 & 1.56 & 1.52 & 1.52 \\
748.flightdm\_r & 0.64 & 0.65 & 0.65 & 0.64 & 0.64 & 0.67 & 0.67 & 0.67 & 0.67 & 0.67 \\
749.fotonik3d\_r & 2.11 & 2.11 & 2.15 & 2.12 & 2.12 & 2.13 & 2.14 & 2.14 & 2.15 & 2.15 \\
750.sealcrypto\_r & 0.93 & 0.93 & 0.96 & 0.93 & 0.95 & 0.96 & 0.96 & 0.96 & 0.96 & 0.96 \\
753.ns3\_r & 0.64 & 0.64 & 0.64 & 0.64 & 0.64 & 0.67 & 0.67 & 0.67 & 0.67 & 0.66 \\
760.rocksdb\_r & 0.77 & 0.77 & 0.75 & 0.76 & 0.76 & 0.79 & 0.77 & 0.77 & 0.78 & 0.78 \\
765.roms\_r & 1.56 & 1.56 & 1.62 & 1.57 & 1.58 & 1.59 & 1.60 & 1.60 & 1.62 & 1.62 \\
766.femflow\_r & 0.73 & 0.71 & 0.69 & 0.71 & 0.72 & 0.75 & 0.72 & 0.72 & 0.75 & 0.75 \\
767.nest\_r & 1.46 & 1.46 & 1.72 & 1.48 & 1.48 & 1.49 & 1.74 & 1.74 & 1.50 & 1.50 \\
772.marian\_r & 1.44 & 1.44 & 1.42 & 1.39 & 1.39 & 1.46 & 1.46 & 1.46 & 1.39 & 1.39 \\
777.zstd\_r & 0.84 & 0.84 & 0.84 & 0.84 & 0.84 & 0.86 & 0.86 & 0.86 & 0.86 & 0.86 \\
782.lbm\_r & 2.03 & 2.03 & 2.03 & 2.03 & 2.04 & 2.05 & 2.05 & 2.05 & 2.06 & 2.06 \\
\hline
\end{tabular}%
}
\end{table*}

\begin{figure*}[t]
    \begin{minipage}{2\columnwidth}
	\centering
	\includegraphics[width=\columnwidth, trim = 2mm 4mm 2mm 2mm, clip=true, page=1]{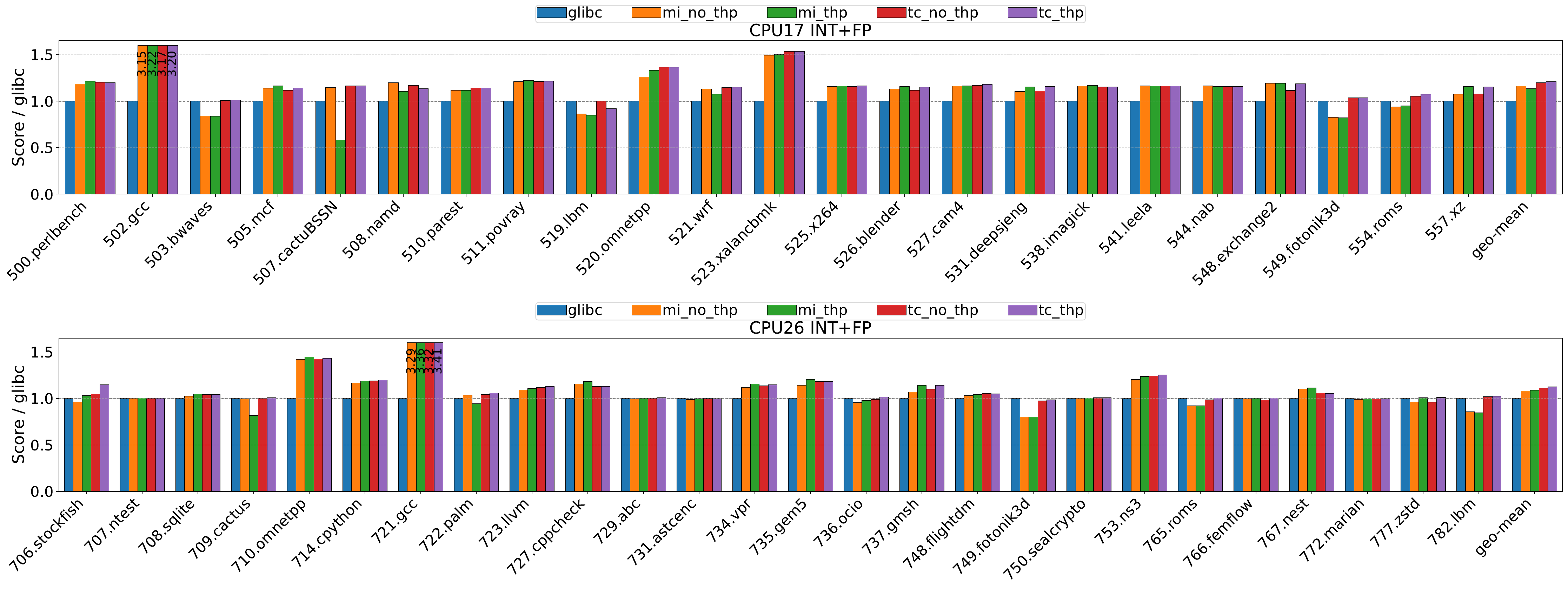}
	\subcaption{Score of CPU-C. }
	\label{fig_thp_appendix_cpu_c_1}
	\end{minipage} 
    \begin{minipage}{2\columnwidth}
	\centering
	\includegraphics[width=\columnwidth, trim = 2mm 4mm 2mm 2mm, clip=true, page=1]{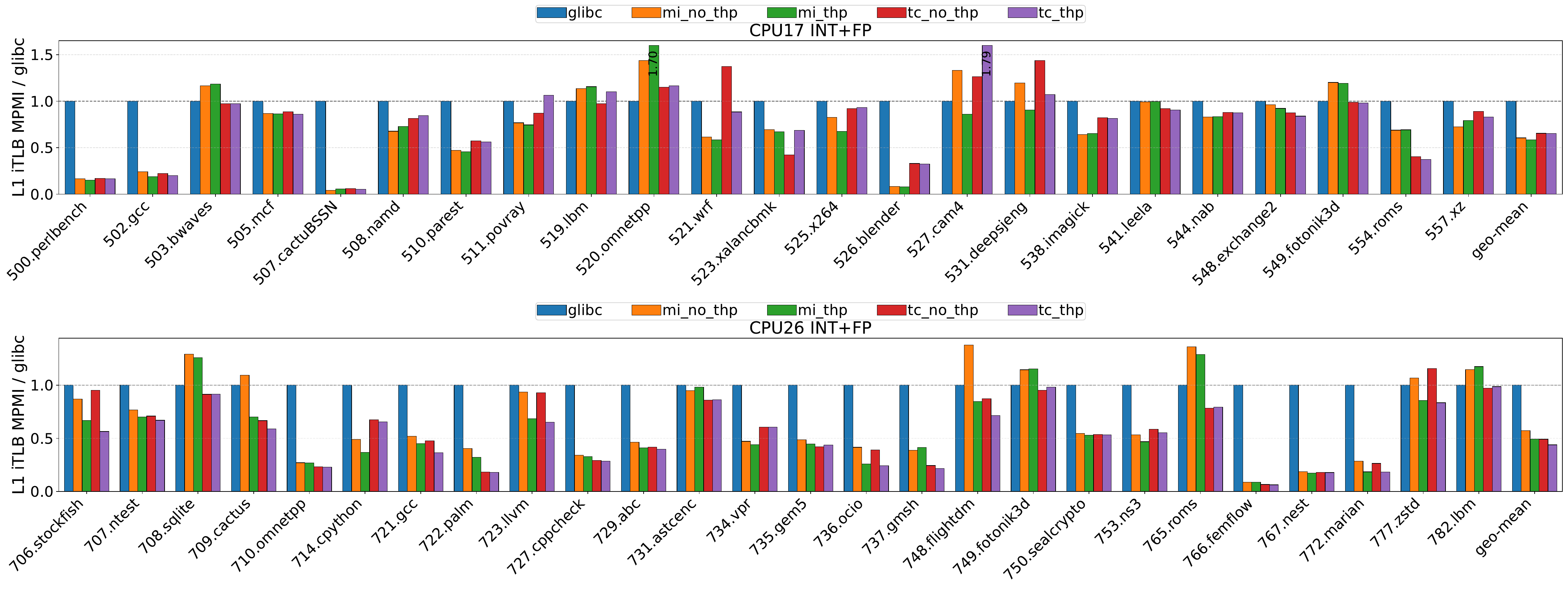}
	\subcaption{L1 iTLB of CPU-C. }
	\label{fig_thp_appendix_cpu_c_2}
	\end{minipage} 
    \begin{minipage}{2\columnwidth}
	\centering
	\includegraphics[width=\columnwidth, trim = 2mm 4mm 2mm 2mm, clip=true, page=1]{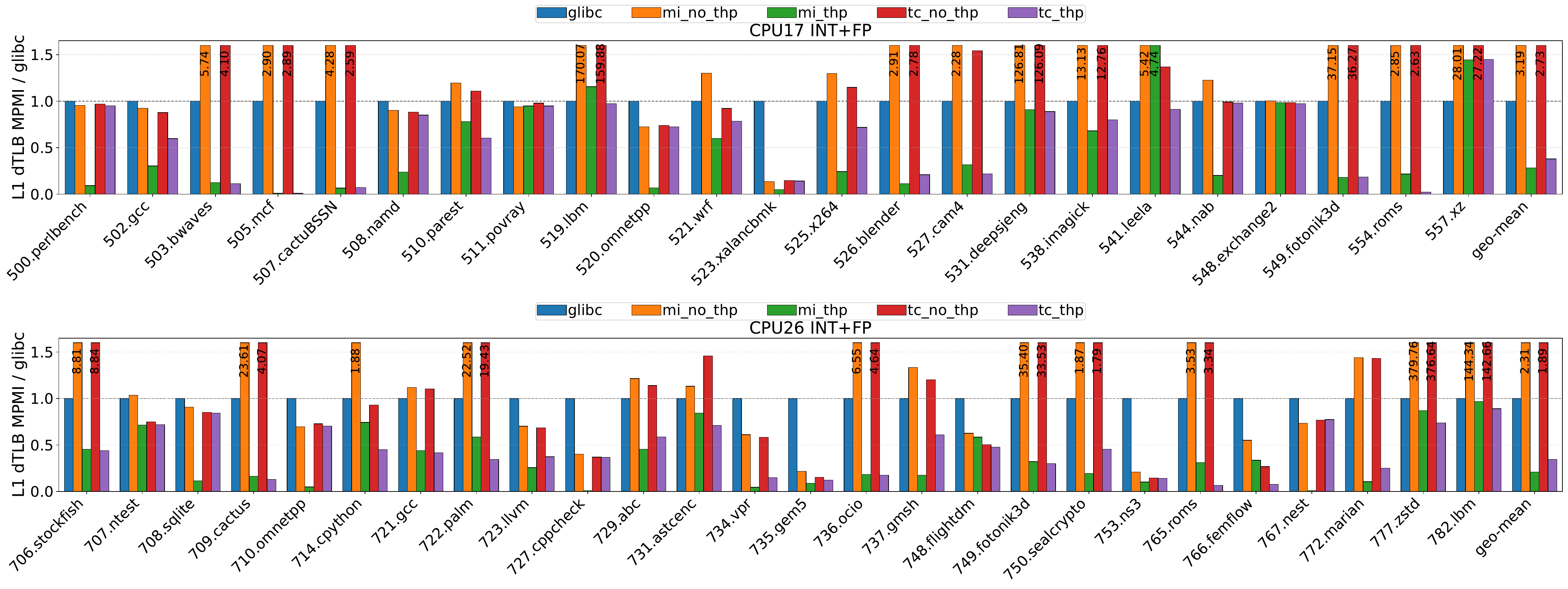}
	\subcaption{L1 dTLB of CPU-C.}
	\label{fig_thp_appendix_cpu_c_3}
	\end{minipage} 
 \caption{Detailed per-workload performance of different allocators on \SPECO and \SPEC (Figure~\ref{fig_thp}, CPU-C).
 }
 \label{fig_thp_appendix_cpu_c}
\end{figure*}

\begin{figure*}[t]
    \begin{minipage}{2\columnwidth}
	\centering
	\includegraphics[width=\columnwidth, trim = 2mm 4mm 2mm 2mm, clip=true, page=1]{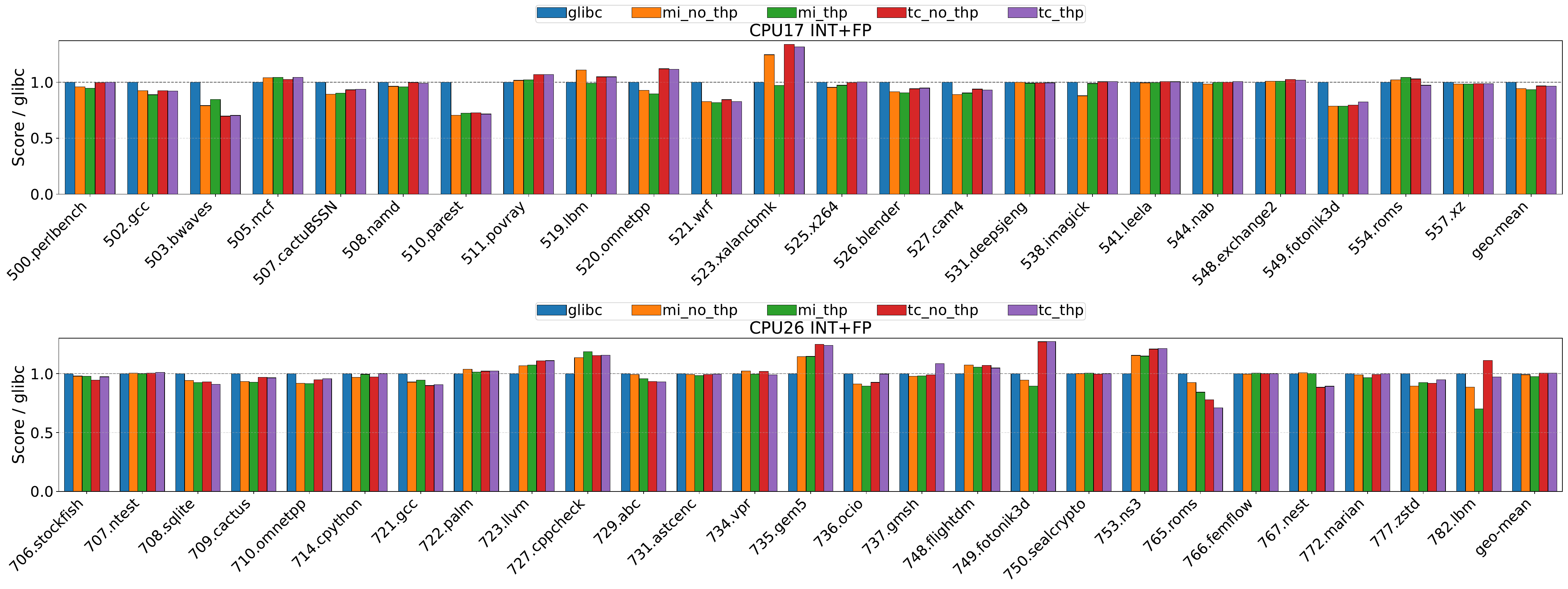}
	\subcaption{Score of CPU-I. }
	\label{fig_thp_appendix_cpu_i_1}
	\end{minipage} 
    \begin{minipage}{2\columnwidth}
	\centering
	\includegraphics[width=\columnwidth, trim = 2mm 4mm 2mm 2mm, clip=true, page=1]{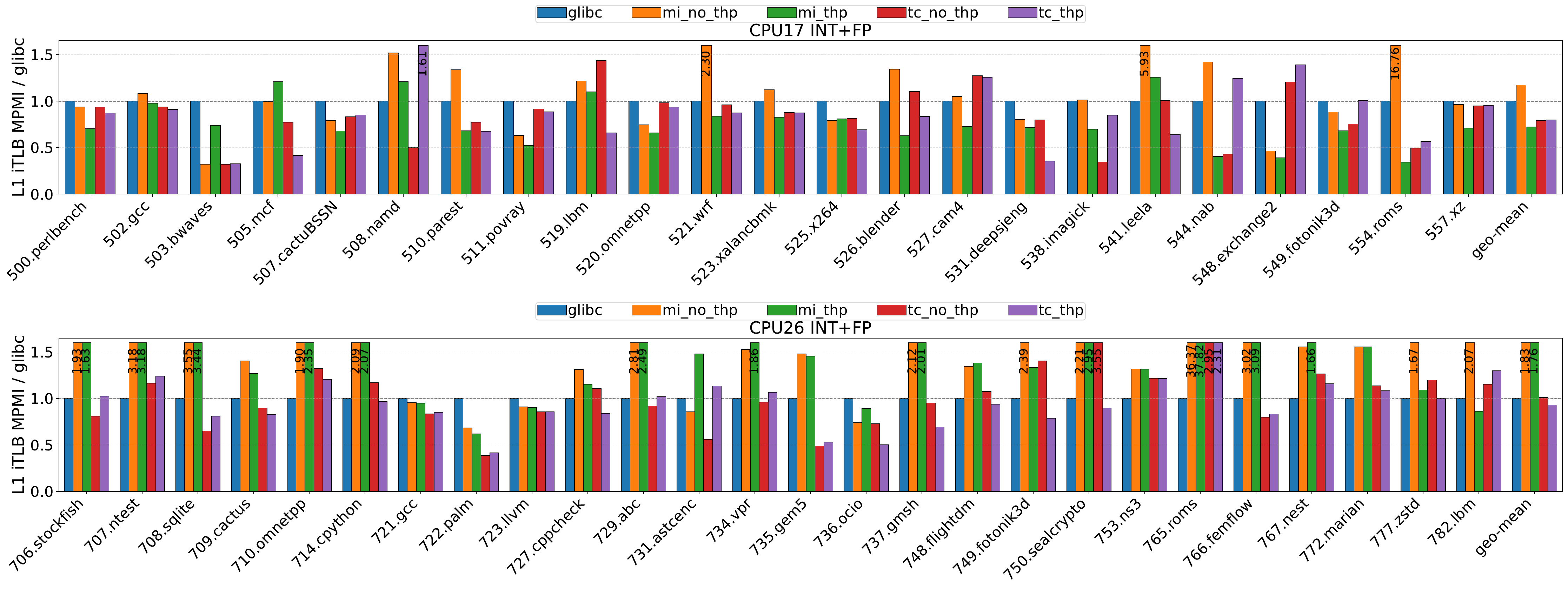}
	\subcaption{L1 iTLB of CPU-I. }
	\label{fig_thp_appendix_cpu_i_2}
	\end{minipage} 
    \begin{minipage}{2\columnwidth}
	\centering
	\includegraphics[width=\columnwidth, trim = 2mm 4mm 2mm 2mm, clip=true, page=1]{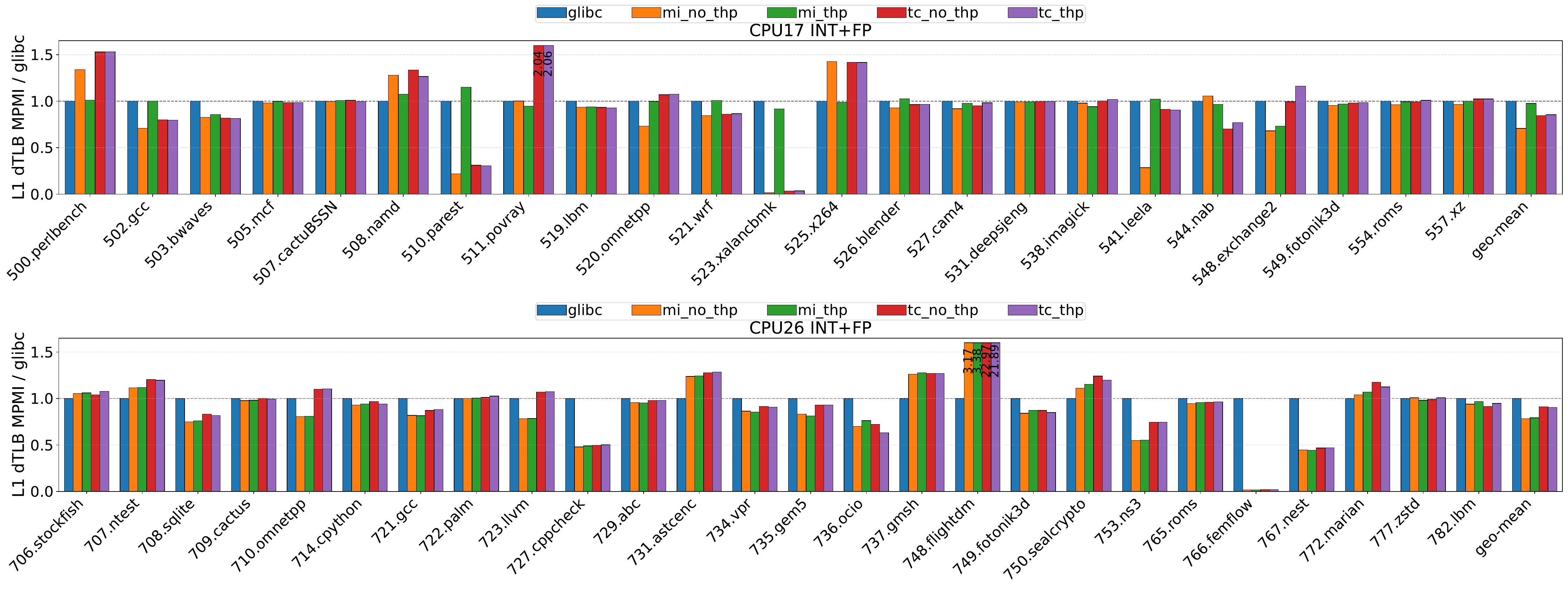}
	\subcaption{L1 dTLB of CPU-I.}
	\label{fig_thp_appendix_cpu_i_3}
	\end{minipage} 
 \caption{Detailed per-workload performance of different allocators on \SPECO and \SPEC (Figure~\ref{fig_thp}, CPU-I).
 }
 \label{fig_thp_appendix_cpu_i}
\end{figure*}
\begin{figure*}[t]
    \begin{minipage}{2\columnwidth}
	\centering
	\includegraphics[width=\columnwidth, trim = 2mm 4mm 2mm 2mm, clip=true, page=1]{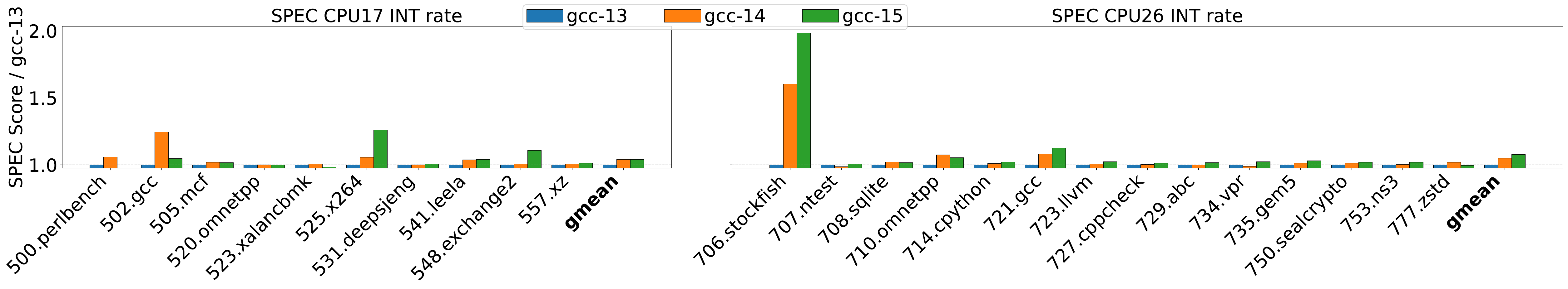}
	\subcaption{Score of INT Rate workloads. }
	\label{fig_compiler_appendix_int_1}
	\end{minipage} 
    \begin{minipage}{2\columnwidth}
	\centering
	\includegraphics[width=\columnwidth, trim = 2mm 4mm 2mm 2mm, clip=true, page=1]{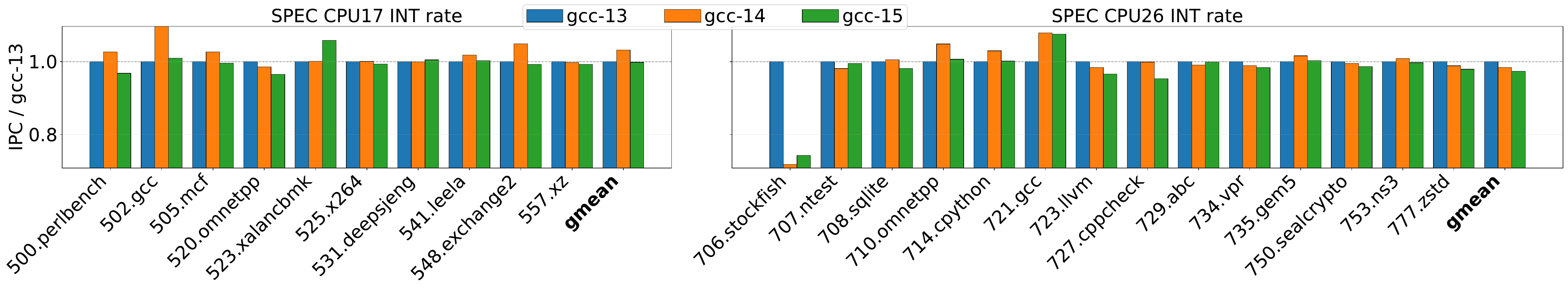}
	\subcaption{IPC of INT Rate workloads. }
	\label{fig_compiler_appendix_int_2}
	\end{minipage} 
    \begin{minipage}{2\columnwidth}
	\centering
	\includegraphics[width=\columnwidth, trim = 2mm 4mm 2mm 2mm, clip=true, page=1]{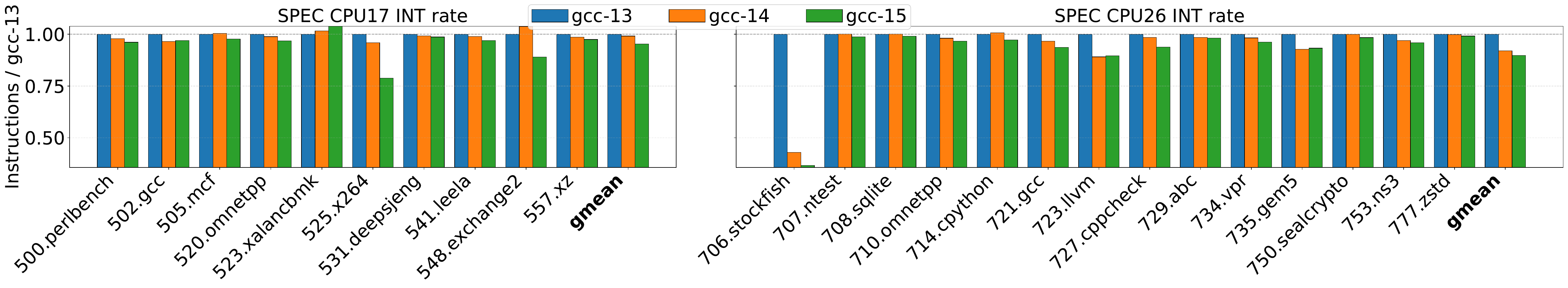}
	\subcaption{Instruction counts of INT Rate workloads.}
	\label{fig_compiler_appendix_int_3}
	\end{minipage} 
 \caption{Detailed per-workload performance of different compilers on \SPECO and \SPEC INT Rate workloads (Figure~\ref{fig_compiler}).
 }
 \label{fig_compiler_appendix_int}
\end{figure*}

\begin{figure*}[t]
    \begin{minipage}{2\columnwidth}
	\centering
	\includegraphics[width=\columnwidth, trim = 2mm 4mm 2mm 2mm, clip=true, page=1]{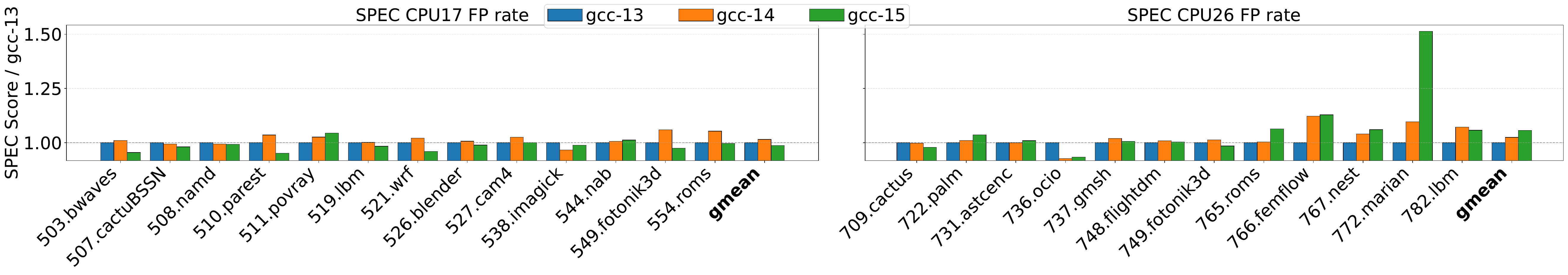}
    \vspace{-2mm}
	\subcaption{Score of FP Rate workloads. }
	\label{fig_compiler_appendix_fp_1}
	\end{minipage} 
    \begin{minipage}{2\columnwidth}
	\centering
	\includegraphics[width=\columnwidth, trim = 2mm 4mm 2mm 2mm, clip=true, page=1]{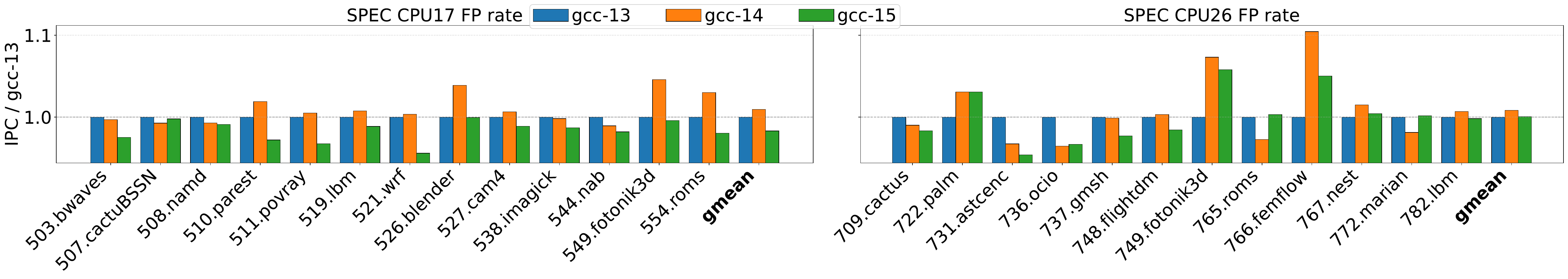}
	\subcaption{IPC of FP Rate workloads.  }
	\label{fig_compiler_appendix_fp_2}
	\end{minipage} 
    \begin{minipage}{2\columnwidth}
	\centering
	\includegraphics[width=\columnwidth, trim = 2mm 4mm 2mm 2mm, clip=true, page=1]{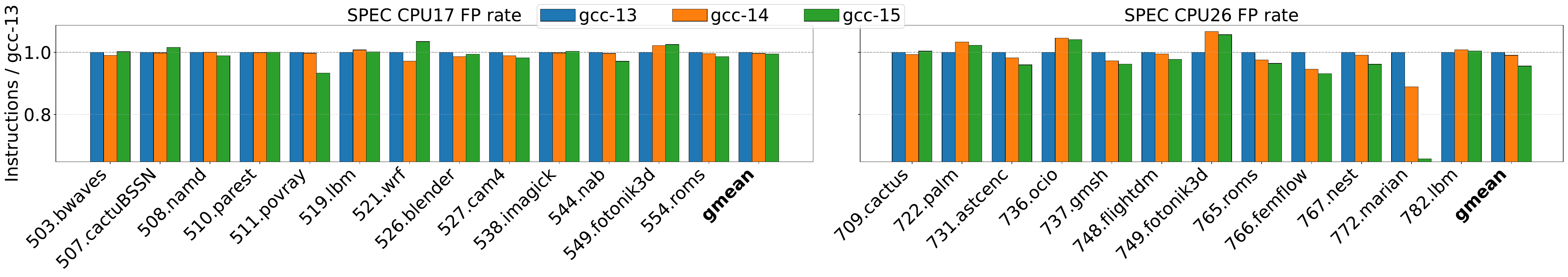}
	\subcaption{Instruction counts of FP Rate workloads. }
	\label{fig_compiler_appendix_fp_3}
	\end{minipage} 
 \caption{Detailed per-workload performance of different compilers on \SPECO and \SPEC FP Rate workloads (Figure~\ref{fig_compiler}).
 }
 \label{fig_compiler_appendix_fp}
\end{figure*}
\begin{figure*}[t]
    \begin{minipage}{2\columnwidth}
	\centering
	\includegraphics[width=\columnwidth, trim = 2mm 4mm 2mm 2mm, clip=true, page=1]{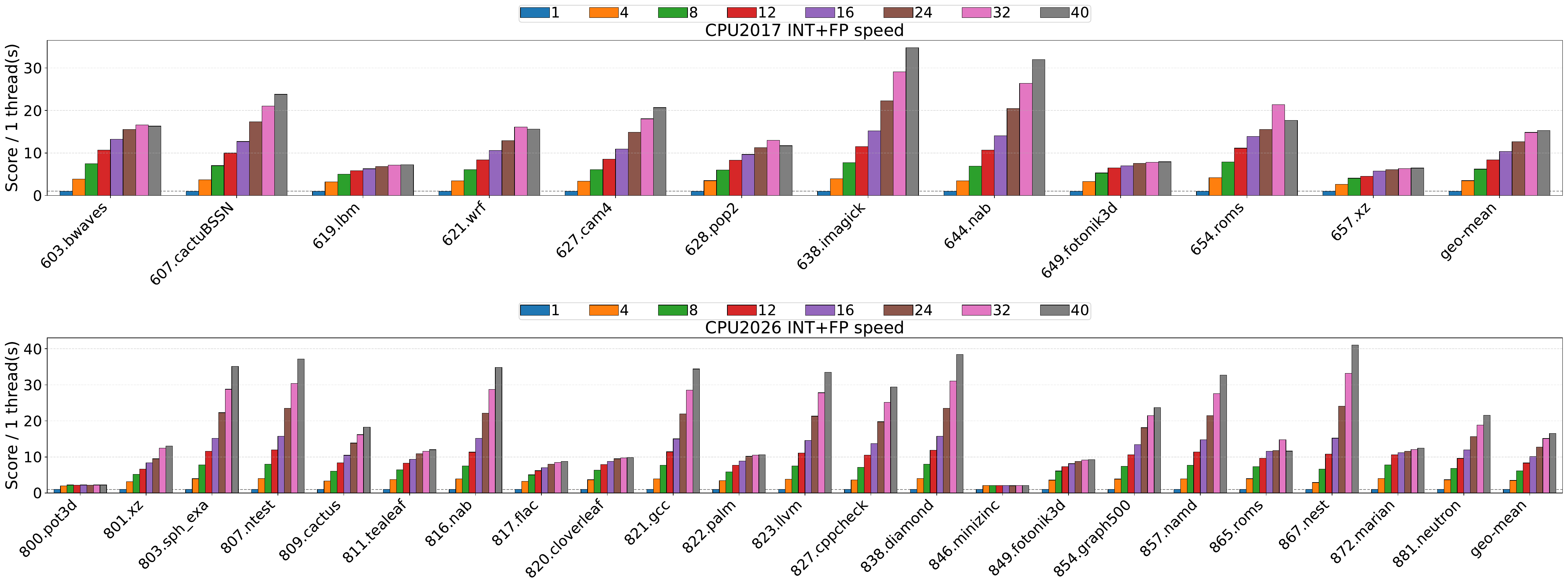}
	\subcaption{Score of CPU-B. }
	\label{fig_scale_appendix_cpu_b_1}
	\end{minipage} 
    \begin{minipage}{2\columnwidth}
	\centering
	\includegraphics[width=\columnwidth, trim = 2mm 4mm 2mm 2mm, clip=true, page=1]{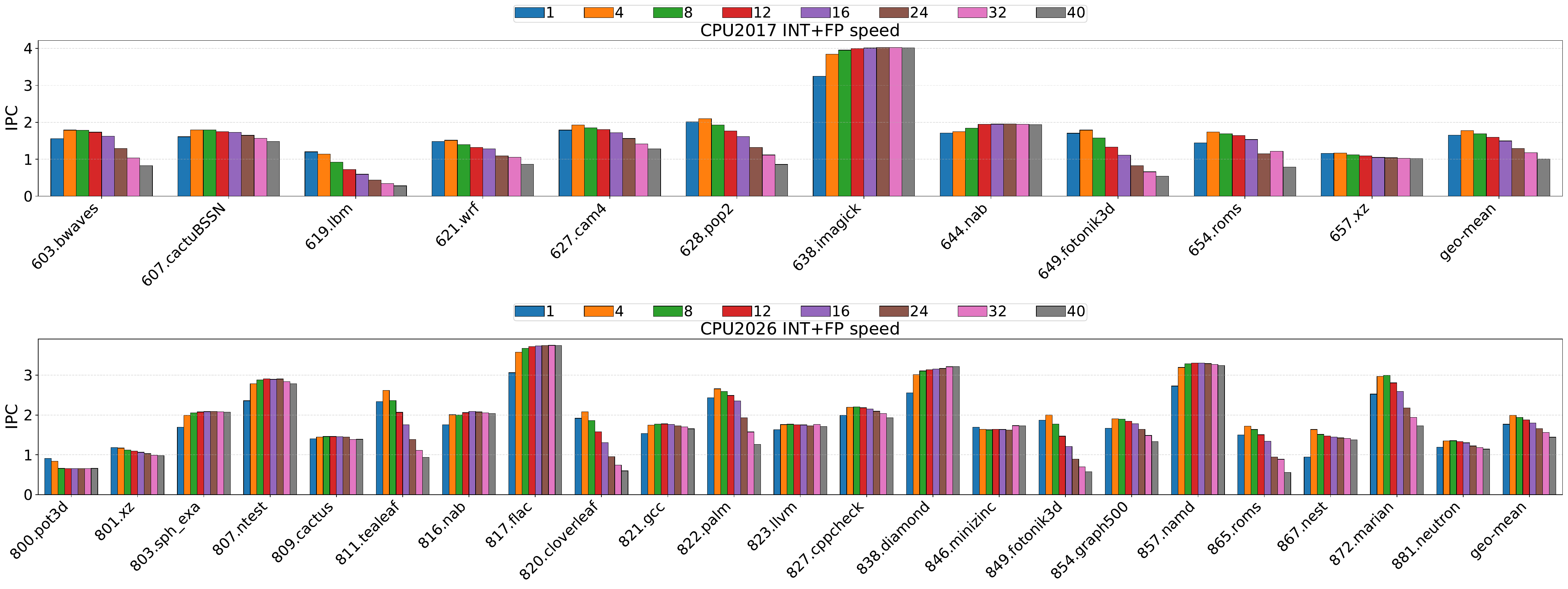}
	\subcaption{IPC of CPU-B. }
	\label{fig_scale_appendix_cpu_b_2}
	\end{minipage} 
    \begin{minipage}{2\columnwidth}
	\centering
	\includegraphics[width=\columnwidth, trim = 2mm 4mm 2mm 2mm, clip=true, page=1]{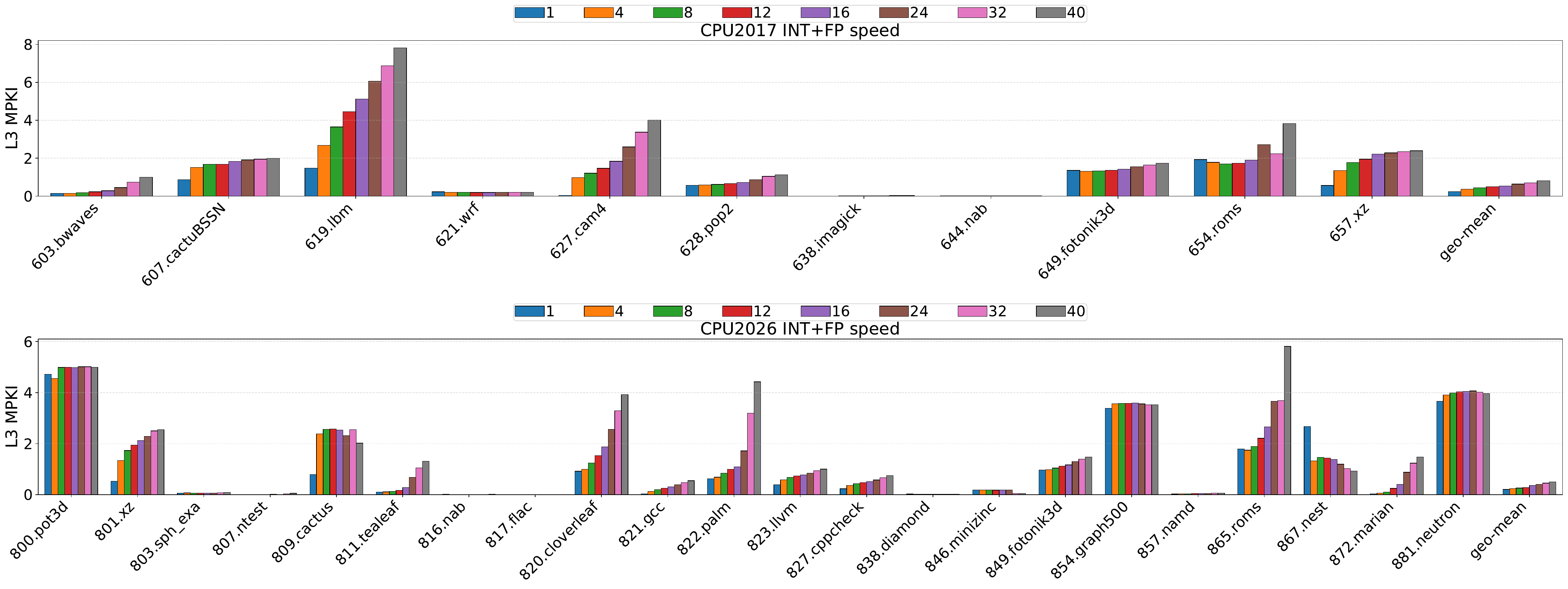}
	\subcaption{L3 cache MPKI of CPU-B.}
	\label{fig_scale_appendix_cpu_b_3}
	\end{minipage} 
 \caption{Detailed per-workload performance of Speed workload scaling on \SPECO and \SPEC (Figure~\ref{fig_speed_scale}, CPU-B).
 }
 \label{fig_scale_appendix_cpu_b}
\end{figure*}

\begin{figure*}[t]
    \begin{minipage}{2\columnwidth}
	\centering
	\includegraphics[width=\columnwidth, trim = 2mm 4mm 2mm 2mm, clip=true, page=1]{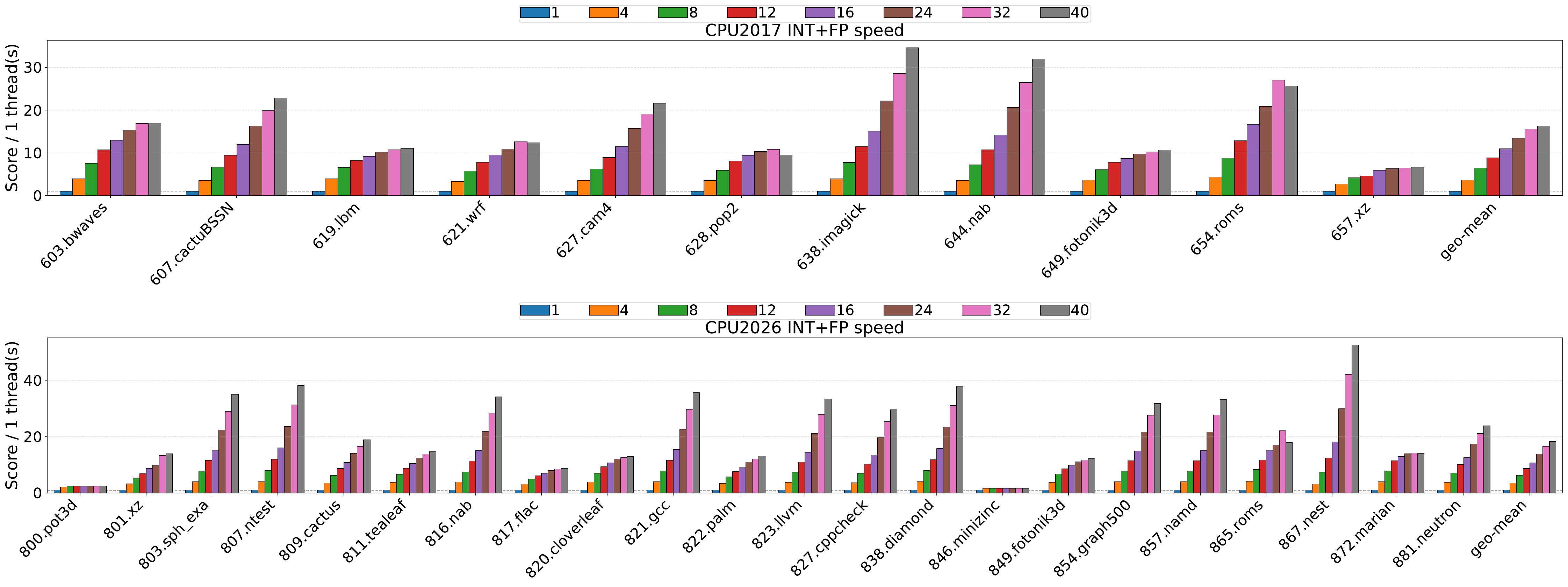}
	\subcaption{Score of CPU-C. }
	\label{fig_scale_appendix_cpu_c_1}
	\end{minipage} 
    \begin{minipage}{2\columnwidth}
	\centering
	\includegraphics[width=\columnwidth, trim = 2mm 4mm 2mm 2mm, clip=true, page=1]{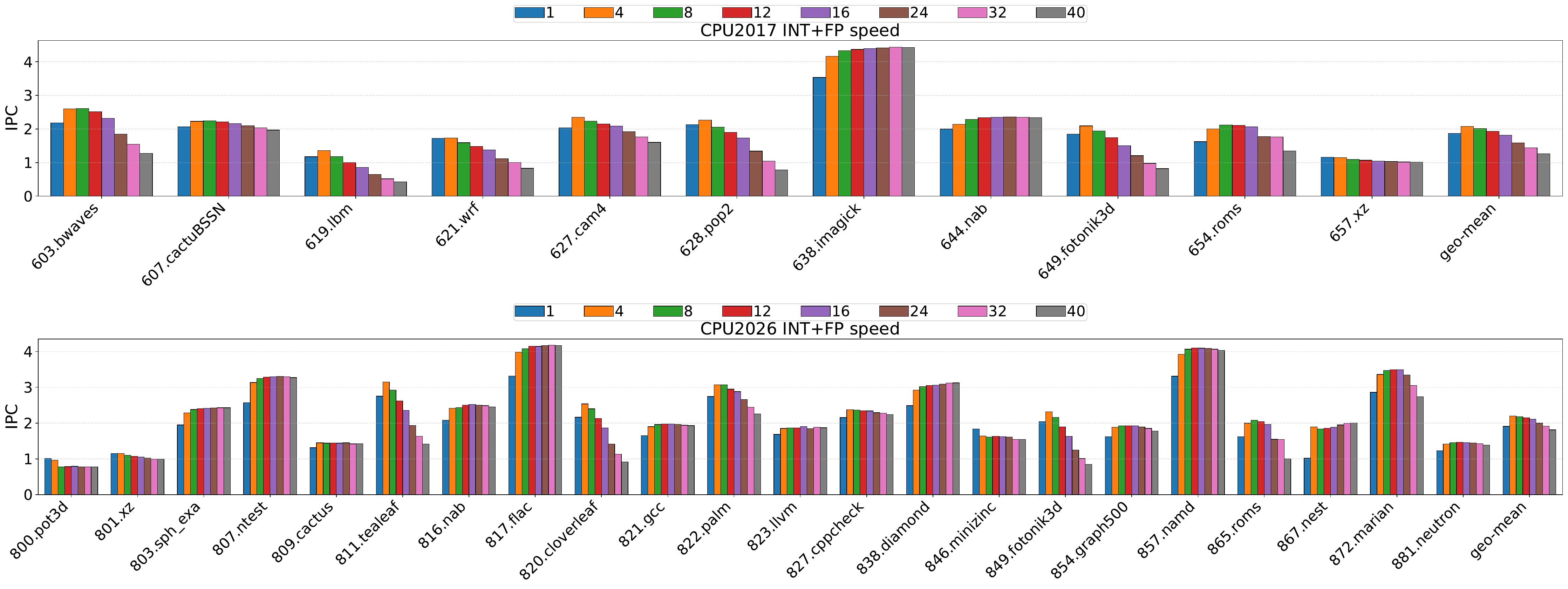}
	\subcaption{IPC of CPU-C. }
	\label{fig_scale_appendix_cpu_c_2}
	\end{minipage} 
    \begin{minipage}{2\columnwidth}
	\centering
	\includegraphics[width=\columnwidth, trim = 2mm 4mm 2mm 2mm, clip=true, page=1]{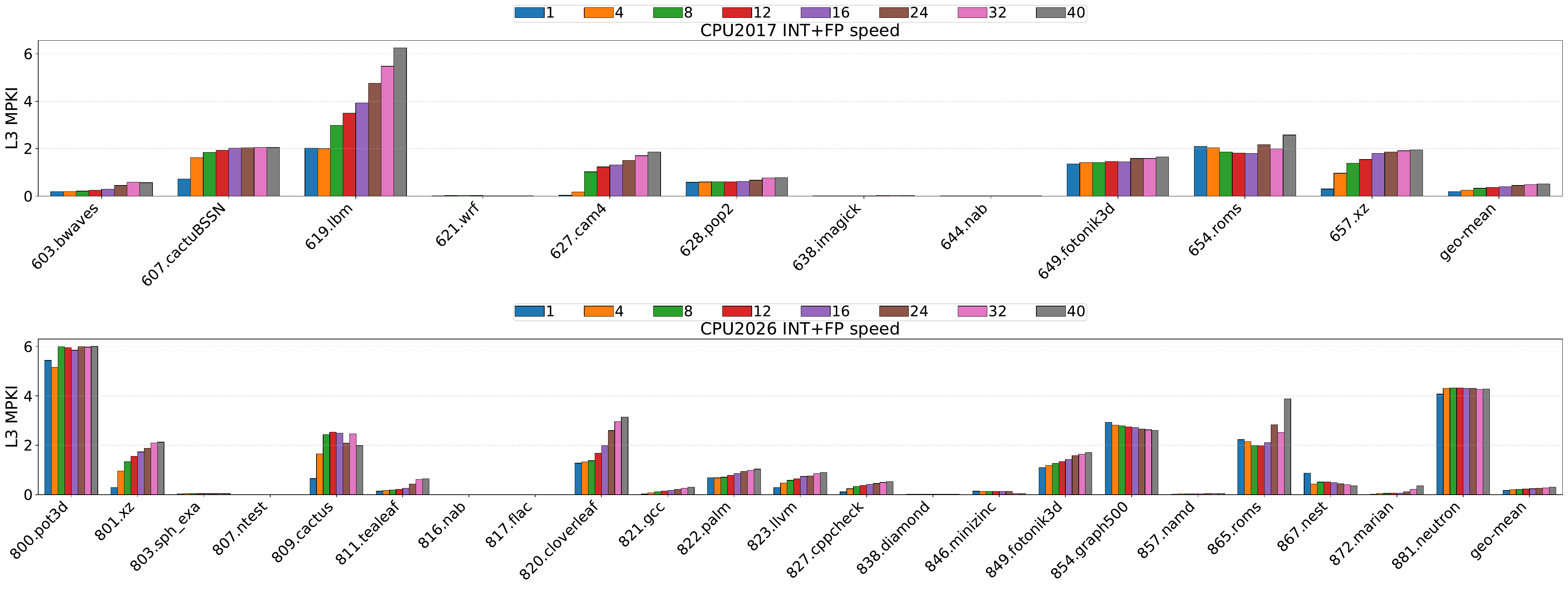}
	\subcaption{L3 cache MPKI of CPU-C.}
	\label{fig_scale_appendix_cpu_c_3}
	\end{minipage} 
 \caption{Detailed per-workload performance of Speed workload scaling on \SPECO and \SPEC (Figure~\ref{fig_speed_scale}, CPU-C).
 }
 \label{fig_scale_appendix_cpu_c}
\end{figure*}

\end{document}